\documentclass[a4paper,aps,prd,superscriptaddress,showkeys,longbibliography,eqsecnum,nofootinbib,11pt]{revtex4-2}
\usepackage[a4paper,margin=1in]{geometry}

\usepackage[english]{babel}
\usepackage[utf8]{inputenc}
\usepackage{amsmath,amssymb,amsthm}
\usepackage{physics,mathtools,tikz,float}
\usepackage{amsfonts}
\usepackage{enumitem}
\usepackage{hyperref}

\usepackage{colortbl}
\definecolor{ddred}{RGB}{198,31,31}
\definecolor{ddgreen}{RGB}{0,160,77}
\hypersetup{
    colorlinks=true,
    linkcolor=ddred,     
    urlcolor=ddred,
    citecolor=ddgreen}

\DeclareMathOperator\arctanh{arctanh}

\theoremstyle{plain}
\newtheorem{theorem}{Theorem}[section]

\newcommand{\supp}
{\operatorname{supp}}
\newcommand{\singsupp}{\operatorname{sing\,supp}}
\newcommand{\WF}{\operatorname{WF}}
\newcommand{\Char}{\operatorname{Char}}

\begin{document}

\title{Bouncing Geodesics, Singularities, \\and the Cavity Thermal Product Formula \\ in  Asymptotically Flat and de Sitter Black Holes}

\author{Sa\v{s}o Grozdanov}
%\email[]{saso.grozdanov@ed.ac.uk}
\affiliation{Faculty of Mathematics and Physics, University of Ljubljana,
Jadranska ulica 19, SI-1000 Ljubljana, Slovenia}
\affiliation{Higgs Centre for Theoretical Physics, University of Edinburgh,
Edinburgh, EH8 9YL, Scotland}

\author{Vita Movrin}
%\email[]{movrin.vita@gmail.com}
\affiliation{Faculty of Mathematics and Physics, University of Ljubljana,
Jadranska ulica 19, SI-1000 Ljubljana, Slovenia}

\author{Samuel Valach}
%\email[]{samuel.valach@fmf.uni-lj.si}
\affiliation{Faculty of Mathematics and Physics, University of Ljubljana,
Jadranska ulica 19, SI-1000 Ljubljana, Slovenia}

\begin{abstract}
We investigate the existence and implications of ``bouncing geodesics'' in asymptotically flat Schwarzschild and Schwarzschild--de Sitter black holes. These trajectories, which probe the high-curvature regions near the black hole singularity, correspond to specific ``bouncing singularities'' in the bulk retarded Green's function. We provide a precise description of these singularities by combining the local Hadamard form with the global propagation of singularities theorem. We then derive the critical times at which the bulk retarded correlator becomes singular, considering all possible anchorings of the bouncing geodesics, including null infinity and the cosmological horizon. Finally, for black holes enclosed in a reflecting cavity, we establish a universal connection between the locations of the bouncing singularities and the spectrum of cavity quasinormal modes (QNMs) by deriving a cavity version of the thermal product formula, analogous to the one known for anti-de Sitter black holes. This relation allows one to extract information about the black hole interior from the asymptotic QNM spectrum measured at a reflecting hypersurface, even when the cosmological constant is zero or positive. We confirm this prediction through explicit examples by computing the cavity QNMs of scalar and electromagnetic fields, as well as gravitational waves, in spacetimes with asymptotically flat and de Sitter black holes.
\end{abstract}

\maketitle

%\newpage

{\hypersetup{linkcolor=black} %force differnt color
\tableofcontents}

%%%%%%%%%%%%%%%%%%%%%%%%%%%%%%%%%%%%%%%%%%%%%%%%%%%%%
%%%%%%%%%%%%%%%%%%%%%%%%%%%%%%%%%%%%%%%%%%%%%%%%%%%%%

\section{Introduction and Summary}
\label{s.intro}

Understanding the physics of black hole interiors, particularly in the vicinity of their curvature singularities, remains one of the major open problems in theoretical physics. Curvature singularities are usually considered to signal the breakdown of classical General Relativity and the standard geometric notions of spacetime, giving way to a more pressing need for a complete theory of quantum gravity. Beyond our desire to theoretically describe such regions of various hypothetical black holes as well as those in our real universe, an even more difficult task may be to devise concrete ``measurable'' signatures of singularities that could be accessible (or inferable) to observers outside their event horizons. 

With the help of AdS/CFT \cite{Maldacena:1997re, Gubser:1998bc, Witten:1998qj}, some progress has been made in answering such questions for black holes in asymptotically anti-de Sitter (AdS) spaces, which are spacetimes with a negative cosmological constant, $\Lambda < 0$. One concrete set of such advances emerged from studying the behaviour of geodesics in two-sided (eternal) AdS black hole geometries, starting with \cite{Fidkowski:2003nf,Festuccia:2005pi}. They observed that certain geodesics can enter the black hole interior, approach and bounce off the curvature singularity, and exit the black hole to the second outer asymptotic region. In the high-energy limit, these geodesics, termed \textit{bouncing geodesics}, become null (lightlike). Their existence was then found to be directly encoded in the analytic structure of the thermal correlation functions in the holographically dual conformal field theory (CFT) \cite{Ceplak:2024bja}. In particular, the existence of such geodesics gives rise to the so-called \textit{bouncing singularities} of retarded thermal correlators located at complexified \textit{bouncing times} \cite{Fidkowski:2003nf,Festuccia:2005pi,Parisini:2023nbd,Ceplak:2024bja,Afkhami-Jeddi:2025wra,Ceplak:2025dds,Dodelson:2025jff,Jia:2025jbi,AliAhmad:2026wem,Jia:2026pmv,Giombi:2026kdz,Arnaudo:2026der,Grozdanov:2026cut,Jia:2026ryl}.

Despite the existence of a number of interesting results, the vast majority of such studies has focused on black holes in AdS. There are of course many reasons for this, most of them related to the absence of a reliable holographic description of black holes in asymptotically flat ($\Lambda= 0$) or asymptotically de Sitter ($\Lambda > 0$) spaces. One reason that a concrete and explicit dual description has been invaluable in such studies is that with the help of AdS holography, the analytic structure of thermal correlation functions in dual large-$N$, strongly-coupled CFTs has been rather thoroughly explored. In particular, we have understood that, in general, (holographic) thermal correlators are meromorphic functions in the complex frequency space \cite{Kovtun:2005ev,Hartnoll:2005ju,Grozdanov:2016vgg,Casalderrey-Solana:2018rle,Grozdanov:2018gfx,Dodelson:2024atp,Dodelson:2023vrw,Grozdanov:2025ulc,Dodelson:2025jff}. Among important advances in understanding the momentum-space structures of holographic thermal correlators dual to wave equations in bulk geometries with $\Lambda < 0$, which rely on this meromorphic structure, is the recently proposed {\em thermal product formula} of \cite{Dodelson:2023vrw}, which then helped establish the connection between bouncing geodesics and the asymptotic (high-overtone) structure of quasinormal modes (QNMs) of momentum space correlators. 

The goal of this work is to extend these studies and establish the existence of bouncing geodesics, explore the analytic structures of correlation functions (bouncing singularities), and derive the thermal product formula for more realistic black holes in asymptotically flat ($\Lambda = 0$) and asymptotically de Sitter ($\Lambda > 0$) black holes. With the help of such a generalised thermal product formula, we can then, in analogy with discussions in AdS, study the signatures of bouncing geodesics as they imprint themselves onto the quasinormal spectra of $\Lambda = 0$ and $\Lambda > 0$ black holes. Clearly, studies of asymptotically flat spaces are relevant in the context of astrophysics. We should note, however, that here, as the first step in this programme, we will only analyse non-rotating, Schwarzschild black holes. De Sitter space, on the other hand, is relevant for any discussion of black holes in the context of cosmology, e.g., primordial black holes, and also for the discussion of the geodesic structure of our late-time (and likely also the early-time) universe. 

More concretely, we show explicitly that, indeed, bouncing geodesics do exist in asymptotically flat and de Sitter black holes. We begin by constructing the relevant families of spacelike geodesics and identifying the null limits corresponding to bouncing trajectories. We then determine the associated bouncing times appearing in bulk retarded propagators for the anchoring configurations relevant to the flat and de Sitter cases, including fixed timelike hypersurfaces, null infinity, and cosmological horizons. We also investigate geodesics bouncing off spatial infinity and stress that these do not generate singularities in the corresponding correlators, highlighting that the analytic structure of (standard thermal) correlators is sensitive specifically to the curvature singularities. By extending our analysis from \cite{Grozdanov:2026cut}, we also show more rigorously that as a result of the Hadamard theory of differential equations and {\em microlocal analysis} that describes the propagation of singularities, bouncing geodesics do indeed imply the existence of bouncing singularities of retarded bulk-to-bulk Green's functions. This framework allows us to analyse the structure of black hole curvature singularities via bouncing geodesics even in spacetimes where no reliable notion of a dual theory exists. 

Unlike in asymptotically AdS spacetimes, here, no canonical (timelike) ``boundary'' exists on which the bouncing geodesics should be anchored. Instead, we may anchor the geodesic at \textit{any} fixed timelike hypersurface. This then naturally leads us to analyse black holes enclosed in reflecting cavities with timelike walls. It is in this setup that we derive a \textit{cavity thermal product formula} analogue of the result from AdS \cite{Dodelson:2023vrw}, but for any value of the cosmological constant. Finally, using this factorisation formula, we establish a general universal relation between the bouncing time $t_*$, which encodes concrete information about the black hole interior, and asymptotic, potentially measurable modes $\omega_n$ of gravitational, electromagnetic, and scalar waves in black hole spacetimes with a reflecting timelike cavity. The result is, in analogy with that in AdS, 
\begin{equation}\label{e.veverica}
    t_* \sim \frac{2\pi n}{\omega_n}, \quad n \to \infty,
\end{equation}
where $\text{Re}(t_*)$ measures twice the coordinate time (at the timelike wall of the cavity) it takes for light to reach from the edge of the spherical cavity to the centre (the singularity) of the black hole. More precisely, up to the overall sign convention, $\text{Re}(t_*)$ is the principal-value part of twice the tortoise-coordinate separation between the cavity wall and the curvature singularity. Moreover, $\text{Im}(t_*)$ measures how many horizons (event and Cauchy horizons) the geodesic crosses on the way to the singularity, measured in units of $\beta/4$, where $\beta = 1/T$ is the inverse Hawking temperature set by the outer (event) horizon. It arises from the residue contributions picked up when the analytically continued geodesic trajectory crosses horizons. 

The paper is organised as follows. In Section~\ref{s.hadsing}, we review the relevant geometric setup governing bouncing geodesics in asymptotically flat Schwarzschild and Schwarzschild–de Sitter spacetimes, and discuss the local Hadamard form and its implications. Sections~\ref{s.AF} and \ref{s.dS} analyse the existence and properties of bouncing geodesics in these spacetimes under all possible anchoring conditions. In Section~\ref{s.QNM}, we derive the cavity thermal product formula and the relation between $t_*$ and asymptotic $\omega_n$. We conclude in Section~\ref{s.disc} with a discussion of the implications of our results. Finally, in Appendix~\ref{a.bending}, we examine the structure of the relevant Penrose diagrams, while in Appendix~\ref{a.propag}, we review the relevant mathematical background needed to state and use the propagation of singularities theorem, which promotes the statement of the local Hadamard theorem to the global structure of thermal correlator singularities.

%%%%%%%%%%%%%%%%%%%%%%%%%%%%%%%%%%%%%%%%%%%%%%%%%%%%%
%%%%%%%%%%%%%%%%%%%%%%%%%%%%%%%%%%%%%%%%%%%%%%%%%%%%%

\section{Bouncing Geodesics and Green's Function Singularities}\label{s.hadsing}

We begin by briefly reviewing the notion of bouncing geodesics and their relation to singularities of the retarded Green's functions.

\subsection{Bouncing Geodesics}

Consider a static black hole metric in $D=d+1$ dimensions of the form
\begin{equation}\label{e.eee}
    \dd s^2=-f(r)\dd t^2+\frac{1}{f(r)}\dd r^2+r^2\dd\mathbf{x}^{\phantom{.}2}.
\end{equation}
In such spacetimes, there exists a Killing vector $\partial_t$ with an associated conserved quantity
\begin{equation}\label{e.endef}
    E\equiv\dot{t}f(r).    
\end{equation}
In addition, any (radial) geodesic has to satisfy
\begin{equation}\label{e.normeq}
    -f(r)\dot{t}^2+\frac{\dot{r}^2}{f(r)}=\epsilon,        
\end{equation}
where $\epsilon =0$, $+1$, $-1$, for null, spacelike and timelike geodesics, respectively. Combining \eqref{e.endef} and \eqref{e.normeq}, we obtain the equation
\begin{equation}\label{e.combinedeq}
    E^2=\dot{r}^2-\epsilon f(r),    
\end{equation}
meaning that we can treat $-\epsilon f(r)$ as a potential $V(r)$ probed by the energy $E^2$. 

The two main quantities of interest related to the geodesics are $t(E)$, which is the elapsed (Schwarzschild-like) coordinate time along a radial geodesic starting at the point $r=r_1$ and ending at $r=r_2$, and, for spacelike (respectively, timelike) geodesics, the corresponding proper length  $\mathcal{L}(E)$. The coordinate-time separation can be read from the ratio between $\dot{t}$ and $\dot{r}$, which leads to
\begin{equation}\label{e.tgeninte}
    t(E)=\pm\int_{r_1}^{r_2}\frac{E\,\dd r}{f(r)\sqrt{E^2+\epsilon f(r)}},
\end{equation}
where the sign is chosen by ``physical considerations''. Whenever the integration contour crosses a singularity of the integrand, $t(E)$ acquires half of the corresponding residue, while the real part is computed as the principal value. This naturally introduces the notion of the complexified Schwarzschild coordinates that cover the whole maximally extended spacetime. 

The second quantity of interest to our analysis is the proper length
\begin{equation}\label{e.lenghtgen}
    \mathcal{L}(E)=\pm\int_{r_1}^{r_2}\frac{\epsilon^2\,\dd r}{\sqrt{E^2+\epsilon f(r)}}=\pm2\int_{r_{i}}^{r_T}\frac{\epsilon^2\,\dd r}{\sqrt{E^2+\epsilon f(r)}}.
\end{equation}
Since we will focus on spacelike geodesics that connect separate exterior regions of two-sided black holes, in the second equality, we assumed that the geodesics are left-right symmetric (in a Penrose diagram), and we introduced the turning point $r_T=r_T(E)$ defined as the largest real root of
\begin{equation}\label{e.solveforrt}
    E^2+\epsilon f(r)=0.
\end{equation}
We denoted the ``anchoring point'' (the initial radial coordinate) of the geodesic by $r_{i}$. Importantly, if the integral \eqref{e.lenghtgen} diverges for some choice of $r_{i}$, one has to renormalise the length, which is conventionally done by using the ``minimal subtraction'' prescription (see \cite{Ceplak:2024bja}). We will employ the same strategy in this work. 

Now, following \cite{Grozdanov:2026cut}, let us state the following definition:\\

\noindent\textit{A bouncing geodesic is a null limit of a spacelike or timelike geodesic that approaches the curvature singularity from a finite distance, comes infinitesimally
close to it, and then moves finitely far away from it (i.e., it ``bounces off
the curvature singularity'').}\\

\begin{figure}[ht]
    \centering
    \includegraphics[scale=1.17]{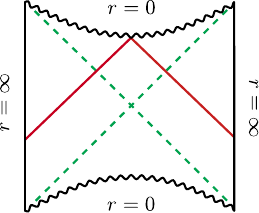}
        \caption{Penrose diagram for the maximally extended AdS-black brane in $D>3$.}
    \label{f.AdS}
\end{figure}

Such geodesics were first discussed in the context of the two-sided AdS-black brane in \cite{Fidkowski:2003nf,Festuccia:2005pi} where it was observed that null limits of a spacelike geodesic can traverse the bulk, ``reflect off'' high-curvature regions, and end at the boundary of the second exterior region of the black brane, see Fig.~\ref{f.AdS}.
Within the large probe mass regime, it was conjectured that such a geodesic would leave certain signatures on the analytic structure of the dual Wightman function -- the so-called ``bouncing singularities'' located on the second sheet of the Wightman function at complex times\footnote{For clarity of notation we will use the superscript indices $\Lambda<0$, $\Lambda=0$ and $\Lambda>0$ for the cases of asymptotically AdS, flat and dS spacetimes.}
\begin{equation}\label{e.needlater}
    t_*^{\Lambda<0}(D)=i\frac{\beta}{2}+\frac{\beta}{2}\cot\left(\frac{\pi}{1-D}\right).
\end{equation}
This is the Schwarzschild coordinate time along the bouncing geodesic, or equivalently, the boundary time difference between the two operator insertions in the boundary CFT. The complex part in \eqref{e.needlater} comes from crossing the black hole horizons.

More recently, in \cite{Parisini:2023nbd,Ceplak:2024bja}, it was found that these ``bouncing singularities'' are directly visible on the principal sheet of a particular sector of the dual propagator for any finite conformal operator dimension (bulk probe mass) and can be effectively analysed using the operator product expansion (OPE) techniques \cite{Parisini:2023nbd,Ceplak:2024bja,Valach:2025saf,Ceplak:2025dds,Araya:2026shz}.\footnote{For more details on these holographic computations and the OPE techniques, see also 
\cite{Fitzpatrick:2019zqz,Karlsson:2022osn,Huang:2022vet,Esper:2023jeq}.} Since then, various other approaches were used to examine these bouncing singularities, using momentum space approaches and the WKB approximation \cite{Afkhami-Jeddi:2025wra,Jia:2025jbi,AliAhmad:2026wem,Jia:2026ryl}, bulk phase-shift computations \cite{Jia:2026pmv} and properties of quasinormal modes \cite{Dodelson:2025jff}, see also \cite{Buric:2025anb,Buric:2025fye,Barrat:2025twb,Giombi:2026kdz,Arnaudo:2026der,Grozdanov:2026cut,futurepap3,futureHONG} for relevant discussion and applications. 

So far, the majority of such investigations have focused on asymptotically AdS spacetimes black holes.\footnote{Past studies of bouncing geodesics and bouncing singularities for a particular fixed anchoring were presented in \cite{Faruk:2023uzs,Faruk:2025bed} for the Schwarzschild-de Sitter black hole, and in \cite{Arnaudo:2026tcy} for the asymptotically flat Schwarzschild black hole.} This is because only AdS holography has been convincingly established. Very recently, however, in our \cite{Grozdanov:2026cut}, we showed that the existence of a dual is not necessary for the notion of bouncing singularities or for utilising these bouncing geodesics as probes of curvature singularities. In fact, using the local Hadamard form of Green’s functions, \cite{Grozdanov:2026cut} provided a general framework in which these signatures of the bulk curvature singularities can be examined for any spacetime. We now review this formalism, make it more rigorous and complete, and then in Sections~\ref{s.AF} and \ref{s.dS} apply it to examinations of asymptotically flat and de Sitter black holes.

\subsection{Singularities of the Retarded Green's Function}

The existence of bouncing geodesics and the relation between null geodesics and singularities of hyperbolic equations in curved spacetimes allow for a  precise identification of the singularities of retarded (bulk-to-bulk) Green's functions $G(x,x')$ in position space \cite{Grozdanov:2026cut}. Since this fact will be used throughout the rest of the paper, we briefly outline the general theory that leads to this relation.

The retarded Green's function $G(x,x')$ can be interpreted as describing the response at $x$ to a point source inserted at $x'$. Understanding where $G(x,x')$ fails to be smooth therefore amounts to identifying how singular signals propagate through spacetime. From the theory of wave propagation, one expects such singularities to travel along lightlike trajectories, i.e., along null geodesics. We will make this expectation precise by showing that $G(x,x')$ is singular whenever $x$ can be reached from $x'$ by a future-directed null geodesic. The argument proceeds in two steps. First, the \textit{local Hadamard form} establishes that singularities occur on the future lightcone of $x'$ in a sufficiently small neighbourhood of $x'$. Second, a result from microlocal analysis, the \textit{propagation of singularities theorem}, ensures that these singularities propagate along null geodesics throughout the whole spacetime.

Consider a Lorentzian manifold $M$ with a metric tensor $g_{\mu\nu}$ (abbreviated by $(M, g)$), and a \textit{normally hyperbolic operator}, that is, a second order differential operator of the form
\begin{equation}
\label{eq: normally hyperbolic}
   P = \Box_g + a^\mu(x) \nabla_\mu + b(x).
\end{equation}
Here, $\Box_g = g^{\mu \nu}(x) \nabla_{\mu}\nabla_{\nu}$ denotes the d'Alembert operator for the metric $g$, while $a^\mu(x)$ is a smooth vector field and $b(x)$ a smooth function on $M$. The \textit{retarded Green's function} for $P$ is a distribution $G(x,x')$ satisfying
 \begin{equation}
     P_x G(x, x') = \frac{\delta(x -x')}{\sqrt{-g}} \quad \text{ with } \quad \supp \left(G(\cdot,x') \right) \subseteq J^+(x'), \quad \forall x' \in M,
 \end{equation}
 where $J^+(x') \subseteq M $ is the \textit{causal future} of $x' \in M$, that is, the set of all points in $M$ that can be reached along future-directed causal curves starting at $x'$.

The seminal work by Hadamard \cite{hadamard1923lectures} showed that the retarded Green's function admits a particular representation on causal domains, known as the \textit{local Hadamard form}. Here, we state this result by rephrasing the Theorem 6.2.1 from \cite{Friedlander:2010eqa}. See also \cite{Grozdanov:2026cut} for a recent (less formal) discussion. First, recall that the \emph{causal domain} is an open set $\Omega \subset M$ for which the following two conditions hold:
\begin{enumerate}[label=\roman*)]
    \item $\Omega$ is contained in a geodesically convex open set $\tilde \Omega \subset M$, meaning that any two points in $\tilde \Omega$ are connected by a unique geodesic which lies entirely in $\tilde \Omega$.

    \item For all $x,y \in \Omega$, the set $J^+(x) \cap J^-(y)$ is a compact subset of $\Omega$. This ensures that causal curves between points remain in a compact subset of $\Omega$.
\end{enumerate}
The following theorem can then be stated.
\begin{theorem}
\label{thm: hadamard}
Let $(M, g)$ be a $D$-dimensional Lorentzian manifold, $P$ a normally hyperbolic operator on $M$ and $G(x, x')$ the retarded Green's function for $P$. Let $\Omega \subseteq M$ be a causal domain and $x' \in \Omega$. Denote by $\sigma(x, x')$ the Synge's world function, which is one half of the squared geodesic distance between two points $x, x' \in \Omega$. Then, for $x\in\Omega$, $G(x,x')$ has the following form:
\begin{enumerate}[label=\roman*)]
\item If $D=2n+2$ with $n \in \mathbb{N}$, $ n \geq 1$, 
\begin{equation}
    G(x, x')
=
\frac{1}{2\pi^n}
\left(
\sum_{\nu=0}^{n-1}
U_\nu(x,x')\,\delta^{(n-\nu-1)}_{+}(\sigma(x,x'))
+
V(x,x')
\right).
\end{equation}
Here, $U_\nu(\cdot, x')$ are smooth functions supported in $\Omega$, determined recursively by transport equations along the unique geodesic connecting $x'$ and $x$, with $U_0(x',x')=1$.\footnote{In case $a^\mu \equiv 0$ in \eqref{eq: normally hyperbolic}, we have $U_0(x, x') = \Delta^{1/2}(x, x')$ where $\Delta(x, x')$ is the \textit{van Vleck–Morette determinant}. See Section~6.2 of \cite{Friedlander:2010eqa} for more details.} We denote by $\delta_+(\sigma(x, x'))$ the Dirac delta (distribution) supported in the future lightcone of $x'$,
\begin{equation}
   \Sigma^+_{x'}\equiv\{x\in \Omega \cap J^+(x') ~|~ \sigma(x,x')=0\},
\end{equation}
and by $\delta^{(k)}_+(\sigma(x, x'))$ its $k$-th derivative in the normal direction to the lightcone. Moreover, $V(\cdot,x')$ is smooth and supported in $\Omega\cap J^+(x')$.

\item If $D=2n+1$ with $n \in \mathbb{N}$, $ n \geq 1$, 
\begin{equation}\label{e.zelena_zelva}
   G(x, x')
=
\frac{1}{2\pi^{n-\frac12}}
\,W(x,x')\,\sigma(x,x')^{\frac12-n}, 
\end{equation}
where $W(\cdot,x')$ is smooth and supported in $\Omega\cap J^+(x')$.
\end{enumerate}
In particular, within the causal domain $\Omega$, the singularities of $G(\cdot,x')$ occur on $\Sigma^+_{x'}$, i.e.~on the future lightcone of $x'$. 
\end{theorem}
The Hadamard form makes the singular structure of $G(x, x')$ completely explicit in a local neighbourhood of $x'$. In particular, since $\sigma(x,x')=0$ characterises points connected to $x'$ by null geodesics within $\Omega$, it follows that, locally, the singular support of the Green's function $G(x,x')$ lies on the future-directed null cone emanating from $x'$. However, this result relies on the existence of a geodesically convex neighbourhood in which geodesics are unique. In the black hole geometries of interest, the relevant null curves can pass through horizons, approach the singularity, and connect different exterior regions. To make a general statement about the singularity structures, we therefore require a similar, but global statement.

This is provided by the propagation of singularities theorem \eqref{thm: propagation of singularities}, which is a result of microlocal analysis due to Duistermaat and Hörmander \cite{Duistermaat:1972}. A precise formulation, together with the necessary background, is reviewed in Appendix~\ref{a.propag}. In the present setting, its physical interpretation is that the singular wave fronts of the Green's function propagate precisely along null geodesics. Equivalently, once the Green's function has a lightcone singularity in a given null direction near the source, that singularity is transported along the corresponding null geodesic. Combining the local Hadamard form with the propagation of singularities theorem then gives the following sought global statement:

\begin{theorem}
\label{eq: thm global}
Let $(M,g)$ be a Lorentzian manifold and $x' \in M$. If there exists a future-directed null geodesic from $x'$ to some $x \in M$, then the retarded Green's function $G(\cdot,x')$ of a normally hyperbolic operator is singular at $x$. That is,
\begin{equation*}
      \{ x \in M \mid \text{there exists a future-directed null geodesic from } x' \text{ to } x \} \subseteq \singsupp G(\cdot,x').
\end{equation*}
\end{theorem}

%%%%%%%%%%%%%%%%%%%%%%%%%%%%%%%%%%%%%%%%%%%%%%%%%%%%%
%%%%%%%%%%%%%%%%%%%%%%%%%%%%%%%%%%%%%%%%%%%%%%%%%%%%%

\section{Bouncing Geodesics in Asymptotically Flat Spacetime}\label{s.AF}

Based on the existence theorems for the bouncing geodesics \cite{Grozdanov:2026cut}, we expect bouncing geodesics to exist in the asymptotically flat Schwarzschild black hole spacetime. Since the asymptotically flat Schwarzschild Penrose diagram, as reviewed in Appendix~\ref{a.bending}, has no relative bending of the singularity with respect to null infinity, the resulting bouncing times take a rather simple form. By the local Hadamard form \eqref{thm: hadamard} together with the propagation of singularities, the limiting null bouncing geodesic therefore gives rise to a corresponding singularity of the bulk retarded Green's function. We discuss this in Subsections~\ref{ss.AFE} and \ref{ss.AFgen}.

Before deriving the concrete prescription for the bouncing geodesic and the corresponding bouncing singularity, let us discuss an important subtlety that one encounters in the asymptotically flat case. Due to the form of the asymptotically flat Penrose diagram, one cannot naturally anchor the geodesic at the ``edges of the Penrose diagram'' in the same way as in asymptotically AdS spacetimes. The reason is that null infinity is a null surface, reached only in a limit in which the Schwarzschild coordinate time diverges ($t=\infty$). Moreover, unlike in the AdS case, pushing spacelike geodesic endpoints to asymptotic infinity at finite coordinate time naturally sends them to spatial infinity $i^0$, which is not the appropriate anchoring point for the bouncing problem. Despite these issues, we will construct a (``holographically'') renormalised bouncing geodesic anchored at the null boundary $\mathcal{J^-}$ in Subsection~\ref{ss.AFinfty}.

\subsection{Existence of Bouncing Geodesics}\label{ss.AFE}

Start with the asymptotically flat Schwarzschild metric in $D$-dimensions, which takes the form \eqref{e.eee} with the blackening factor
\begin{equation}
    f(r)=1-\frac{1}{r^{D-3}},
\end{equation}
where we set the black hole horizon radius to 1. The corresponding Hawking temperature is
\begin{equation}\label{e.hawkiaf}
    T=\frac{D-3}{4\pi} .
\end{equation}
We first explicitly probe and construct the bouncing geodesic in this spacetime and then compute the location of the corresponding bouncing singularity in the bulk Green's function.

Consider a symmetric spacelike geodesic in $D\geq4$ anchored at $r=r_i<\infty$. Its turning point is
\begin{equation}\label{e.flatturpoii}
    r_T=\left(E^2+1\right)^{\frac{1}{3-D}} .
\end{equation}
Since the case $D=4$ is special, one has to consider cases $D=4$ and $D>4$ separately:

\paragraph{$D=4$:}

In four dimensions, the integral \eqref{e.lenghtgen} is
\begin{equation}\label{e.JJL4}
    \mathcal{L}(E,r_i)=\frac{2\left(r_i\sqrt{1+E^2}\sqrt{1+E^2-\frac{1}{r_i}}+\arctanh\left(\frac{\sqrt{1+E^2-\frac{1}{r_i}}}{\sqrt{1+E^2}}\right)\right)}{\left(1+E^2\right)^{3/2}}.
\end{equation}
Taking the large-$E$ limit, the proper length can be expanded as 
\begin{equation}
    \mathcal{L}(E,r_i)=\frac{2 r_i}{E}-\frac{1}{E^3}\left[ 1+r_i-2 \ln(E)-\ln(4 r_i)\right]+\ldots\xrightarrow[]{E\to\infty}0,
\end{equation}
which shows that in the limit, the spacelike (left-right) symmetric geodesic becomes null. Moreover, from the form of the turning point \eqref{e.flatturpoii}, it is clear that, in this limit, the geodesic approaches the curvature singularity at $r=0$. Thus, the limiting null geodesic bounces off the singularity and satisfies the definition of the bouncing geodesic \cite{Grozdanov:2026cut}.

\paragraph{$D>4$:}
Now assume any dimension higher than four, in which case the proper length Eq.~\eqref{e.lenghtgen} yields
\begin{equation}\label{e.JJLH}
\mathcal{L}(E,r_i)=\frac{2 \left[r_i \, _2F_1\left(\frac{1}{2},\frac{1}{3-D};1+\frac{1}{3-D};\frac{r_i^{3-D}}{E^2+1}\right)-\left(E^2+1\right)^{\frac{1}{3-D}} \!\,
   _2F_1\left(\frac{1}{2},\frac{1}{3-D};\frac{D-4}{D-3};1\right)\right]}{\sqrt{E^2+1}},
\end{equation}
with the large-energy expansion given by
\begin{equation}
\begin{split}
    \mathcal{L}(E,r_i)=&\Bigg(\frac{2 r_i}{E}-\frac{\frac{r_i^{4-D}}{D-4}+r_i}{E^3}+\mathcal{O}(E^{-5})\Bigg)+E^{\frac{1-D}{D-3}}\Bigg(-2 \,
   _2F_1\left(\frac{1}{2},\frac{1}{3-D};\frac{D-4}{D-3};1\right)\\
   &+\frac{(D-1) \, _2F_1\left(\frac{1}{2},\frac{1}{3-D};\frac{D-4}{D-3};1\right)}{(D-3)
   E^2}+\mathcal{O}(E^{-4})\Bigg)\xrightarrow[]{E\to\infty}0.
\end{split}
\end{equation}
Together with $r_T \to 0$, this shows the existence of the bouncing geodesic also when $D>4$. 

In summary, we conclude that asymptotically flat Schwarzschild black holes admit bouncing geodesics. We now turn to the investigation of the corresponding bouncing times and Green's function singularities.

\subsection{Connecting Generic Points}\label{ss.AFgen}

To obtain the location of the bouncing singularity, we need to compute the ``bouncing time'', that is, take the large-energy limit of the integral \eqref{e.tgeninte}. We first present the general-$D$ computation and then discuss the most interesting cases of $D=4$ and $D=5$ as special examples.

\paragraph{General Dimension:}

Consider the integral \eqref{e.tgeninte} for general $D$, which gives
\begin{equation}\label{e.Dmore5tintflat}
\begin{split}
    t(E,r_i)=&-\frac{4 \left(E^2+1\right)^{-\frac{D-2}{D-3}} \sqrt{-r_i^{3-D}+E^2+1}}{(D-3) E} \Bigg[\! - E^2 {}_2F_1\bigg(\frac{1}{2},1+\frac{1}{D-3};\frac{3}{2};1-\frac{r_i^{3-D}}{E^2+1}\bigg)\\
   &+\left(E^2+1\right)
   \text{F}_1\bigg(\frac{1}{2};\frac{1}{D-3},1;\frac{3}{2};1-\frac{r_i^{3-D}}{E^2+1},\frac{-r_i^{3-D}+E^2+1}{E^2}\bigg)\Bigg],
\end{split}
\end{equation}
where $\text{F}_1$ is the Appell hypergeometric function. Taking the large-$E$ expansion of Eq.~\eqref{e.Dmore5tintflat}, we can compute the bouncing time $t_*^{\Lambda=0}(r_i)$ as the leading term in $E\to\infty$:
\begin{equation}\label{e.flattstarhidi}
t_*^{\Lambda=0}(r_i)=-\frac{2}{D-2} r_i^{D-2} \, _2F_1\left(1,1+\frac{1}{D-3};2+\frac{1}{D-3};r_i^{D-3}\right).
\end{equation}
By the local Hadamard form \eqref{thm: hadamard}, this is the location of the bouncing singularity in the (bulk) Green's function.
\begin{figure}[ht]
    \centering
    \includegraphics[scale=1.07]{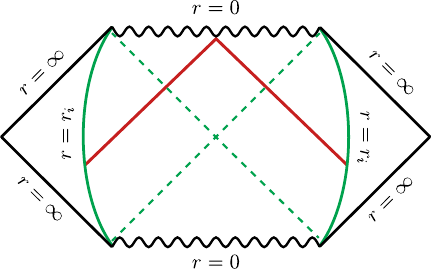}
        \caption{Penrose diagram for the asymptotically flat Schwarzschild black hole with the bouncing geodesic (in red) anchored at $r=r_i$ fixed timelike surfaces (green solid lines).}
    \label{f.Aflat_penrose}
\end{figure}
Also note that, as expected, if $r_i>1$, that is, if the geodesic crosses the two horizons in the maximally extended geometry, one obtains a constant imaginary shift $\Im(t_*^{\Lambda=0}(r_i))=\beta/2$, coming from the hypergeometric function. Note that we conventionally choose the sign of the imaginary term so that it is positive when the horizon is crossed in the anticlockwise direction (as depicted in the Penrose diagram). For clarity, let us now discuss this computation in $D=4$ and $D=5$ in more detail.

\paragraph{Examples:}

In $D=4$, we find
\begin{equation}\label{e.JJt4}
\begin{split}
    t(E,r_i)=&-4\,\text{arccoth}\left(\frac{E}{\sqrt{E^2-\frac{1}{r_i}+1}}\right)+\frac{2E}{\left(E^2+1\right)^{3/2}} \bigg[\sqrt{\left(E^2+1\right) r_i \left(E^2 r_i+r_i-1\right)}\\
    &+\left(2E^2+3\right)\arctanh\left(\sqrt{1-\frac{1}{E^2
   r_i+r_i}}\right)\bigg],
\end{split}
\end{equation}
which, in the large-energy limit, gives
\begin{equation}
    t(E,r_i)=t_*^{\Lambda=0}(r_i)-\frac{r_i}{E^2}+\frac{-12 \ln (E)+6 r_i-6 \ln (r_i)+7-12 \ln 2}{8 E^4}+\ldots,
\end{equation}
where the bouncing time is
\begin{equation}\label{e.tsD4}
    t_*^{\Lambda=0}(r_i)=2 \left(\ln(1-r_i)+r_i\right)\qq{for}D=4.
\end{equation}
If $r_i>1$, we find the imaginary shift $\Im(t_*^{\Lambda=0}(r_i))=\beta/2$, which arises from the logarithmic term.

\begin{figure}[ht]
    \centering
    \includegraphics[scale=1.10]{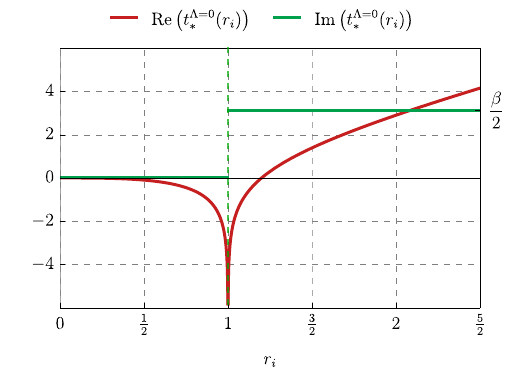}
        \caption{The behaviour of $t_*^{\Lambda=0}$ as a function of $r_i$ in $D=5$ for $\mu=1$. For $r_i\to0$, the bouncing time goes to 0, since the integration interval vanishes. The nontrivial radius for which the real part of $t_*^{\Lambda=0}(r_i)$ vanishes corresponds to the case when the Penrose diagram for the cavity forms a perfect square. The function diverges at the location of the black hole horizon $r_i=1$, since at this point, the anchoring surface becomes null. The imaginary part of $t_*^{\Lambda=0}(r_i)$ is constant and counts how many times the geodesic crosses a horizon.}
    \label{f.AFLATri}
\end{figure}

For $D=5$, one gets
\begin{equation}\label{e.JJt5}
    t(E,r_i)=\frac{2 E \sqrt{\left(E^2+1\right) r_i^2-1}}{E^2+1}-2 \coth ^{-1}\left(\frac{E}{\sqrt{\left(E^2+1\right) r_i^2-1}}\right),
\end{equation}
which gives the large-$E$ expansion of the concrete form
\begin{equation}
    t(E, r_i)=t_*^{\Lambda=0}(r_i)-\frac{r_i}{E^2}+\frac{3 \left(1+r_i^2\right)}{4 E^4 r_i}+\ldots,
\end{equation}
and the bouncing time reads
\begin{equation}
    t_*^{\Lambda=0}(r_i)=2 \left(r_i-\arctanh\left(r_i\right)\right)\qq{for}D=5.
\end{equation}
Again, for $r_i < 1$, the geodesic lies entirely within the black hole interior (i.e., no horizon is crossed) and the bouncing time is real. For $r_i>1$, we get an imaginary shift $\Im(t_*^{\Lambda=0}(r_i))=\beta/2$, as expected. The behaviour of the bouncing time in this case is plotted in Fig.~\ref{f.AFLATri}.

\subsection{Connecting Null Infinities}\label{ss.AFinfty}

To explore all possible anchoring points, we now consider the limit $r_i\to\infty$ and anchor the bouncing geodesic at the past null infinity $\mathcal{J^-}$. Since this is a null surface, we have to renormalise the geodesic integrals; we do this by a direct minimal subtraction. As we will see, the four- and five-dimensional cases are special, hence, we treat the cases with $D=4$, $D=5$ and $D>5$ separately.

\paragraph{$D=4$:}

Taking the $r_i\to\infty$ limit in \eqref{e.JJL4} and \eqref{e.JJt4}, one gets
\begin{align}
    \mathcal{L}(E,r_i\to\infty)&=\frac{2 r_i}{\sqrt{E^2+1}}+\frac{\ln (r_i)}{\left(E^2+1\right)^{3/2}}+\mathcal{L}_{\mathcal{J}}(E)+\mathcal{O}\Big(\frac{1}{r_i}\Big),\\
    t(E,r_i\to\infty)&=\frac{2 E r_i}{\sqrt{E^2+1}}+\frac{\left(2 E^3+3 E\right) \ln (r_i)}{\left(E^2+1\right)^{3/2}}+t_{\mathcal{J}}(E)+\mathcal{O}\Big(\frac{1}{r_i}\Big),
\end{align}
where the finite parts are
\begin{align}
    \mathcal{L}_{\mathcal{J}}(E)&=\frac{\ln \left(E^2+1\right)-1+2\ln 2}{\left(E^2+1\right)^{3/2}},\\
    t_{\mathcal{J}}(E)&=i\frac{\beta}{2}-4 \text{arcsinh}(E)+\frac{E \left(4 E^2 \ln 2-1+6 \ln 2\right)}{\left(E^2+1\right)^{3/2}}+\frac{E \left(2 E^2+3\right) \ln \left(E^2+1\right)}{\left(E^2+1\right)^{3/2}}.
\end{align}
In the large-$E$ limit, we find 
\begin{equation}
    \lim_{E\to\infty}\mathcal{L}_{\mathcal{J}}=0\qq{and}\lim_{E\to\infty}t_{\mathcal{J}}=i\frac{\beta}{2}.
\end{equation}

\paragraph{$D=5$:}

In the five-dimensional case, we find an even simpler large-$r_i$ expansions:
\begin{align}
    \mathcal{L}(E,r_i\to\infty)&=\frac{2 r_i}{\sqrt{E^2+1}}+\mathcal{O}\Big(\frac{1}{r_i}\Big),\\
    t(E,r_i\to\infty)&= \frac{2 E r_i}{\sqrt{E^2+1}}+i\frac{\beta}{2}+\mathcal{O}\Big(\frac{1}{r_i}\Big).\label{e.nut_called_coconut}
\end{align}
Thus, extracting the parts finite in $r_i$, we find $\mathcal{L}_{\mathcal{J}}(E)=0$ and $\text{Re}(t_{\mathcal{J}}(E))=0$ for any value of the energy $E$.

\paragraph{$D>5$:} For any higher dimension, we find that, in general, the finite parts have a non-zero large-$E$ limit. The integrals \eqref{e.JJLH} and \eqref{e.Dmore5tintflat} become
\begin{align}
    \mathcal{L}(E,r_i\to\infty)&=\frac{2 r_i}{\sqrt{E^2+1}}+\mathcal{L}_{\mathcal{J}}(E)+\mathcal{O}\Big(\frac{1}{r_i}\Big),\\
    t(E,r_i\to\infty)&= \frac{2 E r_i}{\sqrt{E^2+1}}+t_{\mathcal{J}}(E)+\mathcal{O}\Big(\frac{1}{r_i}\Big),
\end{align}
where
\begin{align}
    \mathcal{L}_{\mathcal{J}}(E)&=-\frac{2 \sqrt{\pi } \left(E^2+1\right)^{\frac{1}{3-D}-\frac{1}{2}} \Gamma \left(\frac{D-4}{D-3}\right)}{\Gamma \left(\frac{1}{2}+\frac{1}{3-D}\right)},\\
    t_{\mathcal{J}}(E)&=-\frac{2 \pi ^{3/2} \left(E^2\!+\!1\right)^{\frac{1}{3-D}-\frac{1}{2}} \csc \left(\frac{\pi }{D-3}\right) \left[ 2 \left(E^2\!+\!1\right) \,
   _2F_1\left(\frac{1}{2},1;\frac{3}{2}\!+\!\frac{1}{3-D};1\!+\!\frac{1}{E^2}\right)\!+\!(D-5) E^2\right]}{(D-5) E\, \Gamma \left(\frac{1}{2}+\frac{1}{3-D}\right) \Gamma
   \left(\frac{1}{D-3}\right)}.
\end{align}
Taking the large-energy limit, we find $\mathcal{L}_{\mathcal{J}}\to0$ and
\begin{equation}
    t_{\mathcal{J}}(E)=t^{\Lambda=0}_{**}(D)+\frac{1}{E^{\frac{2}{D-3}}}\left[\frac{2 \pi ^{3/2} \csc\! \left(\frac{\pi }{D-3}\right) \left(D-\frac{\pi  (D-5) \csc \left(\frac{\pi }{3-D}\right)}{\Gamma \left(1+\frac{1}{D-3}\right) \Gamma
   \left(\frac{1}{3-D}\right)}-5\right)}{(5-D) \Gamma \left(\frac{1}{2}+\frac{1}{3-D}\right) \Gamma \left(\frac{1}{D-3}\right)}+\mathcal{O}\left(\frac{1}{E^2}\right)\right],
\end{equation}
where $t^{\Lambda=0}_{**}(D)$ is the analogue of the bouncing time for the anchoring at the null infinity $\mathcal{J}^-$ and has the explicit form
\begin{equation}\label{e.sushi_burito}
    t^{\Lambda=0}_{**}(D)=\frac{\beta}{2}\,i+\frac{\beta}{2}\cot \left(\frac{\pi }{3-D}\right).
\end{equation}
Interestingly, we note that this expression for the bouncing time $t_{**}^{\Lambda=0}(D)$ resembles the bouncing time of the asymptotically AdS black brane \eqref{e.needlater}. Formally, one can write the following relation between them:
\begin{equation}\label{e.struklji_is_good_for_you}
    t_{**}^{\Lambda=0}(D+2)=t_*^{\Lambda<0}(D)\qq{for}D>2.
\end{equation}
It would be interesting to explore the possible physical significance of this intriguing relation, if such an interpretation exists.

%%%%%%%%%%%%%%%%%%%%%%%%%%%%%%%%%%%%%%%%%%%%%%%%%%%%%
%%%%%%%%%%%%%%%%%%%%%%%%%%%%%%%%%%%%%%%%%%%%%%%%%%%%%

\section{Bouncing Geodesics in Asymptotically de Sitter Spacetime}\label{s.dS}

In this section, we investigate bouncing geodesics in the Schwarzschild-de Sitter spacetime, which, like in flat space, also give rise to singularities of retarded Green's functions \cite{Grozdanov:2026cut}. Related analyses were initiated in \cite{Faruk:2023uzs} and \cite{Faruk:2025bed} for a particular choice of anchoring. Here, we generalise those analyses and investigate the geodesics and the resulting bouncing times for different possible anchoring choices.

\subsection{Review of the Schwarzschild-de Sitter Black Hole}\label{ss.dSE}

The Schwarzschild-de Sitter (SdS) black hole spacetime in $D$ spacetime dimensions has
\begin{equation}
    f(r)=1-\frac{r^2}{L^2}-\alpha\frac{M}{r^{D-3}},\qq{with}\alpha=\frac{16\pi G}{(D-2)\Omega_{D-2}},
\end{equation}
where $\Omega_{D-2}=2\pi^{\frac{D-1}{2}}/\Gamma(\frac{D-1}{2})$ is the volume of a $(D-2)$-dimensional unit sphere. Henceforth, we set $L=1$ and define $\mu\equiv\alpha M$. Therefore,
\begin{equation}\label{e.genDSdSmetric}
f(r)=1-r^2-\frac{\mu}{r^{D-3}}.   
\end{equation}

For conciseness, we now mainly focus on the $D=5$ case, for which $f(r)=1-r^2-\mu/r^2$. This geometry has two horizons: the black hole event horizon at $r_b$ and the cosmological horizon at $r_c$, which are located at
\begin{equation}\label{e.DJelectro}
    r_{b,c}=\frac{1}{\sqrt{2}}\sqrt{1\mp\sqrt{1-4\mu}}.
\end{equation}
This implies the following useful relations:
\begin{equation}
    r_b r_c = \sqrt{\mu}, \quad r_c^2 + r_b^2 = 1, \quad r_c^2 - r_b^2 = \sqrt{1-4\mu}.
\end{equation}

We now discuss the necessary normalisation of the Killing vector and the associated surface gravities. The static patch admits a timelike Killing vector field $K = \gamma \frac{\partial}{\partial t}$, where $\gamma$ fixes the normalisation of the static time. As explained in the appendix of \cite{Bousso:1996}, to obtain the correct value for surface gravity, 
\begin{equation}
    \kappa = \lim_{r \to r_b} \left( \frac{(K^\mu \nabla_\mu K_\nu) (K^\rho \nabla_\rho K^\nu)}{-K^2}\right)^{1/2},
\end{equation}
one must carefully normalise the Killing vector. Specifically, one needs to find the radius $r_{\mathcal{O}}$ for which the orbit of the Killing vector $K$ coincides with the static geodesic at $r=r_{\mathcal{O}}$ and constant angular variables. Equivalently, $r_{\mathcal{O}}$ is the radius at which a static observer is freely falling. This is the so-called \textit{static sphere} at which the black hole attraction and the cosmological repulsion balance out exactly. One must then normalise the Killing vector by imposing $K^2 = -1$ on this geodesic orbit. 

For a general black hole metric, we have $\gamma = (f(r_{\mathcal{O}}))^{-1/2}$, where the radius $r_{\mathcal{O}}$ of the \textit{static sphere observer} is determined by the stationary point of the blackening factor, $f'(r_{\mathcal{O}})=0$ \cite{Faruk:2023uzs}. Equivalently, the normalised time is
\begin{equation}
    \tilde t=\gamma^{-1}t 
    =\sqrt{f(r_{\mathcal O})}\,t,
\end{equation}
which is the proper time of the static sphere observer (SSO). In what follows, we slightly abuse notation and denote this normalised time simply by $t$.\footnote{Thus, all times computed in the SdS background appearing below, including the bouncing times and the time argument of the corresponding Green's functions, are measured in terms of the normalised static sphere time.} Furthermore, note that this unique value of the
radius $r_{\mathcal{O}}$ also reduces to the special observers in pure de Sitter ($r_{\mathcal{O}}= 0$) and the flat space limit ($r_{\mathcal{O}} \to \infty$). For the SdS black hole in $D=5$, we get 
\begin{equation}
     r_{\mathcal{O}} = \mu^{1/4} \quad \text{and} \quad f(r_{\mathcal{O}}) = 1- 2\sqrt{\mu} = 1-2r_b r_c.
\end{equation}
The surface gravities of the two horizons are then given by 
\begin{equation}
    \kappa_{b,c} = \frac{\gamma}{2} \left | \frac{\partial f}{\partial r} \right| _{r = r_{b,c}} = \frac{1}{\sqrt{1-2r_b r_c}}\frac{r_c^2 - r_b^2}{r_{b,c}} = \frac{r_c + r_b}{r_{b,c}},
\end{equation}
where we have used $r_b^2 + r_c^2 = 1$, which implies $\gamma = (r_c - r_b)^{-1}$. The associated inverse temperatures read
\begin{equation}
    \beta_b=\pi\sqrt{\frac{2-2\sqrt{1-4\mu}}{1+2\sqrt{\mu}}}\qq{and}\beta_c=\pi\sqrt{\frac{2+2\sqrt{1-4\mu}}{1+2\sqrt{\mu}}}.
\end{equation}

Finally, with this normalisation, the surface gravity in the Nariai limit in $D$ spacetime dimensions equals $\kappa_N = \sqrt{D-1}$, which reproduces the temperature of the $dS_2 \times S^{D-2}$ near-horizon geometry that appears in this limit. Note that the choice $\gamma=1$ actually leads to confusing results since the temperatures of both horizons vanish in the Nariai limit, whereas the entropy does not. See \cite{Faruk:2023uzs} for further discussion. 

\subsection{Connecting Generic Bulk Points}\label{ss.dSgen}

\begin{figure}[ht]
    \centering
    \includegraphics[scale=1.00]{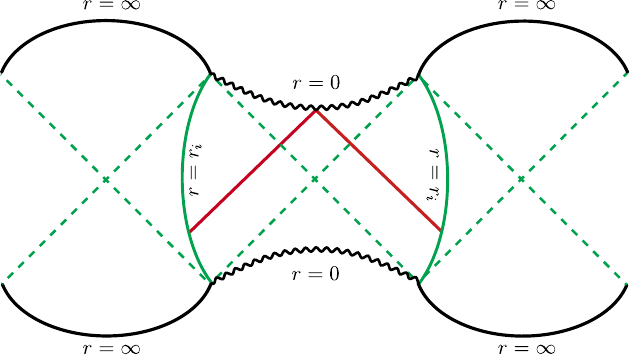}
        \caption{Penrose diagram for the Schwarzschild-de Sitter spacetime. Cosmological and black hole horizons are depicted by green dashed lines while the solid green line corresponds to the choice of a timelike surface on which the bouncing geodesic (red) is anchored.}
    \label{f.SdS_penrose}
\end{figure}

\begin{figure}[ht]
    \centering
    \includegraphics[scale=1.10]{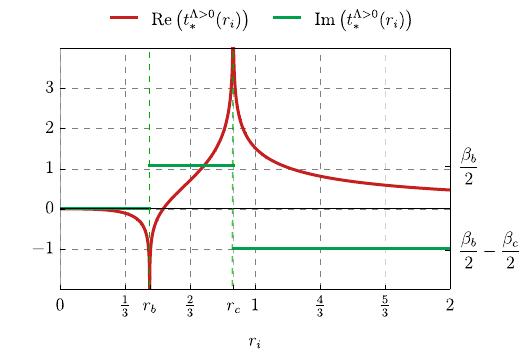}
        \caption{The behaviour of $t_*^{\Lambda>0}$ as a function of $r_i$ in $D=5$ for $\mu=1/8$. For $r_i\to0$, the bouncing time goes to 0, since the integration interval vanishes. The nontrivial radius for which the real part of $t_*^{\Lambda>0}$ vanishes corresponds to the case $r_i=r_{\mathcal{O}}$. For the cases when the hypersurface becomes null-like the bouncing time diverges. The imaginary part is constant and counts how many times the black hole horizons or the cosmological horizons are crossed.}
    \label{f.SDSri}
\end{figure}

We now investigate spacelike geodesics in this spacetime. Unlike in the asymptotically flat case, in the Schwarzschild-de Sitter spacetime, the integrals for proper length and time cannot be computed analytically for a generic spacetime dimension $D$. As was explained in \cite{Faruk:2023uzs}, the case with $D=5$ is the simplest choice, where the Penrose diagram has, in a sense, the most symmetries.\footnote{We will make this argument more precise in Sec.~\ref{ss.dSintfy} and Appendix~\ref{a.bending}.}
For simplicity we thus fix $D=5$, take coincident angular positions $\mathbf{x}=\mathbf{x'}$ and focus on symmetric radial geodesics. For geodesics anchored at a generic bulk location $r_i$ (see Fig~\ref{f.SdS_penrose}),
the integrals we have to compute are \eqref{e.tgeninte} and \eqref{e.lenghtgen} with $\epsilon=+1$ and $t$ denoting the normalised static sphere time introduced above,
\begin{align}
    t(E,r_i)&=\pm2\sqrt{f(r_{\mathcal{O}})}\int_{r_i}^{r_{T_b}}\frac{ E\,\dd r}{f(r)\sqrt{E^2+f(r)}}\label{e.timeinteD5},\\
    \mathcal{L}(E,r_i)&=\pm2\int_{r_i}^{r_{T_b}}\frac{\dd r}{\sqrt{E^2+f(r)}}.\label{e.LinteD5}
\end{align}
The turning points are obtained by solving Eq.~\eqref{e.solveforrt}. For Schwarzschild-de Sitter one generally finds two turning points: one close to the black hole singularity and the other one near spatial infinity,
\begin{equation}\label{e.turnpoinbc}
    r_{T_b,T_c}=\frac{1}{\sqrt{2}}\sqrt{(1+E^2)\mp\sqrt{(1+E^2)^2-4\mu}}\,.
\end{equation}
The integrals \eqref{e.timeinteD5} and \eqref{e.LinteD5} can be computed explicitly, yielding
\begin{align}
    t(E,r_i)&=\!\frac{2E}{r_b\!+\!r_c}\left[\frac{i r_c^2 \arctanh\!\bigg(\frac{\sqrt{r_i^2-r_{T_b}^2} \sqrt{r_c^2-r_{T_c}^2}}{\sqrt{r_i^2-r_{T_c}^2} \sqrt{r_c^2-r_{T_b}^2}}\bigg)}{\sqrt{r_c^2-r_{T_b}^2}
   \sqrt{r_c^2-r_{T_c}^2}}-\frac{r_b^2 \arctanh\!\bigg(\!\frac{\sqrt{r_i^2-r_{T_b}^2} \sqrt{r_{T_c}^2\!-r_b^2}}{\sqrt{r_{T_c}^2\!-r_i^2}
   \sqrt{r_b^2-r_{T_b}^2}}\bigg)}{\sqrt{r_b^2-r_{T_b}^2} \sqrt{r_{T_c}^2-r_b^2}}\right]\!,\label{e.SdSgenUIfint}\\
    \mathcal{L}(E,r_i)&=2 \arctan\left(\sqrt{\frac{r_i^2-r_{T_b}^2}{r_{T_c}^2-r_i^2}}\,\right).\label{e.lienka}
\end{align}
Note that if $r_b<r_i<r_c$, one gets the constant imaginary part $i \frac{\beta_b}{2}$ from crossing of the black hole horizon and the parallel horizon, as expected. Expanding these expressions in large $E$, one finds
\begin{align}
    t(E,r_i)&=t_*^{\Lambda>0}(r_i)-\frac{1}{2 E^2\,
   r_b r_c (r_b\!+\!r_c)}\Bigg[2 r_b r_c r_i \left(r_c^2-r_b^2\right)-r_c \left(\mu -r_b^2+r_b^4\right) \ln \left(\frac{2
   r_b}{r_b\!+\!r_i}-1\right)\nonumber\\
   &+r_b \left(\mu-r_c^2+r_c^4\right) \ln\left(\frac{2 r_c}{r_c+r_i}-1\right)\Bigg]+\mathcal{O}\left(\frac{1}{E^4}\right),\label{e.genie1}\\
    \mathcal{L}(E,r_i)&= \frac{2 r_i}{E}+\frac{-3\mu +r_i^4-3 r_i^2}{3 E^3 r_i}+\mathcal{O}\left(\frac{1}{E^5}\right),\label{e.genie2}
\end{align}
where we have defined the bouncing time for the five-dimensional SdS black hole as
\begin{equation}\label{e.bssdsgent}
    t_*^{\Lambda>0}(r_i)\equiv\lim_{E\to\infty}t(E,r_i)=\frac{r_b\ln\left(\frac{2r_b}{r_b+r_i}-1\right)-r_c\ln\left(\frac{2r_c}{r_c+r_i}-1\right)}{r_b+r_c}.
\end{equation}
The behaviour of this bouncing time is depicted in Fig.~\ref{f.SDSri}

Inverting expression \eqref{e.genie1} in the large-energy limit and substituting the result into the expansion of the proper length \eqref{e.genie2}, we get
\begin{equation}
    \mathcal{L}(t,r_i)\approx\frac{2 \sqrt[4]{2} \sqrt{r_i}\sqrt{t-t^{\Lambda>0}_*(r_i)}}{\sqrt{\sqrt{\sqrt{1-4 \mu }+1}-\sqrt{1-\sqrt{1-4 \mu }}}},
\end{equation}
which, as expected, vanishes when $t\to t_*^{\Lambda>0}(r_i)$. Since the turning point \eqref{e.turnpoinbc} goes to 0 in this limit, the limiting null geodesic satisfies the definition of the \textit{bouncing geodesic} \cite{Grozdanov:2026cut} and the \textit{bouncing time} \eqref{e.bssdsgent} represents the location of the \textit{bouncing singularity} in the bulk retarded Green's function $G$. Concretely, using the local Hadamard form \eqref{e.zelena_zelva}, the retarded propagator has the following behaviour near $t_*^{\Lambda>0}(r_i)$:
\begin{equation}
   G(t)\sim\frac{1}{(t-t_*^{\Lambda>0}(r_i))^{\frac32}},
\end{equation}
up to a prefactor and possible regular terms.

\begin{figure}[ht]
    \centering
    \includegraphics[scale=1.15]{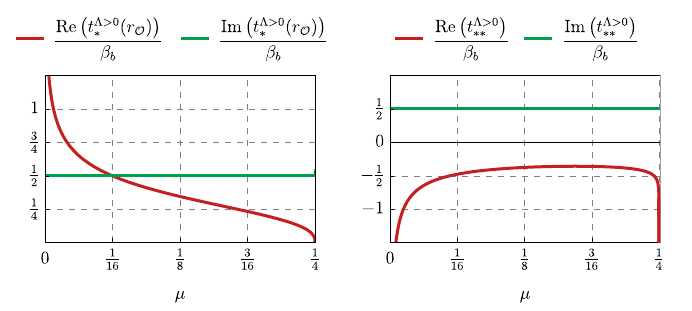}
        \caption{Real and imaginary part of the bouncing time $t^{\Lambda>0}_*(r_{\mathcal{O}})/\beta_b$ (left) and $t^{\Lambda>0}_{**}/\beta_b$ (right) for the Schwarzschild-de Sitter black hole in $D=5$. In the limit where the mass of the black hole vanishes ($\mu\to0$), the real part of the bouncing time diverges. In the Nariai limit ($\mu\to\frac14$), the bouncing singularity for the SSO coincides with the trivial lightcone singularity, while $t_{**}^{\Lambda>0}/\beta_b$ diverges.}
    \label{f.Ret}
\end{figure}

\paragraph{Anchoring at the SSO:} As a special case, consider anchoring the geodesic at the location of the static sphere observer (SSO), $r_i=r_\mathcal{O}$. As discussed above, the static sphere is the timelike hypersurface on which the black-hole attraction is balanced by the cosmological repulsion. In this case, the time integral \eqref{e.SdSgenUIfint} and the proper length \eqref{e.lienka} agree with those of \cite{Faruk:2023uzs} and the corresponding expansions behave as
\begin{align}
    t(E,r_{\mathcal{O}})&=t_*^{\Lambda>0}(r_\mathcal{O})-\frac{\sqrt{r_br_c}(r_c-r_b)}{E^2}+\mathcal{O}\Big(\frac{1}{E^4}\Big),\label{e.somefhjkvsbfv}\\
    \mathcal{L}(t,r_{\mathcal{O}})&=\frac{2\sqrt[4]{\mu}}{\sqrt[4]{r_br_c}\sqrt{(r_c-r_b)}}\sqrt{t-t^{\Lambda>0}_*(r_{\mathcal{O}})}+\mathcal{O}\Big((t-t^{\Lambda>0}_*(r_{\mathcal{O}}))^{\frac32}\Big),\label{e.navierbnpa}
\end{align}
where we have defined\footnote{Note that, as expected, this result is related to the time $\mathcal{T}_b(\mu)$ investigated in \cite{Faruk:2023uzs} via $t_*^{\Lambda>0}(r_\mathcal{O})=\pm\frac{i\beta_b}{2}\pm2\mathcal{T}_b(\mu)$.}
\begin{equation}\label{e.goodnotation}
    t_*^{\Lambda>0}(r_{\mathcal{O}})=\frac{i\beta_b}{2}+2\frac{r_c-r_b}{r_c+r_b}\arctanh{\sqrt{\frac{r_b}{r_{c}}}}.    
\end{equation}
The dependence of $t_*(r_{\mathcal{O}})$ on $\mu$ is shown in the left panel of Fig.~\ref{f.Ret}.

\subsection{Connecting  Cosmological Horizons}\label{ss.dSCH}

We now study what happens when the spacelike geodesic is anchored at the cosmological horizon and its symmetric reflection on the other side of the black hole, i.e., we set $r_i=r_c$ \eqref{e.DJelectro}. The proper length integral can be easily computed as
\begin{equation}
    \mathcal{L}_c(E)\equiv\mathcal{L}(E,r_c)=-i \ln \left(\frac{E^2+i \sqrt{2} E \sqrt{\sqrt{1-4 \mu }+1}-\sqrt{1-4 \mu }}{\sqrt{\left(E^2+1\right)^2-4 \mu }}\right),
\end{equation}
which has the large-$E$ expansion of the following form:
\begin{equation}\label{e.hungryfox}
    \mathcal{L}_c(E)=\frac{\sqrt{2} \sqrt{\sqrt{1-4 \mu }+1}}{E}+\frac{\sqrt{2} \sqrt{\sqrt{1-4 \mu }+1} \left(\sqrt{1-4 \mu }-2\right)}{3 E^3}+\mathcal{O}\Big(\frac{1}{E^5}\Big).
\end{equation}
This expression goes to 0 as one takes $E\to\infty$, thus we have a well-defined bouncing geodesic for this anchoring as well.

However, since the cosmological horizon is a null surface, the Schwarzschild coordinate time is not well defined for $r_i=r_c$, as it diverges. Nevertheless, we can regularise the time integral by introducing a ``stretched horizon'' hypersurface at $r_i=r_c-\epsilon$ and eventually sending $\epsilon\rightarrow0$. Doing so, we find
\begin{equation}
    t(E,t_c-\epsilon)=\frac{r_c}{r_b+r_c}\ln\epsilon+t_c(E)+\mathcal{O}\left(\epsilon\right),
\end{equation}
where we have defined the renormalised time $t_c(E)$ as
\begin{equation}
    t_c(E)=t_{**}^{\Lambda>0}+\frac{\beta_c\left(3-2 r_c^2\right)}{2 \pi  E^2}-\frac{\beta_c \left(2 r_b^2 r_c^2+2 r_b^2+\frac{1}{2}\right)}{2 \pi  E^4}+\mathcal{O}\Big(\frac{1}{E^6}\Big).
\end{equation}
The absolute term $t_{**}^{\Lambda>0}$ in this large-$E$ expansion corresponds to the natural analogue of the bouncing time $t_*^{\Lambda>0}$ for the anchoring at the cosmological horizon. Explicitly, one finds
\begin{equation}\label{e.dirtybunny}
    t_{**}^{\Lambda>0}=\frac{i \beta_b}{2}-\frac{2 \beta_b\tanh ^{-1}\left(\frac{r_b}{r_c}\right)+\frac{1}{2} \beta_c \ln \left(4 r_c^2\right)}{2 \pi }.
\end{equation}
The concrete behaviour of this critical time as a function of $\mu$ is depicted on the right panel of Fig.~\ref{f.Ret}.

\subsection{Bouncing off the Spatial Infinity}\label{ss.dSintfy}

\begin{figure}[ht]
    \centering
    \includegraphics[scale=1.00]{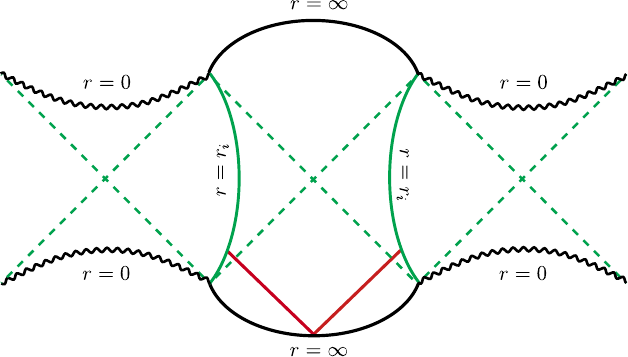}
        \caption{Penrose diagram for the Schwarzschild-de Sitter spacetime depicting a geodesic ``bouncing off the spatial infinity'' (red curve) between fixed timelike surfaces (green solid curves) at $r=r_i$.}
    \label{f.SdS_infty}
\end{figure}

Let us end the analysis of the Schwarzschild-de Sitter black hole by discussing a setup in which the geodesic bounces off the spatial infinity as opposed to the black hole curvature singularity, see Fig.~\ref{f.SdS_infty}.
Indeed, one can simply follow the same steps that were used above and replace the turning point $r_{T_b}$ by $r_{T_c}$ in the geodesic integrals. From \eqref{e.turnpoinbc}, it is clear that in the $E\to\infty$ limit, such a geodesic comes infinitely close to $r=\infty$ and bounces off. As expected, if the geodesic crosses the standard and the parallel cosmological horizon at $r_c$, the corresponding integral of time will have an imaginary part $i\frac{\beta_c}{2}$.

This situation shares many similarities with the case of the standard bouncing geodesic. In particular, as examined in \cite{Faruk:2023uzs}, in $D=5$, the geodesic bouncing off the spatial infinity is a direct continuation of the geodesic bouncing off the black hole singularity, forming a smooth curve that starts and ends at the same anchoring point $r_i$.\footnote{This is due to a special property of the SdS Penrose diagram in $D=5$ where the black hole singularity and the spatial infinity have identical bendings. We discuss this in more detail in Appendix~\ref{a.bending}.} Moreover, \cite{Faruk:2023uzs} showed that in this case,\footnote{In \cite{Faruk:2023uzs}, the authors assumed $r_i=r_\mathcal{O}$. However, as is clear from the geometric setup and the symmetries, this result can be immediately extended to any $0<r_i<\infty$.} 
\begin{equation}\label{e.ananas_is_not_bad}
    \mathcal{L}(E,r_i)+\tilde{\mathcal{L}}(E,r_i)=\pi,
\end{equation}
where $\mathcal{L}(E,r_i)$ corresponds to the geodesic bouncing off the black hole singularity between the radius $r_i$ in the left wedge of the black hole and $r_i$ in the right wedge of the black hole. $\tilde{\mathcal{L}}(E,r_i)$ corresponds to the direct continuation of this geodesic from $r_i$ in the right patch of the black hole, crossing the cosmological horizon, bouncing off the spatial infinity and continuing symmetrically through the second cosmological horizon and ending at radius $r_i$ in the region on the right-hand side of the spatial infinity. The most important feature of Eq.~\eqref{e.ananas_is_not_bad} is that the overall length of this connected geodesic is independent of both $r_i$ and energy. 

The immediate consequence of this analysis is that, although the geodesic bouncing off $r=\infty$  comes arbitrarily close to the spatial infinity, its length does \textit{not} go to 0, instead one finds,\footnote{This simply follows from the following logic: as the left-hand side of \eqref{e.ananas_is_not_bad} is independent of $E$, we can take $E\to\infty$, in which case $\mathcal{L}\to0$, yielding $\tilde{\mathcal{L}}(E\to\infty,r_i)=\pi$ for any $r_i$.}
\begin{equation}\label{e.cat_is_the_best_animal}
    \tilde{\mathcal{L}}(E,r_i)\xrightarrow[]{E\to\infty}\pi.
\end{equation}
This means that the geodesic bouncing off the spatial infinity (see Fig.~\ref{f.SdS_infty}) does \textit{not} correspond to a singularity of the retarded Green's function, since according to the Hadamard theorem, the correlator must be smooth at the corresponding point.

Let us mention that the situation becomes slightly more subtle for $D\neq5$. In those cases, the geodesics corresponding to $\mathcal{L}(E,r_i)$ and $\tilde{\mathcal{L}}(E,r_i)$ cannot be seen as two parts of the same smooth geodesic. Nevertheless, one can investigate these cases separately and find analogous behaviour to \eqref{e.cat_is_the_best_animal}. See \cite{Faruk:2023uzs} for more details. 

Note also that one can compute the analogue of the bouncing time $t_*^{\Lambda>0}(r_i)$ for the geodesic bouncing off the spatial infinity (see \cite{Faruk:2023uzs}, which computed this quantity for $r_i=r_{\mathcal{O}}$).\footnote{Due to symmetry reasons, for $r_i=r_{\mathcal{O}}$ and $D=5$, this specific time is actually equal to $t_*^{\Lambda>0}(r_\mathcal{O})$.} However, since the corresponding geodesic length does not vanish, such a quantity is not interesting for the analysis of singularities we perform in this paper.

In summary, bouncing off the spatial infinity (which is \textit{not} a curvature singularity) is \textit{not} reflected in the singularity structure of the corresponding Green's function. This is in accordance with the general idea that -- if they exist -- bouncing singularities are direct signatures of bulk curvature singularities.

%%%%%%%%%%%%%%%%%%%%%%%%%%%%%%%%%%%%%%%%%%%%%%%%%%%%%
%%%%%%%%%%%%%%%%%%%%%%%%%%%%%%%%%%%%%%%%%%%%%%%%%%%%%

\section{The Thermal Product Formula and Cavity Quasinormal Modes}\label{s.QNM} 

In this section, we derive the thermal product formula and use it to relate the asymptotic spacing of quasinormal frequencies to the bouncing time. We consider a static, spherically symmetric black hole spacetime in $D=d+1$ dimensions with flat, de Sitter, and also anti-de Sitter asymptotics, generally expressed as
\begin{equation}
    ds^2 = -f(r)\,dt^2 + \frac{dr^2}{f(r)} + r^2 d\Omega_{d-1}.
\end{equation}

Crucially, we study a setup in which the black hole is enclosed in a reflecting cavity. By this, we mean that the spacetime is bounded by a timelike hypersurface at fixed radius $r=r_i$, where reflecting boundary conditions are imposed. For the resulting spectra of gravitational, electromagnetic, and other perturbations, we establish a thermal product formula, namely a (meromorphic) factorisation of the two-sided {\em boundary} correlator in terms of the quasinormal frequencies. Such a formula was originally derived in the context of AdS/CFT \cite{Dodelson:2023vrw}. It was also shown that the two-sided correlator in pure de Sitter space admits an analogous factorised form \cite{Grewal:2024jes}. 

After deriving a thermal product formula in a cavity, we then derive a universal asymptotic expression for the quasinormal frequencies and show that, in the cavity setup, their asymptotic spacing is completely determined by the bouncing time associated with the corresponding geodesic with endpoints on the cavity walls. This result thereby generalises the analogous result known in AdS \cite{Fidkowski:2003nf,Festuccia:2005pi,Amado:2008hw,Festuccia:2008zx}.

\subsection{The Cavity Thermal Product Formula}

We assume that the (bulk) theory is in thermal equilibrium at inverse temperature $\beta$, determined by the black hole temperature. In the Schwarzschild-de Sitter black hole setup, this is consistent with our setup because the cosmological horizon is cut off from the bulk once a Dirichlet wall is placed at a fixed radius within the static patch. The ingoing boundary conditions at the black hole event horizon therefore only introduce dependence on the Hawking temperature of the black hole. 

To derive the thermal product formula, we focus on the radial wave equation for a free scalar field. The reason is that the argument uses only the general analytic structure of the corresponding second-order radial problem, and therefore extends to any perturbation sector that admits a Schr\"odinger-type master equation with a real, $\omega$-independent potential that is regular at the wall and has the standard exponential asymptotics at the horizon. In particular, this covers the scalar field and electromagnetic perturbations in the backgrounds considered here. Gravitational perturbations can be a bit more subtle and depend on the specific form of the master equation; see the discussion in Sec.\ 9.4 of  \cite{Dodelson:2023vrw}.

In the derivation below, we first impose the Dirichlet boundary conditions at the cavity wall, since this gives the simplest presentation of the Wronskian construction and of the cavity thermal product formula. However, in the physical examples considered later, the reflecting condition imposed on the original perturbation variables does not necessarily translate into a Dirichlet condition for the master fields. Depending on the perturbation sector and on the chosen gauge-invariant master variable, the corresponding wall condition can be Dirichlet, Neumann, or, more generally, the (mixed) Robin boundary condition. For conciseness, we present in detail the derivation for the Dirichlet boundary conditions and indicate how each step generalises to the Robin boundary condition at the wall. The Neumann case is recovered as the special case of the Robin boundary condition (by setting the parameter $\kappa=0$).

We begin by deriving a convenient representation for the retarded Green's function and then turn to the two-sided Wightman correlator $G_{12}(\omega)$, whose analytic properties make it particularly well suited for our purposes. Once $G_{12}(\omega)$ is known, all other two-point functions can be recovered straightforwardly using the usual thermal field theory relations. 

\subsubsection{Retarded Green's Function}

Let us study the retarded Green's function associated with a minimally coupled scalar field with mass $m$, which gives rise to the Klein--Gordon equation
\begin{equation}
\label{eq: KG equation}
    (-\Box_x + m^2)\phi(x)=0,
\end{equation}
where $x=(t,r,\Omega)$. After separating the angular dependence in spherical harmonics and rescaling the radial field according to
\begin{equation}
\label{eq: angular dependence}
    \phi(t,r,\Omega)=r^{\frac{1-d}{2}}\,Y_\ell(\Omega)\,\psi_\ell(t,r),
\end{equation}
and introducing the tortoise coordinate
\begin{equation}
\label{eq: tortoise coordinate}
    z(r)\equiv-\int_{r_i}^{r}\frac{dr'}{f(r')},
\end{equation}
the Klein--Gordon equation reduces to a family of Schr\"odinger-type equations labelled by $\ell\geq 0$,
\begin{equation}
\label{eq:Schrodinger l}
    \left(\partial_t^2-\partial_z^2+V_\ell(z)\right)\psi_\ell(t,z)=0,
    \qquad 0\leq z<\infty.
\end{equation}
With this convention, the black hole horizon is mapped to $z\to\infty$, while the cavity wall at $r=r_i$ is located at $z=0$. For illustration, in $D=4$ dimensions, the potential is
\begin{equation}
V_\ell(r)=f(r)\left(m^2+\frac{\ell(\ell+1)}{r^2}+\frac{f'(r)}{r}\right),
\end{equation}
whereas in $D=5$ dimensions, it is given by
\begin{equation}
V_\ell(r)=f(r)\left(m^2+\frac{\ell(\ell+2)}{r^2}+\frac{3f'(r)}{2r}+\frac{3f(r)}{4r^2}\right).
\end{equation}
The precise form of the potential will not play an essential role in what follows. 

Let $G_\ell(t,z;t',z')$ denote the retarded Green's function for \eqref{eq:Schrodinger l}, defined by
\begin{equation}
    \left(\partial_t^2-\partial_z^2+V_\ell(z)\right)G_\ell(t,z;t',z')
    =\delta(t-t')\delta(z-z').
\end{equation}
The full retarded Green's function of the Klein--Gordon equation can then be written as
\begin{equation}
    G(t,z,\Omega;t',z',\Omega')
    =
    \left[r(z)r(z')\right]^{\frac{1-d}{2}}
    \sum_{\ell=0}^{\infty} G_\ell(t,z;t',z')\,Y_\ell(\Omega)Y_\ell(\Omega'),
\end{equation}
where we sum over degenerate spherical harmonics.

From now on, we focus on the $\ell=0$ sector, suppress the angular momentum label, set $t'=0$ by time-translation invariance, and write simply $G(t,z;z')$. Instead of $G(t,z;z')$ and $\psi(t,z)$, we will mostly work with their Fourier transform in time, i.e.,
\begin{equation}
    \psi(t,z)=\int_{-\infty}^{\infty}\frac{d\omega}{2\pi}\,e^{-i\omega t}\psi(\omega,z).
\end{equation}
Substituting this into \eqref{eq:Schrodinger l} for $\ell=0$ gives
\begin{equation}
\label{eq:Schrodinger FT}
    \left(-\omega^2-\partial_z^2+V(z)\right)\psi(\omega,z)=0,
    \qquad 0\leq z<\infty.
\end{equation}

The retarded boundary condition is implemented by requiring ingoing behaviour at the black hole horizon $r_b$ (equivalently, as $z\to\infty$), while the cavity setup imposes Dirichlet boundary conditions at $r_i$ (equivalently, at $z=0$). Let $h_\pm(\omega,z)$ denote the solutions to \eqref{eq:Schrodinger FT} satisfying ingoing ($+$ sign) and outgoing ($-$ sign) boundary conditions at the horizon, respectively,
\begin{equation}
    h_\pm(\omega,z)\sim e^{\pm i\omega z},
    \qquad z\to\infty.
\end{equation}
Next, let $g(\omega,z)$ denote the solution satisfying Dirichlet boundary conditions at $z=0$,
\begin{equation}
    g(\omega,0)=0,
    \qquad \partial_z g(\omega,0)=1.
\end{equation}
Using the standard construction of Green's functions for ordinary differential equations \cite{Skinner}, the frequency-space retarded Green's function satisfying Dirichlet boundary condition is written as
\begin{equation}
\label{eq: Green's Wronskian}
G(\omega,z;z')=\frac{g(\omega,z_<)h_+(\omega,z_>)}{\mathcal{F}(\omega)},
\end{equation}
where we have introduced $z_<\equiv\min\{z,z'\}$ and $z_>\equiv\max\{z,z'\}$, while $\mathcal{F}(\omega)$ denotes the Wronskian of the two solutions satisfying the horizon and wall boundary conditions, respectively,
\begin{equation}
    \mathcal{F}(\omega)\equiv W[h_+(\omega,z),g(\omega,z)].
\end{equation}
Recall that $W[\phi_1(z),\phi_2(z)]
    \equiv
    \phi_1(z)\phi_2'(z)-\phi_1'(z)\phi_2(z)$. Since \eqref{eq:Schrodinger FT} contains no first-derivative term, the Wronskian is independent of $z$. In particular, evaluating it at $z=0$ gives
\begin{equation}
    \mathcal{F}(\omega)=h_+(\omega,0).
\end{equation}

Here, we remark that upon replacing the Dirichlet wall condition by the Robin condition
\begin{equation}
\label{eq: Robin BC}
(\partial_z+\kappa)\psi(\omega,z)|_{z=0}=0,
\end{equation}
with fixed $\kappa$, one should replace $g$ by the solution $g_\kappa$ satisfying
\begin{equation}
g_\kappa(\omega,0)=-1,\qquad \partial_zg_\kappa(\omega,0)=\kappa.
\end{equation}
We now find 
$\mathcal F_\kappa(\omega) \equiv W[h_+(\omega,z),g_\kappa(\omega,z)]$, and evaluating the Wronskian at the wall of the cavity gives
\begin{equation}
\mathcal F_\kappa(\omega)=\partial_zh_+(\omega,0)+\kappa h_+(\omega,0).
\end{equation}
The retarded bulk Green's function satisfying the homogeneous Robin wall condition is therefore
\begin{equation}
G_\kappa(\omega,z;z')=\frac{g_\kappa(\omega,z_<)h_+(\omega,z_>)}{\mathcal F_\kappa(\omega)},
\end{equation}
while the Neumann boundary condition case is recovered by setting $\kappa=0$.

\subsubsection{Two-sided Wightman Correlator}

In thermal field theory, various (standard) two-point functions are related to one another. For example, the two-sided Wightman correlator $G_{12}(\omega)$ can be expressed in terms of the retarded two-point function $G(\omega)$ as
\begin{equation}
\label{eq: G12 relation}
    G_{12}(\omega) = e^{\frac{\beta \omega}{2}} G_{W}(\omega) = \frac{G(\omega) - G_A(\omega)}{2 i \sinh(\frac{\beta \omega}{2})}.
\end{equation}
Here, $G_A(\omega)$ denotes the advanced two-point function, while $G_W(t)\equiv\langle \phi(t)\phi(0)\rangle_\beta$ is the Wightman correlator. Recall also that $G_{12}(t)=G_W(t-i\beta/2)$. Further discussion of the various thermal correlators can be found, e.g., in Appendix A of \cite{Dodelson:2023vrw} and Appendix B of \cite{Grewal:2024jes}.

We now make a few observations that simplify the expression for $G_{12}(\omega)$. Since \eqref{eq:Schrodinger FT} depends on $\omega$ only through $\omega^2$, and since the Dirichlet boundary condition at $z=0$ is independent of $\omega$, the solution $g(\omega,z)$ is an everywhere-regular function of $\omega^2$, and therefore $g(\omega,z)=g(-\omega,z)$. By contrast, the boundary condition at the horizon depends explicitly on the sign of $\omega$, so $h_+(-\omega,z)=h_-(\omega,z)$. Since the advanced Green's function is also given by \eqref{eq: Green's Wronskian}, but with $h_+(\omega,z)$ replaced by $h_-(\omega,z)$, it follows, as usual, that $G_A(\omega,z;z')=G(-\omega,z;z')$. 

Next, since $h_\pm(\omega,z)$ form a basis of solutions to \eqref{eq:Schrodinger FT}, $g(\omega,z)$ can be expanded as
\begin{equation}
    g(\omega,z)
    =
    \frac{W[g(\omega,z),h_+(\omega,z)]}{W[h_-(\omega,z),h_+(\omega,z)]}h_-(\omega,z)
    -
    \frac{W[g(\omega,z),h_-(\omega,z)]}{W[h_-(\omega,z),h_+(\omega,z)]}h_+(\omega,z).
\end{equation}
Evaluating the denominator at the horizon gives $W[h_-(\omega,z),h_+(\omega,z)] = 2 i \omega$, while evaluating the numerators at $z=0$ gives $W[g(\omega,z),h_+(\omega,z)] = -h_+(\omega,0) = -\mathcal{F}(\omega)$ and also $W[g(\omega,z),h_-(\omega,z)] = -h_-(\omega,0) = -h_+(-\omega,0) = -\mathcal{F}(-\omega)$. Therefore,
\begin{equation}
\label{eq: g expanded}
    g(\omega,z) = \frac{1}{2 i \omega} \left[ -\mathcal{F}(\omega) h_+(-\omega,z) + \mathcal{F}(-\omega) h_+(\omega,z) \right].
\end{equation}
Finally, combining \eqref{eq: Green's Wronskian} with \eqref{eq: g expanded}, we obtain
\begin{equation}
    G(\omega,z;z') - G_A(\omega,z;z') = 2 i \omega \frac{ g(\omega,z) g(\omega,z')}{\mathcal{F}(\omega)\mathcal{F}(-\omega)},
\end{equation}
and hence, by \eqref{eq: G12 relation}, the two-sided Wightman correlator becomes
\begin{equation}
\label{eq: bulk G12}
    G_{12}(\omega,z;z') = \frac{\omega}{\sinh{\frac{\beta \omega}{2}}} \frac{ g(\omega,z) g(\omega,z')}{\mathcal{F}(\omega)\mathcal{F}(-\omega)}.
\end{equation}

We also note that the same set of arguments applies to the (mixed) Robin boundary condition problem. Indeed, the Robin wall boundary condition \eqref{eq: Robin BC} is independent of $\omega$, and the radial equation depends on $\omega$ only through $\omega^2$. Repeating the Wronskian manipulation above with $g$ and $\mathcal F$ replaced by $g_\kappa$ and $\mathcal F_\kappa$, respectively, then gives
\begin{equation}
G_{12,\kappa}(\omega,z;z')=\frac{\omega}{\sinh{\frac{\beta \omega}{2}}}\frac{g_\kappa(\omega,z)g_\kappa(\omega,z')}{\mathcal F_\kappa(\omega)\mathcal F_\kappa(-\omega)}.
\end{equation}

\subsubsection{The “Boundary” Correlator}

The bulk two-sided correlator \eqref{eq: bulk G12} itself contains the relevant spectral information for linear response in a reflecting cavity. Indeed, for a fixed homogeneous wall condition, its poles are the zeros of $\mathcal F(\omega)$ (or, more generally, of $\mathcal F_\kappa(\omega)$). These zeros are precisely the cavity QNM frequencies. The numerator factors $g(\omega,z)g(\omega,z')$ or $g_\kappa(\omega,z)g_\kappa(\omega,z')$ may introduce additional zeros depending on the insertion points, but they do not change the spectrum of QNMs. To avoid potential complications with these zeros and to make the discussion as close as possible to holographic constructions in AdS, we define a natural ``boundary'' correlator, which automatically removes those zeros and enables a clean statement of the thermal product formula in a cavity. Since we generically do not expect these additional zeros to cancel the poles of the bulk-to-bulk correlators, we therefore expect the boundary correlator to retain the same QNM pole structure as the full bulk retarded two-point function.   

In what is to follow, the “boundary" refers to the timelike wall at $z=0$. By the “boundary" correlator, we mean a function ``measuring'' the response given a certain source applied at the wall of the cavity. For an analogous prescription in AdS/CFT, see, e.g., \cite{Son:2002sd,Kovtun:2005ev}. In particular, we prescribe a wall source $J$, solve the bulk equation for $\psi[J]$ with ingoing boundary conditions at the future horizon, read off the canonically conjugate wall response $\delta \psi$, and define the boundary correlator to be proportional to the ratio of the response to the source, $\delta \psi / J$. While in AdS, the two terms are naturally associated with a normalisable and a non-normalisable mode, at our timelike boundary (which is a regular point of the differential equation), no such clear distinction exists. We note that similar prescriptions for Dirichlet, Neumann and mixed boundary data were discussed in \cite{Solodukhin:1998ec,Minces:1999eg}. The essential point of all such constructions is therefore the choice of how the boundary condition determines which boundary data is held fixed. Passing from fixed Dirichlet data to fixed Neumann or Robin data is then implemented by adding the appropriate boundary term to the action so that the on-shell variation is written in terms of the chosen source. We note that in the AdS/CFT context, changing the boundary condition is also closely related to changing the dual CFT generating functional by double-trace deformations \cite{Witten:2001ua, Minces:2002wp}, which trigger a renormalisation group flow from a UV to an IR CFT (see \cite{Heemskerk:2009pn,Faulkner:2010jy,Grozdanov:2011aa} and more recent \cite{Grozdanov:2025ulc,Wang:2026esp}).

An effective quadratic action that yields the master field equation of motion is
\begin{equation}
\label{eq: quadratic action}
S [\psi]=\frac{N}{2}\int dt\int_0^\infty dz\left[(\partial_t\psi)^2-(\partial_z\psi)^2-V(z)\psi^2\right],
\end{equation}
with $N$ a constant and the horizon, in our coordinates, located at $z = \infty$. Varying the action, integrating by parts and evaluating it for the on-shell $\psi$ that solves the equation of motion then gives the boundary contributions
\begin{equation}
\delta S \big|_{\mathrm{EoM}}  =N\int dt\,\partial_z\psi(t,0)\delta\psi(t,0)-N\int dt\,\partial_z\psi(t,\infty)\delta\psi(t,\infty).
\end{equation}
The first term is the contribution to the ``on-shell action'' from the cavity wall, call it $\delta S^\partial\big|_{\mathrm{EoM}}$, while the second arises from the horizon. Introduce now the following notation at the cavity wall:
\begin{equation}
q(\omega)\equiv\psi(\omega,0),\qquad p(\omega)\equiv\partial_z\psi(\omega,0),
\end{equation}
making the on-shell variation of the quadratic action for the master field proportional, in this language, to $p\delta q$. The action can now be naturally adapted to fixed Dirichlet boundary conditions. We write the (regular) expansion of the ingoing solution near the wall as
\begin{equation}
h_+(\omega,z)=A(\omega)+B(\omega)z+\mathcal O(z^2).
\end{equation}
Then, the source is given by $J=q$, while the conjugate response is proportional to $p$. Given a source $J$, let $\psi[J]$ denote
the solution of the bulk equation satisfying ingoing horizon conditions and $q=J$. In frequency space,
this solution is
\begin{equation}
\psi[J](\omega,z)
=
J(\omega)\frac{h_+(\omega,z)}{A(\omega)},
\quad
A(\omega)\neq0,
\end{equation}
and 
\begin{equation}
 p[J](\omega)
\equiv \partial_z\psi[J](\omega,0) = \frac{B(\omega)}{A(\omega)}J(\omega).  
\end{equation}
Hence, it is natural to define the on-shell ``response function'' with Dirichlet boundary conditions as
\begin{equation}
G^\partial(\omega)\equiv\frac{\delta^2 S^\partial\big|_{\rm EoM}[J]}{\delta J^2}\propto
\frac{B(\omega)}{A(\omega)}=\frac{\partial_zh_+(\omega,0)}{h_+(\omega,0)}=\frac{\partial_zh_+(\omega,0)}{\mathcal F(\omega)}.
\end{equation}
Equivalently, using \eqref{eq: Green's Wronskian}, this can be written as
\begin{equation}
G^\partial(\omega)\propto\partial_z\partial_{z'}G(\omega,z;z')|_{z=z'=0}.
\end{equation}
Using Eq.~\eqref{eq: bulk G12}, we note that in the $z,z'\to 0$ limit, 
\begin{equation}
G_{12}(\omega,z;z')\sim\frac{\omega}{\sinh(\beta\omega/2)}\frac{zz'}{\mathcal F(\omega)\mathcal F(-\omega)},
\end{equation}
which then finally gives the two-sided ``boundary'' correlator
\begin{equation}
\label{eq: boundary G12}
G_{12}^\partial(\omega)\propto\frac{\omega}{\sinh(\beta\omega/2)}\frac{1}{\mathcal F(\omega)\mathcal F(-\omega)}.
\end{equation}
Indeed, as discussed above, this definition removes the numerator factors from the bulk two-point functions \eqref{eq: bulk G12}, which results in an expression with an unambiguous pole structure set by $\mathcal{F}$.

Let us now also state the prescription for the more general Robin boundary conditions. Consider the Robin wall variable
\begin{equation}
r_\kappa(\omega) \equiv p(\omega)+\kappa q(\omega).
\end{equation}
To obtain a variational problem with fixed $r_\kappa$, one has to add the appropriate wall term to the action \eqref{eq: quadratic action}. In particular, we perform a boundary Legendre transform and add a quadratic Robin term,
\begin{equation}
S_\kappa \equiv S -N\int dt\,q p-\frac{N\kappa}{2}\int dt\,q^2,
\end{equation}
giving the on-shell boundary variation
\begin{equation}
\delta S^\partial_\kappa|_{\mathrm{EoM}} = -N\int dt\,q \delta r_\kappa.
\end{equation}
Thus, for Robin data, the source is $J_\kappa=r_\kappa$, while the conjugate response is proportional to $-q$. For each prescribed source $J_\kappa$, let $\psi[J_\kappa]$ now denote
the solution of the bulk equation satisfying the ingoing condition at
the future horizon and the cavity wall boundary condition $r_\kappa=J_\kappa$. In frequency space, we now have
\begin{equation}
\psi[J_\kappa](\omega,z)
=
J_\kappa(\omega)\frac{h_+(\omega,z)}{B(\omega) + \kappa A(\omega)},
\quad
B(\omega) + \kappa A(\omega) \neq0.
\end{equation}
At the boundary, then,
\begin{equation}
 q[J_\kappa](\omega)
\equiv
\psi[J_\kappa](\omega,0)
 =
\frac{A(\omega)}{B(\omega) + \kappa A(\omega)}J_\kappa(\omega).
\end{equation}
Therefore, the boundary response function is now given by the ratio
\begin{equation}
G_\kappa^\partial(\omega)\equiv\frac{\delta^2 S_\kappa^\partial\big|_{\rm EoM}[J_\kappa]}{\delta J_\kappa^2}\propto -\frac{A(\omega)}{B(\omega)+\kappa A(\omega)}=-\frac{h_+(\omega,0)}{\mathcal F_\kappa(\omega)},
\end{equation}
or, equivalently, 
\begin{equation}
G_\kappa^\partial(\omega)\propto G_\kappa(\omega,z;z')|_{z=z'=0}.
\end{equation}
The corresponding two-sided Robin “boundary” correlator is therefore given by
\begin{equation}
G_{12,\kappa}^\partial(\omega)\propto\frac{\omega}{\sinh(\beta\omega/2)}\frac{1}{\mathcal F_\kappa(\omega)\mathcal F_\kappa(-\omega)}.
\end{equation}
The Neumann case is again recovered by setting $\kappa=0$. Defined in this manner, the Dirichlet, Neumann, and Robin boundary conditions therefore give rise to the same universal form of the two-sided “boundary” correlator.

\subsubsection{Cavity Thermal Product Formula for the ``Boundary'' Correlator}

We now show that imposing reflecting boundary conditions at a timelike surface, and thereby enclosing the black hole in a reflecting cavity, allows us to generalise the AdS/CFT thermal product formula of \cite{Dodelson:2023vrw}. As discussed above, to avoid potential issues with additional zeros, we derive the statement for the two-sided boundary correlators defined above. In doing so, we arrive at a universal formula valid in asymptotically flat, de Sitter, and anti-de Sitter spacetimes, with the reason for this generality being the asymptotic form of the potential $V(z)$ in \eqref{eq:Schrodinger FT} near the black hole horizon and near the wall. In particular, in general, we have
\begin{equation}
\label{eq: potential asymp}
    V(z) \sim \begin{cases}
        \sum_{n=0}^{\infty} a_n z^n, & z \to 0 \qquad \text{(timelike boundary at $r=r_i$)}, \\
        \sum_{n=1}^{\infty} b_n e^{-\frac{4 \pi n}{\beta}z}, & z \to \infty \qquad \text{(black hole horizon)}.
    \end{cases}
\end{equation}
In fact, the derivation of the thermal product formula is in some respects simpler in the present cavity setup than in the AdS/CFT case of \cite{Dodelson:2023vrw}, since the potential is regular at $z=0$.

The key input behind the factorisation formula for $G_{12}^\partial(\omega)$ is Hadamard's factorisation theorem: an entire function of finite order is determined by its zeros up to multiplication by a simple exponential factor. We recall the theorem here for convenience together with the relevant definitions (see, for example, Theorem 5.1 in \cite{book}).
\begin{theorem}
\label{thm: hadamard factorisation}
    Let $f$ be an entire function of growth order $\rho$, meaning that $\rho$ is the infimum of all $r>0$ such that
    \begin{equation}
        \exists A,B>0 \colon |f(w)| \leq A e^{B|w|^r},
        \quad \text{for all } w\in\mathbb{C}.
    \end{equation}
    Define the canonical factors by
    \begin{align}
        &E_0(w)=1-w, \quad &&\text{for } k=0, \\
        &E_k(w)=(1-w)e^{w+\frac{w^2}{2}+\cdots+\frac{w^k}{k}}, \quad &&\text{for } k\in\mathbb{Z}^+.
    \end{align}
    If $w_1,w_2,\ldots$ denote the non-zero zeros of $f$, then $f$ admits a factorisation of the form
    \begin{equation}
        f(w)=w^k e^{q(w)} \prod_{n=1}^{\infty} E_p\!\left(\frac{w}{w_n}\right),
    \end{equation}
    where $k$ is the order of the zero of $f$ at $w=0$, $q(w)$ is a polynomial of degree at most $\rho$, and $p=[\rho]$, i.e., the integer satisfying $p\leq \rho < p+1$.
\end{theorem}

We now show that the assumptions of Theorem \eqref{thm: hadamard factorisation} are satisfied by the boundary correlator $G_{12}^\partial(\omega)$ defined in \eqref{eq: boundary G12}. More precisely, we will argue that $1/G_{12}^\partial(\omega)$, regarded as a function of the complex variable $\omega$, is an entire function of order $1$.
Our discussion will mostly focus on the Wronskian $\mathcal{F}(\omega)$, since the analytic properties of the boundary correlator can be inferred from those of $\mathcal{F}(\omega)$.\footnote{In the language of scattering theory \cite{Newton:1982qc}, $\mathcal{F}(\omega)$ is usually referred to as the \textit{Jost function}.} Our derivation closely follows the arguments of \cite{Newton:1982qc} and Appendix C of \cite{Dodelson:2023vrw}. In particular, we proceed in three steps: we show that $1/G_{12}^\partial(\omega)$ is entire, that it is of order $1$, and then use the structure of its zeros to simplify the resulting product formula. 
We present the derivation for Dirichlet boundary conditions. However, the argument is not special to Dirichlet boundary conditions. For the Robin condition $(\partial_z+\kappa)\psi|_{z=0}=0$, with fixed, $\omega$-independent $\kappa$, the same derivation goes through with the replacement
\begin{equation}
\mathcal{F}(\omega)\longmapsto \mathcal{F}_\kappa(\omega),
\qquad
\mathcal{F}_\kappa(\omega)=\partial_z h_+(\omega,0)+\kappa h_+(\omega,0).
\end{equation}
The Neumann case is obtained by setting $\kappa=0$.

First, we argue that $1/G_{12}^\partial(\omega)$ is entire. To do so, we closely follow Section~12.1.1 of \cite{Newton:1982qc}. See also Appendix C of \cite{Dodelson:2023vrw} for a review of the relevant methods. We begin by multiplying the potential $V(z)$ in \eqref{eq:Schrodinger FT} by a parameter $\gamma$, to be set equal to $1$ at the end, and rewriting \eqref{eq:Schrodinger FT} as an integral equation, namely the \textit{Volterra equation},
\begin{equation}
\label{eq: Volterra}
    h_+(\omega, z) = e^{i \omega z} - \frac{\gamma}{\omega} \int_z^\infty dz'\, \sin(\omega(z-z')) V(z') h_+(\omega, z').
\end{equation}
It is shown in \cite{Newton:1982qc} that \eqref{eq: Volterra} can be solved iteratively, and that the resulting solution is given by an absolutely convergent series in $\gamma$, provided that
\begin{equation}
    \alpha \equiv \int_0^\infty dz \, |V(z)| \, e^{(|\!\Im \omega|-\Im \omega) z} < \infty .
\end{equation}
Hence, $h_+(\omega,z)$ is well defined throughout the upper half $\omega$-plane. In the lower half-plane, no such conclusion follows unless stronger assumptions are imposed on the potential. For example, if the potential decays exponentially as $z\to\infty$ so that
\begin{equation}
\label{eq: cond exp}
    \int_0^\infty dz \, |V(z)| \, e^{2cz} < \infty,
\end{equation}
for some $c>0$, then $\alpha$ remains finite for $\Im(\omega)<0$ down to $\Im(\omega)=-c$. Therefore, $h_+(\omega,z)$ is well defined for $\Im(\omega)\ge -c$. Moreover, by differentiating \eqref{eq: Volterra} with respect to $\omega$ and repeating the iteration argument, one finds \cite{Newton:1982qc} that if both
\begin{equation}
\label{eq: cond z^2}
    \int_0^\infty dz\, z^2 |V(z)| < \infty ,
\end{equation}
and \eqref{eq: cond exp} hold, then for each fixed $z$, the function $h_+(\omega,z)$ is analytic in $\omega$ for $\Im(\omega)\ge -c$.

Finally, the iterative construction can be used to analytically continue $h_+(\omega,z)$ step by step into the lower half-plane. Suppose first that, for sufficiently large $z$,
\begin{equation}
    V(z)\sim e^{-a z}, \quad z\to\infty,
\end{equation}
with $a >0$. Then the condition \eqref{eq: cond exp} is satisfied for every
$c<a/2$, and hence we have analyticity in the strip $\Im(\omega)>-a/2$.  However, the first Volterra iterate, obtained by inserting the zeroth approximation into the integral equation, can be continued beyond $\Im(\omega)=-a/2$, except for a simple pole at $\omega=-\frac{i}{2}a$. After this first possible pole is isolated, the remainder has improved exponential decay, and the Volterra iteration can be continued one half-strip further, so that $h_+(\omega,z)$ is meromorphic for $\Im(\omega)>-a$. Repeating this argument, the second iterate produces the next possible pole at $\omega=-ia$ and extends the continuation to $\Im(\omega)>-3a/2$. Inductively, one obtains a meromorphic continuation of $h_+(\omega,z)$ into the lower half-plane, with possible simple poles at
\begin{equation}
    \omega_n=-\frac{i}{2} a n, \quad n\in\mathbb{N}.
\end{equation}
See Section~12.1.1 of \cite{Newton:1982qc} for details. Using now the asymptotic expansion \eqref{eq: potential asymp}, we have
\begin{equation}
    V(z)\sim \sum_{n=1}^\infty b_n e^{-4\pi n z/\beta},
    \quad z\to\infty,
\end{equation}
so that $a=4\pi/\beta$. Therefore, the singularities of $h_+(\omega,z)$ occur precisely at the (negative, imaginary) Matsubara frequencies
\begin{equation}
    \omega_n=-i\frac{2\pi n}{\beta}, \quad n\in\mathbb{N}.
\end{equation}
In particular, the same analytic properties hold for $\mathcal{F}(\omega)\equiv h_+(\omega,0)$. It then follows from \eqref{eq: boundary G12} that $1/G_{12}^\partial(\omega)$ is entire. Indeed, $\mathcal{F}(\omega)\mathcal{F}(-\omega)$ is meromorphic, with simple poles at $\omega=2\pi i n/\beta$ for every non-zero integer $n$, and these poles are precisely cancelled by the simple zeros of the factor $\sinh(\beta\omega/2)/\omega$. 

We remark that the same conclusion holds for the Robin boundary conditions case with $\mathcal{F}_\kappa(\omega)$. Indeed, 
for fixed, $\omega$-independent $\kappa$, taking a $z$-derivative at the regular wall cannot introduce new singularities in the $\omega$-plane. Hence, $\mathcal{F}_\kappa(\omega)$ has the same meromorphic continuation and the same Matsubara pole structure as $\mathcal{F}(\omega)$. What changes with $\kappa$ are the zeros of $\mathcal{F}_\kappa$, namely the Robin cavity quasinormal frequencies.

Second, we need to show that $1/G_{12}^\partial(\omega)$ has growth order $1$. The essential point is the large-$|\omega|$ behaviour of $\mathcal{F}(\omega)$. We again rely on the results of Section~12.1 from \cite{Newton:1982qc}. Since the potential is regular near $z=0$, we may, unlike\footnote{In the Schwarzschild-AdS case, the AdS boundary is located at $z=0$ and the effective potential behaves as $V(z)\sim 1/z^2$ near the boundary. In particular, the potential does not satisfy \eqref{eq: cond z}, so the large-$|\omega|$ scattering-theory result quoted here does not directly apply. For this reason, the derivation in \cite{Dodelson:2023vrw} relies instead on the large-$\omega$ expansion of equation \eqref{eq:Schrodinger FT} and insights from dual conformal field theory, using the results of \cite{festucciathesis}.} in \cite{Dodelson:2023vrw}, use the large-$|\omega|$ asymptotics stated in Eq.\ 12.26 of \cite{Newton:1982qc}. For potentials satisfying
\begin{equation}
\label{eq: cond z}
    \int_0^\infty dz\, z \, |V(z)| < \infty,
\end{equation}
one has
\begin{equation}
    h_+(\omega,z)=e^{i\omega z}+\mathcal{O}(e^{-\Im(\omega)z}),
    \quad \text{as } |\omega|\to\infty \text{ with } \Im(\omega)\geq 0,
\end{equation}
uniformly in $z$. In particular, for $\Im(\omega)\geq 0$,
\begin{equation}
\label{eq: F limit}
    \lim_{|\omega|\to\infty}\mathcal{F}(\omega)=1.
\end{equation}
Furthermore, if the potential $V(z)$ admits an analytic continuation to complex $z$, one can show that \eqref{eq: F limit} also holds in the lower half-plane, except possibly along the line containing poles \cite{Newton:1982qc}. It follows that
\begin{equation}
\label{eq: G12 asymptotics}
    \frac{1}{G_{12}^\partial(\omega)}
    \propto
    \frac{1}{\omega}\sinh\!\left(\frac{\beta\omega}{2}\right)\mathcal{F}(\omega)\mathcal{F}(-\omega)
    \sim \frac{1}{\omega}e^{\frac{\beta\omega}{2}},
    \quad |\omega|\to\infty,
\end{equation}
along any ray in the complex plane asymptotically avoiding the poles. Note that the same conclusion holds for the Robin correlator. Since the wall is regular, replacing $\mathcal{F}(\omega)$ by $\mathcal{F}_\kappa(\omega)$ changes the large-$|\omega|$ behaviour only by a polynomial factor. More explicitly, $\partial_z h_+(\omega,0)$ can contribute one additional power of $\omega$, so $\mathcal{F}_\kappa(\omega)\mathcal{F}_\kappa(-\omega)$ can generically differ from $\mathcal{F}(\omega)\mathcal{F}(-\omega)$ by a factor of order $\omega^2$. However, this does not affect the exponential type relevant for the growth order.

To conclude that $1/G_{12}^\partial(\omega)$ is an entire function of order $1$, it remains to control the growth along the remaining directions, namely those asymptotically approaching the pole line, i.e., the imaginary axis. Here we appeal to the expected large-imaginary-$\omega$ behaviour of the correlator. In particular, it was shown in \cite{festucciathesis} that for the Schwarzschild black hole in AdS, the large-imaginary-$\omega$ behaviour of the two-sided Wightman correlator is controlled by a bouncing geodesic. As a result, one finds the asymptotics
\begin{equation}
\label{eq: i omega asymptotics}
    G^\partial_{12}(\omega) \sim e^{\pm i \omega \tau},
    \quad \omega\to \pm i\infty,
\end{equation}
up to polynomial prefactors, where $\tau\equiv\Re(t_*)$ is the real part of the complex bouncing time. The assumptions underlying that analysis, most importantly, the exponential tortoise-coordinate asymptotics of the potential in the Schr\"odinger-type equation \eqref{eq:Schrodinger FT} and the existence of a bouncing geodesic, are also satisfied in our cavity setup, whether asymptotically flat, de Sitter, or anti-de Sitter. We therefore expect an analogous complex-WKB argument to apply here as well. The numerically observed asymptotic spectral spacing $2\pi/t_*$, discussed in Section~\ref{ss.examples}, provides further independent support for this. Altogether, this allows us to conclude that $1/G_{12}^\partial(\omega)$ grows at most exponentially in $|\omega|$ in all directions in the complex plane, and hence is of order $1$.

Finally, we use the fact that the poles of $G^\partial_{12}(\omega)$ are isolated and symmetric with respect to the real and imaginary $\omega$-axes. Indeed, since $\mathcal{F}(\omega)$ is analytic away from the Matsubara poles, its zeros are necessarily discrete. Moreover, the relations $h_+(-\omega,z)=h_-(\omega,z)$ and $h_+(\omega,z)^*=h_-(\omega^*,z)$ imply that
\begin{equation}
    \mathcal{F}(\omega)^*=\mathcal{F}(-\omega^*).
\end{equation}
We therefore conclude that the poles of $G^\partial_{12}(\omega)$, equivalently the zeros of $1/G^\partial_{12}(\omega)$, come in families
\begin{equation}
\label{eq:G12_qnms}
    (\omega_n,-\omega_n,\omega_n^*,-\omega_n^*).
\end{equation}
Altogether, we have shown that $1/G^\partial_{12}(\omega)$ is an entire function of order $\rho=1$. Applying Theorem \eqref{thm: hadamard factorisation} to $1/G^\partial_{12}(\omega)$, using the evenness of $G^\partial_{12}(\omega)$, and noting that the order of the zero at $\omega=0$ is $k=0$, together with
\begin{equation}
    E_1(w)E_1(-w)=E_0(w)E_0(-w)=1-w^2,
\end{equation}
we arrive at the product formula
\begin{equation}
\label{eq: CTPF}
    G^\partial_{12}(\omega)
    = \frac{G^\partial_{12}(0)}
    {\displaystyle\prod_{n=1}^\infty
    \left(1-\frac{\omega^2}{\omega_n^2}\right)
    \left(1-\frac{\omega^2}{(\omega_n^*)^2}\right)}.
\end{equation}
The formula \eqref{eq: CTPF} is written for Dirichlet wall data, for which the poles $\omega_n$ are the zeros of $\mathcal{F}(\omega)$. For the Robin wall boundary condition, the frequencies $\omega_n$ in \eqref{eq: CTPF} should be understood as zeros of $\mathcal{F}_\kappa(\omega)$.

\subsection{Relation Between the Bouncing Time and Cavity QNMs}

As in \cite{Dodelson:2023vrw}, we can use the asymptotics
\eqref{eq: G12 asymptotics} to place constraints on the spectrum of QNMs. Throughout this argument, we will be discussing the
poles of $G^\partial_{12}(\omega)$, whereas usually, by QNMs, we mean the poles of the retarded propagator, or equivalently, the zeros of $\mathcal{F}(\omega)$. Their respective sets of poles are related by \eqref{eq: G12 relation}. The stable retarded cavity QNMs lie in the lower half of the complex $\omega$-plane with our Fourier convention $e^{-i\omega t}$. However, the two-sided correlator $G_{12}^{\partial}(\omega)$ also has poles at the reflected points in the upper half-plane (see \eqref{eq:G12_qnms}). We denote these upper-half-plane representatives by $\hat{\omega}_n$, so that the corresponding retarded cavity QNMs are $\omega_n=\hat{\omega}_n^*$. Following \cite{Dodelson:2023vrw} and intuition from the numerical investigations below, we assume that the upper-half-plane poles are asymptotically organised along a ray at an angle $\theta$,
\begin{equation}
\label{eq: qnm ansatz}
    \hat{\omega}_n \sim r e^{i\theta} n^\alpha,
    \qquad n\to\infty,
    \qquad r>0,
    \qquad 0<\theta<\frac{\pi}{2}.
\end{equation} 
Inserting it into the thermal product formula \eqref{eq: CTPF}, it is useful to first rewrite the product as
\begin{equation}
\label{eq: logG12}
    \partial_\omega \ln G^\partial_{12}(\omega)
    = - \sum_{n=1}^{\infty}\left(
    \frac{1}{\omega - \omega_n}
    + \frac{1}{\omega + \omega_n}
    + \frac{1}{\omega - \omega^*_n}
    + \frac{1}{\omega + \omega^*_n} \right).
\end{equation}
Substituting the asymptotic form \eqref{eq: qnm ansatz} and its complex conjugate into \eqref{eq: logG12}, and using the Euler--Maclaurin formula, one finds \cite{Dodelson:2023vrw}
\begin{equation}
    \partial_\omega \ln G^\partial_{12}(\omega)\sim -\frac{2 \pi}{\alpha \omega} \left( \frac{\omega}{r}\right)^{1/\alpha} \frac{ \cos\!\left(\frac{\pi - 2 \theta}{2 \alpha}\right)}{ \sin\!\left(\frac{\pi}{2 \alpha}\right) }, \quad \omega\to\infty .
\end{equation}
On the other hand, \eqref{eq: G12 asymptotics} gives
\begin{equation}
    \partial_\omega \ln G^\partial_{12}(\omega) \sim -\frac{\beta}{2} + \frac{1}{\omega}, \quad \omega\to\infty .
\end{equation}
Matching the leading terms therefore implies that $\alpha = 1$, so the spectrum is asymptotically linearly spaced. Furthermore, the matching gives
\begin{equation}
    \beta = \frac{4 \pi \sin \theta}{r}.
\end{equation}
To extract the first subleading constraint, we refine the ansatz \eqref{eq: qnm ansatz} to
\begin{equation}
\label{eq: subleading ansatz}
    \hat{\omega}_n
    = \Omega n + s e^{i\phi} + \cdots,
    \qquad
    \Omega \equiv r e^{i\theta},
    \qquad n\to\infty .
\end{equation}
Inserting this into \eqref{eq: logG12}, one finds\footnote{We can again directly use the
results of \cite{Dodelson:2023vrw} by noting that their asymptotics
$\partial_\omega \ln G^\partial_{12}(\omega) \sim -\frac{\beta}{2} + \frac{2\Delta-d}{\omega}$ agree with ours after setting \(2\Delta-d=1\).}
\begin{equation}
    \frac{4s \cos(\theta-\phi) + 2r}{r} = 1.
\end{equation}

Having the thermal product formula \eqref{eq: CTPF}, we can also use the
result of \cite{Dodelson:2025jff}. They insert the ansatz
\eqref{eq: subleading ansatz} into the product formula, expand the result for large real $\omega$, and then Fourier transform to the time domain.
After setting $2\Delta-d=1$, appropriate to our case, we find
\begin{equation}
    G^\partial_{12}(t) \approx \sum_{n,m=0}^{\infty} \frac{ e^{\frac{2\pi i s}{r} \left( m e^{i(\phi-\theta)}- n e^{i(\theta-\phi)} \right)}}{\left(i(t - \hat{t}_{nm})\right)^{2}} + (t \to -t) .
\end{equation}
In particular, $G^\partial_{12}(t)$ has singularities at the lattice points\footnote{As discussed in \cite{Grozdanov:2026cut}, the lattice derived this way may (at least in some cases) overcount the number of position space singularities. The additional points, known as the ``phantom singularities'', appear due to the Fourier transform over the remaining spatial coordinates and appear, e.g., in the case of the BTZ black hole or the self-dual axion model black holes. Here, we for simplicity assume this phenomenon is not present for a generic higher-dimensional black hole.}
\begin{equation}
\label{eq: hattnm}
    \hat{t}_{nm}
    = \frac{i\beta}{2}
      + \frac{2\pi}{r}
        \left(
            n e^{i\theta}
            + m e^{i(\pi-\theta)}
        \right)
    = \frac{i\beta}{2}
      + 2\pi
        \left(
            \frac{n}{\Omega^*}
            - \frac{m}{\Omega}
        \right),
    \qquad n,m\in\mathbb{Z}_{\geq0},
\end{equation}
together with the reflected points $-\hat{t}_{nm}$. Here,
$\Omega\equiv r e^{i\theta}$ is the leading asymptotic spacing of the
upper-half-plane poles, so that
\begin{equation}
    \hat{\omega}_n \sim \Omega n,
    \qquad
    \omega_n=\hat{\omega}_n^* \sim \Omega^* n,
    \qquad n\to\infty .
\end{equation}
The above singularity structure of $G_{12}(t)$ straightforwardly extends to the case of the retarded propagator $G(t)$, for which one gets\footnote{Note that despite the fact that the retarded Green's function does not have a natural analytic extension due to the presence of the Heaviside step function, it may be analytically extended separately for its negative or positive argument. See Appendix~B of \cite{Grozdanov:2026cut} for more details.}
\begin{equation}
\label{eq: tnm}
    t_{nm}
    = 2\pi\left(
        \frac{n}{\Omega^*}
        - \frac{m}{\Omega}
      \right),
    \qquad n,m\in\mathbb{Z}.
\end{equation}
The point $t_{00}=0$ simply corresponds to the trivial coincident
singularity of the correlator, i.e., the point where both insertions
are at the same location. As was shown in concrete examples, e.g.,
in \cite{Dodelson:2025jff,Grozdanov:2026cut}, the lowest-lying
non-trivial lattice point $t_{10}$ (respectively, $t_{01}$)
corresponds to the bouncing time $t_*$ (respectively, its symmetric
reflection).\footnote{Since, in this section, we discuss the
asymptotically flat, AdS and dS cases simultaneously, we drop the
superscript index in $t_*$.} More generally, the local Hadamard form
implies that every singularity of $G(t)$ must correspond to a null
geodesic (or a null limit of a spacelike or timelike geodesic). For
vanishing angular displacement, the first such non-trivial null
geodesic is precisely the bouncing geodesic. We therefore identify
\begin{equation}
    t_* = t_{10} = \frac{2\pi}{\Omega^*}.
\end{equation}
Since the retarded cavity QNMs have the leading asymptotic behaviour
$\omega_n\sim\Omega^* n$, we obtain
\begin{equation}
\label{e.centralformula}
    t_* \sim \frac{2\pi n}{\omega_n},
    \qquad n\to\infty .
\end{equation}
This is the central connection formula between the momentum space physics and the position space singularities that bear information about the black hole interior. 

Several comments are in order. First, changing the wall boundary condition changes the Wronskian $\mathcal{F}(\omega)$, and hence changes the detailed quasinormal spectrum. However, for fixed $\omega$-independent Robin data at a regular wall, the replacement $\mathcal{F}(\omega)\mapsto\mathcal{F}_\kappa(\omega)$ only changes polynomial prefactors and subleading phase data in the large-$|\omega|$ asymptotics. It does not change the exponential type entering the thermal product formula. Consequently, the leading asymptotic spacing remains controlled by the same bouncing time, with $\omega_n$ now denoting the quasinormal frequencies associated with the chosen wall condition. In particular, Dirichlet, Neumann and Robin boundary conditions are expected to give spectra with the same leading asymptotic spacing governed by $t_*$.

Before turning to concrete examples and numerical checks of \eqref{e.centralformula} for asymptotically flat and de Sitter black holes, let us make a few further remarks. The same reasoning applies to any perturbation sector whose equations can be written in terms of a Schrödinger-type master equation with a regular, frequency-independent reflecting condition at the wall. In particular, the discussion below includes scalar, electromagnetic and gravitational perturbations.

Although the derivation used the two-sided correlator, the final relation \eqref{e.centralformula} only involves data of the ordinary one-sided cavity problem. The cavity QNMs $\omega_n$ are obtained by solving the wave equation in the region $r\in(r_b,r_i)$ with ingoing boundary conditions at the black hole horizon and a reflecting boundary condition at the wall. This region lies entirely in the physically accessible Lorentzian wedge. On the other hand, the bouncing time $t_*$ is the elapsed (complexified) Schwarzschild time along the two-sided bouncing geodesic. Equivalently, since radial null rays satisfy $dt=\pm dz$, where $z$ is the tortoise coordinate, $t_*$ is twice the analytically continued tortoise-coordinate separation between the cavity wall and the curvature singularity. Its real part corresponds to the principal-value contribution, while its imaginary part corresponds to the residue contributions from crossings of the event and (potential) Cauchy horizons. 

The reflecting wall should be viewed as an idealised spherical mirror enclosing the black hole. For electromagnetic perturbations, this interpretation is particularly direct: imposing a perfect conductor condition at the wall gives a vanishing normal energy flux and leads to the corresponding cavity quasinormal spectrum. We will discuss this example explicitly below. As will also be demonstrated in the next section, the cavity QNMs $\omega_n$ typically approach their asymptotic form \eqref{e.centralformula} rather quickly, at least for the range of wall radii $r_i$ considered here. It is therefore possible to determine $t_*$ with reasonably good numerical accuracy by measuring only the first few overtones.

Finally, the formula \eqref{e.centralformula} relates the, potentially measurable, cavity QNM spectrum to the location of the bouncing singularity. Since $t_*$ is determined by the geometry of null propagation into the black hole interior, it is sensitive to the internal structure of the spacetime. For charged, rotating, asymptotically flat, de Sitter or anti-de Sitter black holes, and also for geometries modified by higher-derivative corrections or by a proposed resolution of the singularity, the precise value of $t_*$ changes. This leads to different predictions for the asymptotic QNM spectrum and thus provides a possible way of probing the black hole interior through exterior spectral data.

\subsection{Examples of Scalar, Electromagnetic and Gravitational Cavity QNMs}
\label{ss.examples}
We now numerically compute several examples of cavity quasinormal frequencies for scalar, electromagnetic, and gravitational perturbations. In each case, the corresponding equation of motion can be written in terms of a master Schr\"odinger-type equation
\begin{equation}
\label{eq: master eq form}
    \left(\frac{d^2}{dz^2}+\omega^2-V(r)\right)\psi(r)=0,
\end{equation}
where $z$ is the tortoise coordinate, defined in \eqref{eq: tortoise coordinate} so that $dr/dz=-f(r)$, and $\psi(r)$ is the appropriate master field. For simplicity, we solve \eqref{eq: master eq form} using a Frobenius expansion. To this end, we introduce the coordinate $x(r)\equiv(r-r_b)/(r_i-r_b)$, which maps the black hole horizon $r=r_b$ to $x=0$ and the wall location $r=r_i$ to $x=1$. The solution satisfying the ingoing boundary condition at the horizon can then be written as a power series around $x=0$,
\begin{equation}
\label{eq: frob ansatz}
    \psi(x)=x^{-i\omega/f'(r_b)}\sum_{n=0}^\infty a_n(\omega,\ell)x^n.
\end{equation}
The quasinormal spectrum is obtained by substituting the ansatz \eqref{eq: frob ansatz} into the relevant master equation, deriving recursion relations for the coefficients $a_n(\omega,\ell)$, and imposing the appropriate boundary condition at the wall $r=r_i$ (equivalently, at $x=1$). As discussed below, the precise form of this boundary condition depends on the perturbation sector and may be of Dirichlet, Neumann, or Robin type. For example, after truncating the series at some sufficiently large order $N$ in the Dirichlet case, one imposes $\psi(1) = 0$. In examples below, we choose $N$ by increasing the truncation order until the computed roots are stable under $N\mapsto N+10$, with changes smaller than $10^{-10}$. For the electromagnetic and gravitational modes presented below, this criterion is satisfied at $N=110$. In the scalar examples, however, the convergence rate depends on different anchoring choices and we therefore apply the same criterion separately in each case, increasing $N$ as needed. For the cases shown below we used truncation orders up to $N=230$. We also note that the expansion \eqref{eq: frob ansatz} is guaranteed to converge in the complex $x$-plane only inside a circle of radius $\rho$ centred at $x=0$ (corresponding to $r=r_b$), where $\rho$ is the distance to the nearest singular point of the corresponding master equation, viewed as a Fuchsian ODE (see \cite{Kovtun:2005ev} or \cite{Horowitz:1999jd} for more details). 

In what is to follow, we present the results for scalar, electromagnetic, and gravitational perturbations in four-dimensional asymptotically flat Schwarzschild spacetime and for scalars also in the Schwarzschild-de Sitter spacetime. The spectra are qualitatively similar in all considered cases. Most importantly, the spacing between the overtones (rapidly) asymptotically approaches the predicted relation given in \eqref{e.centralformula}.

\subsubsection{Scalar Perturbations}

\begin{figure}[ht]
    \centering
    \includegraphics[width=\linewidth]{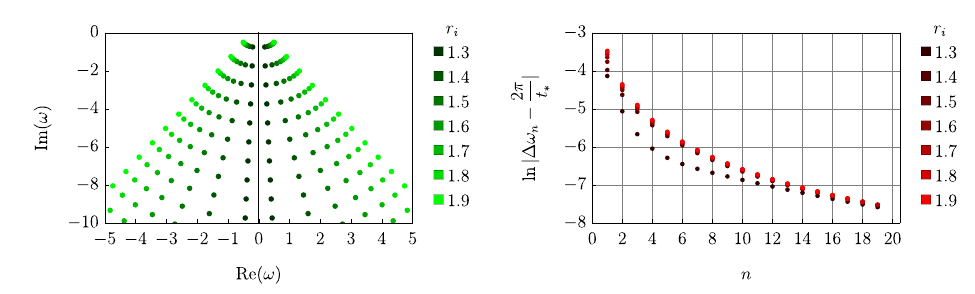}
    \caption{Left: Numerically computed scalar quasinormal frequencies $\omega_n$ for four-dimensional asymptotically flat Schwarzschild spacetime with a Dirichlet wall at different radii $r_i$, in units where the black hole horizon radius is equal to $1$. Right: Comparison with the predicted asymptotic spacing. Here, $\Delta\omega_n \equiv \omega_{n+1}-\omega_n$ denotes the numerically computed spacings, and $t_*$ is the bouncing time for a geodesic anchored at $r_i$; see Eq.~\eqref{e.tsD4}.}
    \label{fig:flat QNMs}
\end{figure}

\begin{figure}[ht]
    \centering
    \includegraphics[width=\linewidth]{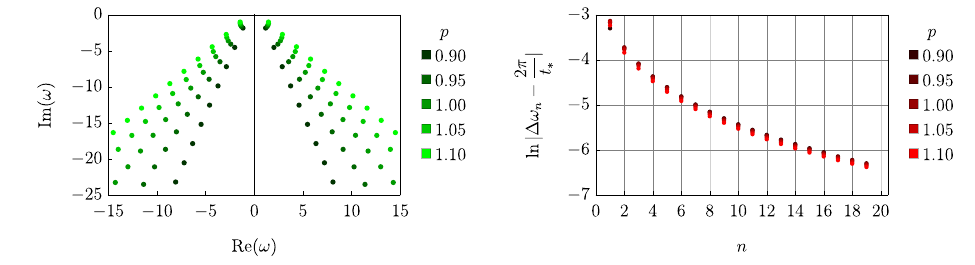}
    \caption{Left: Numerically computed scalar quasinormal frequencies $\omega_n$ for five-dimensional Schwarzschild-de Sitter spacetime with $\mu=0.15$ and a Dirichlet wall at radii $r_i=p\,r_{\mathcal O}$. The values of the parameter $p$ are displayed in the legend. Right: Comparison with the predicted asymptotic spacing. As above, $\Delta\omega_n \equiv \omega_{n+1}-\omega_n$ denotes the numerically computed spacings, and $t_*$ is the bouncing time for geodesics anchored at $r_i$; see Eq.~\eqref{e.bssdsgent}}

\label{fig:SdS QNMs}
\end{figure}

Scalar perturbations are governed by the Klein--Gordon equation \eqref{eq: KG equation}. After separating the angular dependence in spherical harmonics and rescaling the field (see Eq.~\eqref{eq: angular dependence}), the equation reduces to the Schr\"odinger-type equation \eqref{eq:Schrodinger l}. We impose Dirichlet boundary conditions at $r=r_i$ and restrict to the $s$-wave sector, $\ell=0$. The results for four-dimensional asymptotically flat Schwarzschild spacetime are then presented in Fig.~\ref{fig:flat QNMs}, while those for five-dimensional Schwarzschild--de Sitter spacetime are shown in Fig.~\ref{fig:SdS QNMs}. In both cases, the spacing between overtones rapidly approaches $2\pi/t_*$, where $t_*$ is the corresponding bouncing time, in agreement with \eqref{e.centralformula}.

More generally, the natural choice in the cavity setup is to impose reflecting boundary conditions at $r=r_i$, or, more precisely, to require that the radial flux through the wall vanish. We will discuss this condition in more detail in Sec.~\ref{ss.EM pert} in the context of electromagnetic perturbations. For scalar perturbations, this physically motivated requirement can be implemented by Dirichlet, Neumann, or, more generally, Robin boundary conditions. Numerically, we find that the asymptotic QNM spacing determined by $t_*$ is robust under these different choices. Although the detailed spectrum changes slightly, we have verified numerically that the asymptotic spacing remains unchanged.

\subsubsection{Electromagnetic Perturbations}
\label{ss.EM pert}

Next, we consider electromagnetic perturbations of an uncharged four-dimensional asymptotically flat Schwarzschild black hole, which are governed by the Maxwell equations,
\begin{equation}
\label{eq: Maxwell}
    \nabla_\nu F^{\mu \nu}=0,
\end{equation}
with $F_{\mu \nu} = \partial_\mu A_\nu - \partial_\nu A_\mu$. Using the fact that the background is spherically symmetric, we can work in the Regge-Wheeler-Zerilli formalism \cite{Regge:1957td, Zerilli:1970se} and expand $A_\mu$ in terms of four-dimensional spherical harmonics \cite{BH:1973, Cardoso:2001bb, Lei:2021kqv},
\begin{equation}
\label{eq: A_mu expansion}
A_\mu = e^{-i\omega t}
\sum_{\ell,m}
\left[
\begin{pmatrix}
0 \\
0 \\
b^{\ell m}(r)\mathbf{X}^{\ell m}
\end{pmatrix}
+
\begin{pmatrix}
j^{\ell m}(r)Y^{\ell m} \\
h^{\ell m}(r)Y^{\ell m} \\
k^{\ell m}(r)\mathbf{Y}^{\ell m}
\end{pmatrix}
\right],
\end{equation}
with $Y^{\ell m}$ being the usual scalar spherical harmonics, while $\mathbf{X}^{\ell m}$ and $\mathbf{Y}^{\ell m}$ are defined as
\begin{equation}
\label{eq: vector harmonics}
\mathbf{X}^{\ell m}
=
\begin{pmatrix}
\dfrac{1}{\sin\theta}\partial_\varphi Y^{\ell m} \\
-\sin\theta\,\partial_\theta Y^{\ell m}
\end{pmatrix},
\qquad
\mathbf{Y}^{\ell m}
=
\begin{pmatrix}
\partial_\theta Y^{\ell m} \\
\partial_\varphi Y^{\ell m}
\end{pmatrix}.
\end{equation}
Here, $\ell$ is the angular momentum quantum number and $m$ is the azimuthal number. The first term in the square brackets on the right-hand side of \eqref{eq: A_mu expansion} has parity $(-1)^{\ell+1}$, while the second has parity $(-1)^\ell$. Accordingly, we call the former odd (or axial) modes, and the latter even (or polar) modes. Substituting \eqref{eq: A_mu expansion} into \eqref{eq: Maxwell}, and using the harmonic time dependence $e^{-i\omega t}$, one obtains two radial master equations of the form \eqref{eq: master eq form}. The corresponding gauge-invariant master fields are
\begin{equation}
\label{eq: EM master fields}
    \psi_{-}(r) = b^{\ell m}(r) \quad \text{and} \quad \psi_{+}(r) = \frac{r^2}{\ell (\ell +1)} \left(- i \omega h^{\ell m}(r) - \partial_r j^{\ell m}(r) \right),
\end{equation}
where $\psi_-$ describes the odd sector and $\psi_+$ describes the even sector \cite{BH:1973, Cardoso:2001bb}. In both cases, the potential is given by
\begin{equation}
    V_{\text{EM}}(r) = \ell (\ell +1)\frac{f(r)}{r^2}.
\end{equation}

Next, we need to impose the appropriate physically motivated boundary conditions. At the horizon, we impose the usual ingoing wave boundary condition. Imposing boundary conditions at $r=r_i$ is, however, slightly more subtle than for a scalar field, mainly because the fundamental fields $A_\mu$ are gauge-dependent. In particular, it is not meaningful to require $A_\mu$ itself to vanish at $r=r_i$. The gauge-invariant object is instead the field strength $F=dA$. Moreover, in the cavity, the wall at $r=r_i$ is interpreted as a reflecting ``mirror'' in the sense that no energy flux should cross it. We therefore require the energy flux of the Maxwell field to vanish at $r_i$. This follows the proposal from \cite{Wang:2015goa} for asymptotically AdS black holes, which was applied to Schwarzschild black holes in a cavity in \cite{Lei:2021kqv}. One should note that these \textit{vanishing energy flux} (VEF) boundary conditions should be distinguished from field-vanishing boundary conditions. In particular, Dirichlet-type boundary conditions usually imply VEF, whereas the converse is generally not true, as we already noted in the case of scalar perturbations.

We now write the VEF condition explicitly. Let $\Sigma_i\equiv\{r=r_i\}$ be the timelike wall, and let $\iota:\Sigma_i\hookrightarrow\mathcal M$ denote the embedding into the spacetime manifold $\mathcal M$. The electromagnetic stress tensor is
\begin{equation}
   T_{\mu\nu}=g^{\rho\sigma}F_{\mu\rho}F_{\nu\sigma}-\frac{1}{4}g_{\mu\nu}F_{\rho\sigma}F^{\rho\sigma}.
\end{equation}
Since the background is static, the Killing vector field $\xi=\partial_t$ defines a conserved energy current
\begin{equation}
    J^\mu=-T^{\mu\nu}\xi_\nu,\qquad \nabla_\mu J^\mu=0.
\end{equation}
The energy flux through $\Sigma_i$ is therefore
\begin{equation}
\label{eq: flux}
     \mathcal F_i=J^\mu n_\mu=-T^{\mu\nu}\xi_\nu n_\mu,
\end{equation}
where $n^\mu$ is the outward-pointing unit normal to $\Sigma_i$. Explicitly, we may take
\begin{equation}
    n^\mu=f(r_i)^{1/2}(\partial_r)^\mu.
\end{equation}
We also define the electric and magnetic fields measured by static observers at $\Sigma_i$ by
\begin{equation}
    E^\mu=F^{\mu\nu}u_\nu,\qquad B^\mu={*F}^{\mu\nu}u_\nu=\frac{1}{2}\epsilon^{\mu\nu\rho\sigma}F_{\rho\sigma}u_\nu,\quad \text{with} \quad u^\mu=f(r_i)^{-1/2}(\partial_t)^\mu.
\end{equation}
Using $g_{\mu\nu}n^\mu u^\nu=0$ and $g_{\mu\nu}E^\mu u^\nu=0$, together with the decomposition
\begin{equation}
    F_{\nu\sigma}=u_\nu E_\sigma-u_\sigma E_\nu+\epsilon_{\mu\nu\sigma\rho}u^\mu B^\rho,
\end{equation}
the flux $\mathcal F_i$ in \eqref{eq: flux} can be written as
\begin{equation}
  \mathcal F_i=-f(r_i)^{1/2}g^{\rho\sigma}F_{\mu\rho}F_{\nu\sigma}u^\mu n^\nu=-f(r_i)^{1/2}n_\nu\epsilon^{\mu\nu\sigma\rho}u_\mu E_\sigma B_\rho.
\end{equation}
Thus, a necessary condition for a perfectly reflecting wall,
\begin{equation}
     \mathcal F_i\big|_{r=r_i}=0,
\end{equation}
in Schwarzschild coordinates becomes
\begin{equation}
     \sqrt{f(r_i)}\frac{1}{r_i^2}\left(F_{\theta t}F_{\theta r}+\frac{1}{\sin^2\theta}F_{\varphi t}F_{\varphi r}\right)\Bigg|_{r=r_i}=0.
\end{equation}
This condition is gauge-invariant, but by itself it does not uniquely determine the boundary condition for the Maxwell field. A natural stronger condition is the \textit{perfect electric conductor} (PEC) condition \cite{Brito:2015oca}. This means that the electric field measured by an observer at rest with respect to the conductor has no tangential components, while the magnetic field has no normal component:
\begin{equation}
\label{eq: PEC}
   E_\theta\propto F_{\theta t}\big|_{r_i}=0,\qquad E_\varphi\propto F_{\varphi t}\big|_{r_i}=0,\qquad B_r\propto F_{\theta\varphi}\big|_{r_i}=0.
\end{equation}
Note that $F_{ij}$ with $i,j\in\{t,\theta,\varphi\}$ are precisely the components of the pullback $\iota^*F$ of the two-form $F$ to the timelike surface $\Sigma_i$. Therefore, the PEC condition is simply the statement that
\begin{equation}
    \iota^*F=0.
\end{equation}
Using \eqref{eq: EM master fields} together with the Maxwell equations \eqref{eq: Maxwell}, the conditions \eqref{eq: PEC} become \cite{Brito:2015oca}
\begin{equation}
\label{eq: EM master BC}
    \psi_{-}(r_i)=0\quad \text{and}\quad \partial_r\psi_{+}(r_i)=0.
\end{equation}

Let us also note that there also exists a dual electromagnetic reflecting condition, namely, the \textit{perfect magnetic conductor} (PMC) condition, $\iota^*{*F}=0$. It corresponds to the vanishing of the tangential components of the magnetic field and the normal component of the electric field,
\begin{equation}
\label{eq: PMC}
   B_\theta \propto F_{r\varphi}\big|_{r_i}=0,\qquad B_\varphi \propto F_{r\theta}\big|_{r_i}=0,\qquad E_r\propto F_{rt}\big|_{r_i}=0.
\end{equation}

Finally, observe that both PEC and PMC imply VEF. Indeed, the electromagnetic energy flux through the wall is proportional to the normal component of the Poynting vector. The PEC condition eliminates the tangential electric field, while the PMC condition eliminates the tangential magnetic field. In either case, the normal energy flux through $\Sigma_i$ vanishes, so the cavity wall acts as a perfectly reflecting surface.

\begin{figure}[ht]
    \centering
    \includegraphics[width=0.85\linewidth]{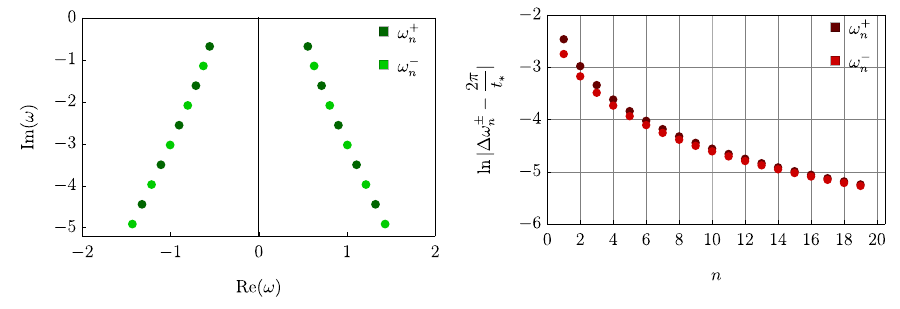}
    \caption{Left: Numerically computed electromagnetic $\ell=1$ quasinormal frequencies $\omega_n^\pm$ for four-dimensional asymptotically flat Schwarzschild spacetime with a reflecting wall at $r_i=1.5$, in units where the black hole horizon radius is equal to $1$. We impose the perfect electric conductor (PEC) boundary condition at the wall, corresponding to Dirichlet boundary conditions on the odd master field, $\psi_-(r_i)=0$, and Neumann boundary conditions on the even master field, $\partial_r\psi_+(r_i)=0$. The frequencies $\omega_n^-$ and $\omega_n^+$ denote the odd and even channel quasinormal modes, respectively. Right: Comparison with the predicted asymptotic spacing. Here, again, $\Delta\omega_n^\pm\equiv\omega_{n+1}^\pm-\omega_n^\pm$ denotes the numerically computed spacings, and $t_*$ is the bouncing time for a geodesic anchored at $r_i$; see Eq.~\eqref{e.tsD4}.}
    \label{fig:EM flat QNMs}
\end{figure}

The numerically computed QNM spectra for the odd and even channels, with the boundary conditions \eqref{eq: EM master BC}, are presented in Fig.~\ref{fig:EM flat QNMs}. Again, we observe that the spacing between overtones quickly approaches $2\pi/t_*$, in agreement with \eqref{e.centralformula}.

\subsubsection{Gravitational Perturbations}

We continue with gravitational perturbations, again considering the case of a four-dimensional asymptotically flat Schwarzschild black hole in a cavity. We now write its metric as
\begin{equation}
\label{eq: 4d bh metric}
ds^2 = g_{ab} dx^a dx^b + r^2 \gamma_{AB} dx^A dx^B,
\end{equation}
so that the coordinates $x^a$ span the $(t, r)$-plane of Schwarzschild spacetime, with lower-case Latin indices $a, b \in \{ 0, 1 \}$. The coordinates $x^A$ span the two-spheres $x^a = \mathrm{const.}$, with upper-case Latin indices $A, B \in \{ 2, 3 \}$, and $\gamma_{AB} = \mathrm{diag}(1, \sin^2 \theta)$ the metric on the unit two-sphere. To consider perturbations, we write
\begin{equation}
\label{eq: metric pert}
g_{\mu \nu} (x) = g^{(0)}_{\mu \nu} (x) + \delta g_{\mu \nu} (x),
\end{equation}
where $g^{(0)}_{\mu \nu} (x)$ is the background black hole metric and $\delta g_{\mu \nu} (x)$ is a small perturbation. The metric $g_{\mu \nu} (x)$ satisfies the vacuum Einstein equations,
\begin{equation}
\label{eq: Einstein eq}
G_{\mu \nu} = 0.
\end{equation}
Substituting \eqref{eq: metric pert} into \eqref{eq: Einstein eq} and keeping only terms linear in $\delta g_{\mu\nu}$, one obtains differential equations for the perturbations. After a decomposition in terms of tensorial spherical harmonics, these equations again fall into two distinct channels, odd and even, with parities $(-1)^{\ell + 1}$ and $(-1)^{\ell}$, respectively.

First, let us define the scalar, vector, and tensor spherical harmonics that are used in the decomposition of the metric perturbation. The scalar harmonics are again the usual spherical harmonics $Y^{\ell m}(x^A)$. Vector spherical harmonics come in two types. Denoting by $D_A$ the covariant derivative on the unit two-sphere, the even-parity harmonics are defined by
\begin{equation}
Y^{\ell m}_A = D_A Y^{\ell m} =
\begin{pmatrix}
\partial_\theta Y^{\ell m} \\
\partial_\varphi Y^{\ell m}
\end{pmatrix},
\end{equation}
and the odd-parity harmonics are
\begin{equation}
X^{\ell m}_A = -\epsilon_A{}^B D_B Y^{\ell m} = 
\begin{pmatrix}
\dfrac{1}{\sin\theta}\partial_\varphi Y^{\ell m} \\
-\sin\theta\,\partial_\theta Y^{\ell m}
\end{pmatrix}.
\end{equation}
These are the same as $\mathbf{Y}^{\ell m}$ and $\mathbf{X}^{\ell m}$ introduced in \eqref{eq: vector harmonics}. Tensorial spherical harmonics also come in two types. The even-parity harmonics are $\gamma_{AB}Y^{\ell m}$ and
\begin{equation}
Y^{\ell m}_{AB} = \left(D_A D_B + \frac{1}{2}\ell(\ell+1)\gamma_{AB}\right)Y^{\ell m},
\end{equation}
while the odd-parity harmonics are
\begin{equation}
X^{\ell m}_{AB} = -\frac{1}{2}\left(\epsilon_A{}^C D_B + \epsilon_B{}^C D_A \right)D_C Y^{\ell m}.
\end{equation}
For more details, such as orthogonality relations and explicit components of the tensor spherical harmonics, see \cite{Martel:2005ir}. We can now decompose the linearised perturbation as
\begin{align}
\label{eq: metric pert 1}
\delta g_{ab} &= \sum_{\ell m} h_{ab}^{\ell m} Y^{\ell m}, \\
\label{eq: metric pert 2}
\delta g_{aB} &= \sum_{\ell m} \left( j_a^{\ell m} Y_B^{\ell m} + h_a^{\ell m} X_B^{\ell m} \right), \\
\label{eq: metric pert 3}
\delta g_{AB} &= \sum_{\ell m}
\left(
r^2 w^{\ell m} \gamma_{AB} Y^{\ell m}
+
r^2 k^{\ell m} Y_{AB}^{\ell m} + h_2^{\ell m} X_{AB}^{\ell m}
\right),
\end{align}
and define the following gauge-invariant combinations:
\begin{align}
\label{eq: gauge inv metric pert 1}
\tilde{h}_{ab} &= h_{ab} - \nabla_a \varepsilon_b - \nabla_b \varepsilon_a,  \\
\label{eq: gauge inv metric pert 2}
\tilde{h}_a &= h_a - \frac{1}{2}\nabla_a h_2 + \frac{1}{r}r_a h_2,  \\
\label{eq: gauge inv metric pert 3}
\tilde{w} &= w + \frac{1}{2}\ell(\ell+1)k - \frac{2}{r}r^a\varepsilon_a,  
\end{align}
where we have defined
\begin{equation}
\varepsilon_a = j_a - \frac{1}{2}r^2\nabla_a k \quad \text{and} \quad r_a = \frac{\partial r}{\partial x^a}.
\end{equation}
Here and below, we suppress the harmonic labels $\ell$ and $m$. Calculations simplify in the \textit{Regge--Wheeler} (RW) \textit{gauge}, in which $j_a = k = h_2 =0$. Relations \eqref{eq: gauge inv metric pert 1}--\eqref{eq: gauge inv metric pert 3} imply that, in the RW gauge, $\tilde{h}_{ab} = h_{ab}$, $\tilde{h}_{a} = h_{a}$ and $\tilde{w} = w$. See, for example, \cite{Martel:2005ir,Grozdanov:2023txs} for more details.

Substituting the perturbed metric \eqref{eq: metric pert 1}--\eqref{eq: metric pert 3} into the Einstein equation \eqref{eq: Einstein eq}, and using the harmonic time dependence $e^{-i\omega t}$, one obtains two radial master equations of the form \eqref{eq: master eq form}. The odd-parity sector can be described by the \textit{Regge--Wheeler function} \cite{Regge:1957td,Martel:2005ir,Grozdanov:2023txs}, whose covariant and gauge-invariant definition is 
\begin{equation}
\label{eq: RW master field}
\psi_{-}(r) = \frac{1}{r}r^a \tilde{h}_a(r).
\end{equation}
It satisfies the master equation \eqref{eq: master eq form} with potential
\begin{equation}
\label{eq: RW potential}
V_-(r) = f(r) \left(\frac{\ell(\ell+1)}{r^2} - \frac{6M}{r^3} \right).
\end{equation}
Similarly, the even-parity sector can be described by the \textit{Zerilli--Moncrief function} \cite{Zerilli:1970se,Martel:2005ir}. Denoting $\lambda = (\ell-1)(\ell+2)/2$, its covariant and gauge-invariant definition is
\begin{equation}
\label{eq: Zerilli master field}
\psi_{+}(r) = \frac{r}{\lambda + 1}\left[\tilde{w}(r) + \frac{r}{\lambda r + 3M} \left(r^a r^b \tilde{h}_{ab}(r) - r r^a \nabla_a \tilde{w}(r)\right)\right].
\end{equation}
The corresponding potential is 
\begin{equation}
\label{eq: Zerilli potential}
V_+(r) =\frac{2f(r)}{r^3(\lambda r+3M)^2}\left(\lambda^2(\lambda+1)r^3+3\lambda^2Mr^2+9\lambda M^2r+9M^3\right).
\end{equation}
Note that in the RW gauge, \eqref{eq: RW master field} reduces to
\begin{equation}
\label{eq: RW master field RW gauge}
\psi_{-}(r) = \frac{f(r)}{r}h_r(r),
\end{equation}
while \eqref{eq: Zerilli master field} becomes
\begin{equation}
\label{eq: Zerilli master field RW gauge}
\psi_{+}(r) = \frac{r}{\lambda + 1}\left[w(r) + \frac{r-2M}{\lambda r + 3M} \left(f(r) h_{rr}(r) - r w'(r)\right)\right].
\end{equation}

Next, we need to impose the appropriate physically motivated boundary conditions. At the horizon, we again impose the usual ingoing wave boundary condition. Imposing boundary conditions at $r=r_i$ is slightly more subtle. Let $\Sigma_i\equiv\{r=r_i\}$ denote the timelike wall. In the cavity problem, the wall is treated as a fixed timelike boundary of the spacetime region under consideration, rather than as a dynamical surface. Thus the embedding $\iota:\Sigma_i\hookrightarrow\mathcal M$ is part of the prescribed background structure. The natural condition is to fix the induced metric $\iota^*g$. At linear order, this gives
\begin{equation}
\iota^*\delta g=0.
\end{equation}
This condition is not meant to be invariant under arbitrary infinitesimal diffeomorphisms of the extended spacetime. Rather, the allowed gauge transformations are those which preserve the boundary and the prescribed boundary metric. In a gauge where the wall remains at $r=r_i$, the condition becomes
\begin{equation}
\delta g_{tt}\big|_{r_i}=0,
\qquad
\delta g_{tA}\big|_{r_i}=0,
\qquad
\delta g_{AB}\big|_{r_i}=0.
\end{equation}
For modes with $\ell\geq 2$, the corresponding conditions can be written
in terms of gauge-invariant combinations as
\begin{equation}
\tilde w(r_i)=0,
\qquad
\tilde{h}_t(r_i)=0,
\qquad
\tilde{h}_{tt}(r_i)=0.
\end{equation}
Since in the RW gauge we have $\tilde{h}_{ab}=h_{ab}$, $\tilde{h}_{a}=h_{a}$
and $\tilde{w}=w$, these boundary conditions are equivalent to
\begin{equation}
\label{eq: RW gauge BC}
w(r_i)=0,
\qquad
h_t(r_i)=0,
\qquad
h_{tt}(r_i)=0.
\end{equation}
It is therefore convenient to work in the RW gauge and impose
\eqref{eq: RW gauge BC} directly.

Finally, we need to translate these boundary conditions into boundary conditions on the master fields $\psi_\pm$. Using the linearised vacuum
Einstein equations, see for example Appendix C of \cite{Martel:2005ir},
we find that, in the odd sector, the condition $h_t(r_i)=0$ corresponds
to the following Robin boundary condition:
\begin{equation}
\label{eq: odd/RW bc}
    \left. f(r) \frac{d}{dr}(r\psi_-)\right|_{r=r_i}=0 .
\end{equation}
In the even sector, we use the vacuum reconstruction formula
\cite{Zerilli:1970se, Cardoso:2001bb}, 
\begin{equation}
    w = f(r)\frac{d\psi_+}{dr}
    + \frac{6M^2+3\lambda Mr+\lambda(\lambda+1)r^2}
    {r^2(\lambda r+3M)}\psi_+,
    \qquad
    \lambda=\frac{(\ell-1)(\ell+2)}{2}.
\end{equation}
Then the condition $w(r_i)=0$ gives the following Robin boundary condition:
\begin{equation}
\label{eq: even/Z bc}
 \left.\left[ f(r)\frac{d\psi_+}{dr}
+ \frac{ 6M^2+3\lambda
Mr+\lambda(\lambda+1)r^2
}{r^2(\lambda r+3M)}\psi_+
\right]\right|_{r=r_i}=0 .   
\end{equation}

\begin{figure}[ht]
    \centering
    \includegraphics[width=0.85\linewidth]{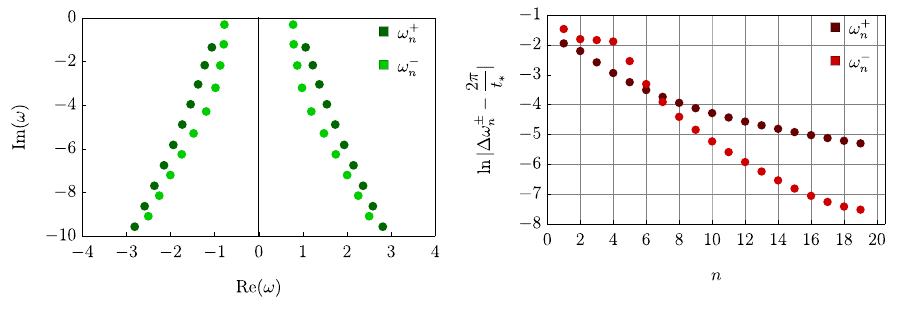}
    \caption{Left: Numerically computed gravitational $\ell=2$ quasinormal frequencies $\omega_n^\pm$ for four-dimensional asymptotically flat Schwarzschild spacetime with a reflecting wall at $r_i=1.5$, in units where the black hole horizon radius is equal to $1$. We impose boundary condition \eqref{eq: odd/RW bc} on the odd master field $\psi_-$, and condition \eqref{eq: even/Z bc} on the even master field $\psi_+$. The frequencies $\omega_n^-$ and $\omega_n^+$ denote the odd and even channel quasinormal modes, respectively. Right: Comparison with the predicted asymptotic spacing. As in all cases above, $\Delta\omega_n^\pm\equiv\omega_{n+1}^\pm-\omega_n^\pm$ denotes the numerically computed spacings, and $t_*$ is the bouncing time for a geodesic anchored at $r_i$; see Eq.~\eqref{e.tsD4}.} 
    \label{fig:grav flat QNMs}
\end{figure}

The numerically computed QNM spectra for the odd and even channels, with the boundary conditions \eqref{eq: odd/RW bc} and \eqref{eq: even/Z bc}, are presented in Fig.~\ref{fig:grav flat QNMs}. Again, we see that the spacing between overtones approaches $2\pi/t_*$ for sufficiently high overtones, in agreement with \eqref{e.centralformula}.

%%%%%%%%%%%%%%%%%%%%%%%%%%%%%%%%%%%%%%%%%%%%%%%%%%%%%
%%%%%%%%%%%%%%%%%%%%%%%%%%%%%%%%%%%%%%%%%%%%%%%%%%%%%

\section{Discussion}\label{s.disc}

In this work, we have extended the study of bouncing geodesics and their corresponding position-space Green's function singularities from asymptotically anti-de Sitter geometries to asymptotically flat ($\Lambda = 0$) and de Sitter ($\Lambda > 0$) spacetimes. By moving away from standard holographic dualities and instead leveraging Hadamard's local representation together with the global propagation of singularities theorem, we have demonstrated that the geometric footprints of black hole curvature singularities are robustly encoded in retarded position-space correlators, independently of any underlying gauge/gravity duality. Furthermore, by enclosing these black holes inside a reflecting cavity, we have established a precise mapping between the ``bouncing times'' $t_*$ and the asymptotic spectrum of cavity quasinormal modes $\omega_n$ through the derivation of the cavity thermal product formula. It is also important to note that, as shown in a number of examples, the spacings between QNMs converge very quickly to the spacing set by $t_*$, already for low-lying modes.

Our results leave a number of open questions for future work. In particular, it would be interesting to understand whether the relation between $t_*^{\Lambda<0}$ and $t_{**}^{\Lambda=0}$ in Eq.~\eqref{e.struklji_is_good_for_you} has a deeper geometric meaning and may offer concrete insight into the relation between asymptotically flat and anti-de Sitter black holes. Moreover, the thermal product formula in AdS has played an essential role in the recent development of the {\em spectral duality relation} \cite{Grozdanov:2024wgo,Grozdanov:2025ner,Grozdanov:2025ulc} and novel QNM sum rules \cite{Grozdanov:2024wgo,Grozdanov:2025ner,Dodelson:2023vrw}. It is therefore natural to ask, now that a thermal product formula is available for black holes with arbitrary $\Lambda$, whether analogous relations could provide interesting new constraints on the quasinormal spectra of black holes in a cavity. Furthermore, given the relation between pole-skipping \cite{Grozdanov:2017ajz,Blake:2017ris,Blake:2018leo,Grozdanov:2018kkt} and the notion of algebraically special frequencies \cite{Grozdanov:2020koi,Grozdanov:2023tag,Grozdanov:2023txs}, which set the structure of the spectral duality relation, it is likely that the framework developed in this paper can provide further insights into the nature of quantum chaos and its connection to pole-skipping in a hypothetical dual description of black holes in de Sitter space, extending the results of \cite{Ahn:2025exp}, or even in asymptotically Minkowski space.

Beyond these questions, it will also be important to understand how gravitational higher-derivative corrections change the nature of bouncing geodesics and singularities, and how they affect the asymptotic behaviour of cavity QNMs. Another interesting direction is the extension of our results to rotating Kerr black holes, which are, in fact, the types of black holes present in our universe. There, the presence of non-vanishing angular momentum breaks spherical symmetry and introduces complex frame-dragging effects, which will presumably significantly enrich the trajectories of the bouncing geodesics. A further, more technical but important question is what the necessary modifications of the cavity thermal product formula are in cases where the physics is described by coupled wave equations. This was discussed in the context of AdS holography in \cite{Bhattacharya:2025vyi}. From the point of view of holography, it also seems important to understand the potential relation between our setup and holography with a cutoff, where one imposes the so-called conformal boundary conditions \cite{Shahbazi-Moghaddam:2024emr,Banihashemi:2025qqi}. Finally, one may also wonder whether our framework could capture certain aspects of the physics of Belinski--Khalatnikov--Lifshitz-type singularities and the corresponding mixmaster dynamics (see, e.g., \cite{DeClerck:2023fax}).

In conclusion, our results suggest that, as in anti-de Sitter space, the information about the interior geometry of black holes in cosmological and flat universes is not entirely isolated from the outside world. If a hypothetical reflecting cavity were to be ``constructed'' surrounding a realistic astrophysical black hole -- a scenario we further discuss in \cite{futurepaper2} -- the curvature singularity in the interior would leave an explicit universal imprint on the exterior spectrum of linearised perturbations. Hence, in a setup that closely resembles established holography in AdS, this would offer a probe, and with it, a potential new perspective on the most extreme environments in our universe. In particular, extracting the bouncing time $t_*$ from the spectrum of QNMs could provide a concrete ``observational'' window into the interiors of black holes. By mapping deviations from the predicted $t_*$, we may thus, in principle, even be able to distinguish between various proposed scenarios, beyond classical General Relativity, for the eventual resolution of singularities. Although this would clearly be difficult in practice, this empirical bridge offers a concrete conceptual way in which we may eventually gain novel insights into the physics of black hole interiors in our universe.

%%%%%%%%%%%%%%%%%%%%%%%%%%%%%%%%%%%%%%%%%%%%%%%%%%%%%
%%%%%%%%%%%%%%%%%%%%%%%%%%%%%%%%%%%%%%%%%%%%%%%%%%%%%

\section*{Acknowledgements}

We would like to thank M.~Bajec, A.~Matevc and G.~Policastro for useful discussions. The work of S.G.\ is supported by the STFC Ernest Rutherford Fellowship ST/T00388X/1. The work is also supported by the research programme P1-0402 and the project J7-60121 of Slovenian Research Agency (ARIS). S.G.\ would also like to thank the Physics Laboratory of the \'{E}cole Normale Sup\'{e}rieure (LPENS) for its hospitality, support, and stimulating conversations during the final stages of this project. V.M.\ is supported by the project J7-60121 of Slovenian Research Agency (ARIS). S.V.\ is supported by the Marie Skłodowska-Curie Actions programme GA-101177446 and the Slovenian Research and Innovation Agency (ARIS), contract number 5110-18/2025-5.

%%%%%%%%%%%%%%%%%%%%%%%%%%%%%%%%%%%%%%%%%%%%%%%%%%%%%
%%%%%%%%%%%%%%%%%%%%%%%%%%%%%%%%%%%%%%%%%%%%%%%%%%%%%

\appendix

\section{Bending of the Penrose Diagrams}\label{a.bending}

\subsection{\texorpdfstring{$\Lambda<0$}{Lambda<0}}
First, we consider the large black hole limit of the Schwarzschild--AdS geometry in $D=5$, with the AdS radius set to one. The blackening factor is
\begin{equation}
f(r)=r^2-\frac{1}{r^2},
\end{equation}
so that an ingoing radial null geodesic satisfies $dt=-dr/f(r)$. If the geodesic starts at the asymptotic boundary $r=\infty$ at Schwarzschild time $t=t_0$, then
\begin{equation}
t(r)-t_0=\int_r^\infty\frac{dr'}{f(r')}=-\frac{1}{2}\left(\arctan r-\frac{\pi}{2}\right)+\frac{1}{2}\left(\operatorname{arctanh}r-\frac{i\pi}{2}\right),
\end{equation}
where the imaginary contribution, which equals $-i\beta/4$, comes from crossing the horizon. A null ray emitted from the boundary at $t=0$ reaches the curvature singularity at
\begin{equation}
t(0)=\frac{\pi}{4}(1-i).
\end{equation}
In particular, the real part of this time is non-zero. Thus, if the asymptotic boundaries are drawn as straight vertical lines in the Penrose diagram, the singularity cannot be represented by a horizontal straight line: the null ray does not hit it at the centre of the diagram. Equivalently, the singularity must bend inwards.

This bending can also be seen directly from the explicit Kruskal construction, as reviewed in Appendices A and B of \cite{Fidkowski:2003nf}. We define the tortoise coordinate by
\begin{equation}
r_*(r)=\int_0^r\frac{dr'}{f(r')}+C=\frac{1}{2}\arctan r-\frac{1}{4}\ln\frac{1+r}{1-r}+C.
\end{equation}
When the integration contour crosses the horizon at $r=1$, the logarithm acquires an imaginary contribution. Choosing $C= i \pi/4$ makes $r_*$ real in the exterior region $r>1$. With this convention,
\begin{equation}
e^{4r_*}=\frac{r-1}{r+1}e^{2\arctan r}.
\end{equation}
Introducing the Eddington--Finkelstein null coordinates $u=t-r_*$ and $v=t+r_*$ and then Kruskal coordinates
\begin{equation}
U=-e^{-\kappa u},\qquad V=e^{\kappa v},
\end{equation}
removes the coordinate singularity at the horizon. Here $\kappa=\frac{1}{2}f'(r_h)$ is the surface gravity. In the present case $r_h=1$ and $\kappa=2$. In these coordinates, suppressing the angular directions, the metric becomes
\begin{equation}
ds^2=-\frac{f(r)}{4}e^{-4r_*}dUdV,
\end{equation}
where $r$, and hence $f(r)$ and $r_*(r)$, is understood as a function of $UV$. Letting $U=T-X$ and $V=T+X$, the metric takes the conformally flat form
\begin{equation}
ds^2=g(T,X)(-dT^2+dX^2),
\end{equation}
where the conformal factor $g(T,X)$ is nonsingular at the horizon. We may therefore extend the exterior region $U<0$, $V>0$ to the full Kruskal plane with $U,V\in(-\infty,\infty)$, subject to the constraint $0<r<\infty$. The Kruskal and Schwarzschild coordinates are related by
\begin{equation}
-e^{4r_*(r)}=T^2-X^2=UV,
\end{equation}
and
\begin{equation}
\tanh(2t)=\frac{T}{X}=\frac{V+U}{V-U}.
\end{equation}
Thus, surfaces of constant $r$ correspond to hyperbolas of constant $T^2-X^2$. In particular, we have
\begin{align}
    &\text{singularity } (r=0): \quad T^2 - X^2 = 1 \\
    &\text{boundary } (r=\infty): \quad T^2 - X^2 = -e^{\pi} 
\end{align}
Already at this stage, one can see the relative bending of the singularity. Since $e^\pi>1$, the singularity lies closer to the origin of the Kruskal plane than the asymptotic boundary. Consequently, a radial null geodesic emitted from the boundary at $t=0$, for which $T=0$, cannot hit the singularity at the midpoint $X=0$. To obtain the Penrose diagram, we compactify the Kruskal coordinates. We define
\begin{equation}
V=e^{\pi/2}\tan\frac{\tilde v}{2},\qquad U=e^{\pi/2}\tan\frac{\tilde u}{2},
\end{equation}
so that the spacetime is mapped into a finite region with $\tilde u,\tilde v\in(-\pi,\pi)$. Introducing $\tilde v = \tau + \rho$, $\tilde u = \tau - \rho$, and using the relation
\begin{equation}
    \tan(\frac{\tau + \rho}{2}) \tan(\frac{\tau - \rho}{2}) = \frac{\cos{\rho}- \cos{\tau}}{\cos{\rho} + \cos{\tau}},
\end{equation}
we can check that the boundaries
\begin{equation}
    UV = -e^{\pi} \iff \frac{\cos{\rho}- \cos{\tau}}{\cos{\rho} + \cos{\tau}} = - 1 \iff \cos{\rho}=0
\end{equation}
are represented by straight lines $\rho = \pm \frac{\pi}{2}$, $\tau \in (-\frac{\pi}{2}, \frac{\pi}{2})$. However, the singularities,
\begin{equation}
    UV = 1 \iff \frac{\cos{\rho}- \cos{\tau}}{\cos{\rho} + \cos{\tau}} = e^{-\pi} \iff \cos{\tau}= \frac{1-e^{-\pi}}{1+e^{-\pi}} \cos{\rho},
\end{equation}
are not straight horizontal lines in these compactified coordinates. Rather, they bend inwards. Indeed, at $\rho=\pm\pi/2$ one finds $\cos\tau=0$, hence $\tau=\pm\pi/2$, while at $\rho=0$ one obtains
\begin{equation}
\cos\tau=\frac{1-e^{-\pi}}{1+e^{-\pi}}\approx0.9,
\end{equation}
which gives $\tau\approx\pm0.4$. One could instead choose a compactification that straightens the singularity, for example by defining $V=\tan\frac{\tilde v}{2}$ and $U=\tan\frac{\tilde u}{2}$. With this choice, however, the asymptotic boundary bends outwards. There is in fact no compactification in which both the boundary and the singularity are represented by straight lines. See \cite{Fidkowski:2003nf} for further discussion.

\subsection{\texorpdfstring{$\Lambda>0$}{Lambda>0}}

Next, consider the Schwarzschild--de Sitter spacetime. In $D=5$, setting the de Sitter radius to one, the blackening factor is
\begin{equation}
f(r)=1-r^2-\frac{\mu}{r^2}.
\end{equation}
We fix the observer at the static sphere $r=r_{\mathcal O}$ and use its proper time
$\tilde t=\gamma^{-1}t=\sqrt{f(r_{\mathcal O})}t$. Computing the time $\tilde t(r)$ along an ingoing/outgoing radial null geodesic which starts from $r=r_{\mathcal O}$ at $\tilde t_0=0$, we find
\begin{equation}
\tilde t(0)=\frac{1}{\gamma}\int_0^{r_{\mathcal O}}\frac{dr'}{f(r')}=\frac{r_c-r_b}{r_b+r_c}\operatorname{arctanh}\sqrt{\frac{r_b}{r_c}}-\frac{i\beta_b}{4},
\end{equation}
and
\begin{equation}
\tilde t(\infty)=-\frac{1}{\gamma}\int_{r_{\mathcal O}}^\infty\frac{dr'}{f(r')}=-\frac{r_c-r_b}{r_b+r_c}\operatorname{arctanh}\sqrt{\frac{r_b}{r_c}}+\frac{i\beta_c}{4}.
\end{equation}
Since the real parts of $\tilde t(0)$ and $\tilde t(\infty)$ are non-zero, a geodesic starting at $r_{\mathcal O}$ reaches the singularity and spacelike infinity off-centre in the Penrose diagram. Moreover, since $\Re\tilde t(0)>0$ and $\Re\tilde t(\infty)<0$, the singularity must bend inwards, while spacelike infinity must bend outwards, assuming that the static sphere observer is drawn as a straight vertical line. In general, this bending occurs in any dimension $D>3$, but only in $D=5$ is the inwards bending of the singularity equal to the outwards bending of spacelike infinity \cite{Faruk:2023uzs}. 

We can again see this bending by explicitly constructing the Schwarzschild--de Sitter black hole Penrose diagram. We start by defining the tortoise coordinate with respect to the static-sphere proper time,
\begin{align}
\label{e.tortoise}
r_*(r)&=\frac{1}{\gamma}\int_{r_{\mathcal O}}^r\frac{dr'}{f(r')} \\
&=\frac{1}{\gamma}\frac{1}{r_c^2-r_b^2}\left(r_c\operatorname{arctanh}\frac{r}{r_c}-r_b\operatorname{arctanh}\frac{r_b}{r}\right)
-\frac{1}{\gamma}\frac{1}{r_c+r_b}\operatorname{arctanh}\sqrt{\frac{r_b}{r_c}},
\nonumber
\end{align}
so that $r_*=0$ at $r=r_{\mathcal O}$ and $r_*$ is real in the static patch $r\in(r_b,r_c)$. The factor $1/\gamma$ appears because we are using the normalised static time $\tilde t=\gamma^{-1}t$; equivalently, the corresponding surface gravities are those associated with the normalised Killing vector. Next, we introduce the Eddington--Finkelstein coordinates $u=\tilde t-r_*$ and $v=\tilde t+r_*$. Since $\kappa_b\neq\kappa_c$, there is no single Kruskal coordinate system which is nonsingular at both horizons. We therefore introduce two Kruskal charts, separately regular at the black hole and cosmological horizons \cite{Aguilar-Gutierrez:2024rka},
\begin{equation}
U_b=-e^{-\kappa_b u},\qquad V_b=e^{\kappa_b v},
\qquad
U_c=e^{\kappa_c u},\qquad V_c=-e^{-\kappa_c v}.
\end{equation}
Here $\kappa_b$ and $\kappa_c$ are the normalised surface gravities of the black hole and cosmological horizons, respectively. Both coordinate systems are well defined in the central static patch $r\in(r_b,r_c)$, where one can pass from one chart to the other.

As before, we extend each pair $(U_b,V_b)$ and $(U_c,V_c)$ to $U,V\in(-\infty,\infty)$, subject to the constraint $0<r<\infty$, and set
\begin{equation}
U_{b,c}=T_{b,c}-X_{b,c},\qquad V_{b,c}=T_{b,c}+X_{b,c}.
\end{equation}
The relation to the original coordinates is then
\begin{equation}
\label{e.SdS coord relations}
T_b^2-X_b^2=U_bV_b=-e^{2\kappa_b r_*(r)},
\qquad
T_c^2-X_c^2=U_cV_c=-e^{-2\kappa_c r_*(r)}.
\end{equation}
It follows that constant-$r$ surfaces correspond to curves of constant $T^2-X^2$ in the appropriate Kruskal chart. In particular, inserting \eqref{e.tortoise} into \eqref{e.SdS coord relations}, we obtain
\begin{align}
\text{singularity }(r=0):&\qquad T_b^2-X_b^2=\left(\frac{\sqrt{r_c}-\sqrt{r_b}}{\sqrt{r_c}+\sqrt{r_b}}\right)^{\frac{r_c-r_b}{r_b}}\equiv k_0,\\
\text{static sphere }(r=r_{\mathcal O}):&\qquad T_{b,c}^2-X_{b,c}^2=-1,\\
\text{spacelike infinity }(r=\infty):&\qquad T_c^2-X_c^2=\left(\frac{\sqrt{r_c}+\sqrt{r_b}}{\sqrt{r_c}-\sqrt{r_b}}\right)^{\frac{r_c-r_b}{r_c}}\equiv k_\infty.
\end{align} 
To obtain the Penrose diagram, we compactify each Kruskal chart separately by defining
\begin{equation}
V_{b,c}=\tan\frac{\tilde v_{b,c}}{2},\qquad U_{b,c}=\tan\frac{\tilde u_{b,c}}{2},
\end{equation}
with $\tilde u_{b,c},\tilde v_{b,c}\in(-\pi,\pi)$. We then introduce
\begin{equation}
\tilde v_{b,c}=\tau_{b,c}+\rho_{b,c},\qquad \tilde u_{b,c}=\tau_{b,c}-\rho_{b,c}.
\end{equation}
Using the identity
\begin{equation}
\tan\frac{\tau+\rho}{2}\tan\frac{\tau-\rho}{2}
=\frac{\cos\rho-\cos\tau}{\cos\rho+\cos\tau},
\end{equation}
we can identify the images of the static sphere, the singularity, and spacelike infinity. First, the static sphere satisfies $U_{b,c}V_{b,c}=-1$. Therefore,
\begin{equation}
U_{b,c}V_{b,c}=-1
\iff
\frac{\cos\rho_{b,c}-\cos\tau_{b,c}}{\cos\rho_{b,c}+\cos\tau_{b,c}}=-1
\iff
\cos\rho_{b,c}=0.
\end{equation}
Thus the static sphere observers are represented by the straight lines
\begin{equation}
\rho_{b,c}=\pm\frac{\pi}{2},\qquad \tau_{b,c}\in\left(-\frac{\pi}{2},\frac{\pi}{2}\right).
\end{equation}
On the other hand, the black hole singularities satisfy $U_bV_b=k_0$. Hence
\begin{equation}
U_bV_b=k_0
\iff
\frac{\cos\rho_b-\cos\tau_b}{\cos\rho_b+\cos\tau_b}=k_0
\iff
\cos\tau_b=\frac{1-k_0}{1+k_0}\cos\rho_b.
\end{equation}
They are therefore not represented by straight horizontal lines, but bend inwards. Indeed, at $\rho_b=\pm\pi/2$ the above relation gives $\cos\tau_b=0$, hence $\tau_b=\pm\pi/2$. At $\rho_b=0$, one obtains
\begin{equation}
\cos\tau_b=\frac{1-k_0}{1+k_0},
\end{equation}
which corresponds to $\tau_b\in(-\pi/2,\pi/2)$ since $0<k_0<1$. Similarly, spacelike infinities satisfy $U_cV_c=k_\infty$. Thus
\begin{equation}
U_cV_c=k_\infty
\iff
\frac{\cos\rho_c-\cos\tau_c}{\cos\rho_c+\cos\tau_c}=k_\infty
\iff
\cos\tau_c=\frac{1-k_\infty}{1+k_\infty}\cos\rho_c.
\end{equation}
They therefore bend outwards. In particular, at $\rho_c=0$ one has
\begin{equation}
\cos\tau_c=\frac{1-k_\infty}{1+k_\infty},
\end{equation}
which corresponds to $|\tau_c|\in(\pi/2,\pi)$ since $k_\infty>1$.

At first sight, the two bendings do not appear to agree. Indeed, if the two compactifications are compared independently, equality of the bending at $\rho=0$ would require $k_\infty=1/k_0$, which is not true. This is not a contradiction, however, because the compactified time coordinates in the two Kruskal charts are not normalised in the same way on the overlap of the charts. Along the static sphere $r=r_{\mathcal O}$, the black-hole chart depends on $\kappa_b$, whereas the cosmological chart depends on $\kappa_c$. Since $\kappa_b\neq\kappa_c$, the variables $\rho_b, \tau_b$ and $\rho_c, \tau_c$ do not represent the same parametrisation of the static sphere observer. To compare the two bendings, one must first match this parametrisation on the static sphere. Taking the black-hole chart as reference amounts to replacing $k_\infty$ by $k_\infty^{\kappa_b/\kappa_c}$. In $D=5$ we have $\kappa_b/\kappa_c=r_c/r_b$, and therefore
\begin{equation}
k_\infty^{\kappa_b/\kappa_c}
=
\left[
\left(\frac{1+\sqrt{r_b/r_c}}{1-\sqrt{r_b/r_c}}\right)^{\frac{r_c-r_b}{r_c}}
\right]^{\frac{r_c}{r_b}}
=
\left(\frac{1+\sqrt{r_b/r_c}}{1-\sqrt{r_b/r_c}}\right)^{\frac{r_c-r_b}{r_b}}
=
\frac{1}{k_0}.
\end{equation}
Consequently,
\begin{equation}
\frac{1-k_0}{1+k_0}
=
-\frac{1-k_\infty^{\kappa_b/\kappa_c}}{1+k_\infty^{\kappa_b/\kappa_c}}.
\end{equation}
Thus, after matching the parametrisation of the static sphere in the two charts, the inwards bending of the singularity equals the outwards bending of spacelike infinity in $D=5$. This statement is to be understood relative to the convention used throughout this construction: the static sphere observer is drawn as a straight vertical line. One could choose a different compactification, for example one which straightens spacelike infinity, but then the static sphere observer would no longer remain a straight vertical line in general. The bending is therefore not an invariant feature of either curve separately; it is a statement about their relative shape with respect to the chosen reference worldline.

\subsection{\texorpdfstring{$\Lambda=0$}{Lambda=0}}

Finally, consider the asymptotically flat Schwarzschild geometry in $D=d+1$ dimensions, with the horizon radius set to one. The blackening factor is
\begin{equation}
f(r)=1-\frac{1}{r^{d-2}}.
\end{equation}
We again focus on the case $D=5$, so that
\begin{equation}
f(r)=1-\frac{1}{r^2}.
\end{equation}
For an ingoing radial null geodesic starting at $r=r_i$ and $t=t_0$, one finds
\begin{equation}
t(r)-t_0=\int_r^{r_0}\frac{dr'}{f(r')}=(r_0-r)-\left(\operatorname{arctanh}r_0-\operatorname{arctanh}r\right).
\end{equation}
However, this expression should not be used by simply setting $r_0=\infty$. In the asymptotically flat case, the relevant boundary is null infinity rather than a timelike conformal boundary. Points on $\mathcal{I}^\pm$ are obtained by taking $r\to\infty$ and $t\to\pm\infty$ while keeping the appropriate null coordinate finite. Thus the Schwarzschild coordinate time diverges along $\mathcal{I}^\pm$, and the naive coordinate-time argument used in the AdS case does not directly determine the bending of the Penrose diagram. Instead, the bending is most cleanly read off from the explicit construction of the Penrose diagram. Define the tortoise coordinate by
\begin{equation}
r_*(r)=\int_0^r\frac{dr'}{f(r')}+C=r-\operatorname{arctanh}r+C.
\end{equation}
The integral acquires an imaginary contribution when continued across the horizon at $r=1$. Choosing $C=\frac{i\pi}{2}$ makes $r_*$ real in the exterior region $r>1$. We then again introduce the null Eddington-Finkelstein coordinates $u=t-r_*$ and $v=t+r_*$ and the Kruskal coordinates
\begin{equation}
U=-e^{-\kappa u},\qquad V=e^{\kappa v}.
\end{equation}
For the present metric, $\kappa=\frac{1}{2}f'(1)=1$. Hence,
\begin{equation}
UV=-e^{2r_*(r)}=-e^{2r}e^{-2\operatorname{arctanh}r+i\pi}=\frac{1-r}{1+r}e^{2r}.
\end{equation}
We compactify the Kruskal coordinates by defining
\begin{equation}
V=\tan\frac{\tilde v}{2},\qquad U=\tan\frac{\tilde u}{2},
\end{equation}
with $\tilde u,\tilde v\in(-\pi,\pi)$. Introducing $\tilde v=\tau+\rho$ and $\tilde u=\tau-\rho$, the boundary $r=\infty$ corresponds to
\begin{equation}
UV=-\infty
\iff
\tan\frac{\tilde v}{2}\tan\frac{\tilde u}{2}=-\infty
\iff
\tilde v=\pm\pi\quad\text{or}\quad \tilde u=\pm\pi.
\end{equation}
Thus null infinity is represented by the four straight lines
\begin{equation}
\tau+\rho=\pm\pi,\qquad \tau-\rho=\pm\pi.
\end{equation}
On the other hand, the singularities at $r=0$ correspond to $UV=1$. Using
\begin{equation}
\tan\frac{\tau+\rho}{2}\tan\frac{\tau-\rho}{2}
=\frac{\cos\rho-\cos\tau}{\cos\rho+\cos\tau},
\end{equation}
we find
\begin{equation}
UV=1
\iff
\frac{\cos\rho-\cos\tau}{\cos\rho+\cos\tau}=1
\iff
\cos\tau=0.
\end{equation}
Therefore the future and past singularities are represented by the straight lines
\begin{equation}
\tau=\pm\frac{\pi}{2}.
\end{equation}

Hence, in the asymptotically flat Schwarzschild case, the singularities and null infinities can be made straight simultaneously in the compactified Kruskal diagram. In this sense, the asymptotically flat Schwarzschild diagram exhibits no inwards bending of the singularity relative to null infinity.

%%%%%%%%%%%%%%%%%%%%%%%%%%%%%%%%%%%%%%%%%%%%%%%%%%%%%
%%%%%%%%%%%%%%%%%%%%%%%%%%%%%%%%%%%%%%%%%%%%%%%%%%%%%

\section{Propagation of Singularities}
\label{a.propag}

In this appendix, we briefly summarise the microlocal terminology needed to state the propagation of singularities theorem \cite{Hormander:2007,Hormander:2009,Reed:1975uy}. Throughout, $M$ denotes a smooth $n$-dimensional manifold, $\mathcal{D}(M) \equiv C_c^\infty(M)$ denotes the space of smooth compactly supported test functions, and $\mathcal{D}'(M)$  the corresponding space of distributions.

We first fix the regularity terminology for distributions. A distribution $T\in\mathcal{D}'(M)$ is called \textit{regular} if there exists a locally integrable function $T(x)$ on $M$ such that
\begin{equation}
\label{eq: regular distrib}
    T[\phi]=\int T(x)\phi(x)\sqrt{|g|}\,d^n x,
    \quad \phi\in\mathcal{D}(M).
\end{equation}
If $T(x)$ can be chosen smooth, we call $T$ a \textit{smooth distribution}. More locally, $T$ is said to be \textit{smooth at} $x\in M$ if there exists $\psi\in\mathcal{D}(M)$ with $\psi(x)\neq 0$ such that $\psi T$ is smooth. The \textit{singular support} $\singsupp(T)$ is the complement of the largest open subset of $M$ on which $T$ is smooth. Colloquially, it is the set on which $T$ is ``singular''. We will also use the standard distributional extension of differential operators. For $T\in\mathcal{D}'(\mathbb{R}^n)$, its derivative is defined by
\begin{equation}
    (\partial_\mu T)[\phi]\equiv -T[\partial_\mu\phi],
    \quad \phi\in\mathcal{D}(\mathbb{R}^n).
\end{equation}
We note that this definition is motivated by the following identity, which follows from a simple integration by parts: If $T$ is represented by a smooth function $T(x)$, then for every $\phi\in\mathcal D(\mathbb R^n)$,
\begin{equation}
    \int (\partial_\mu T)(x)\phi(x) \, \sqrt{|g|} \, d^nx = -\int T(x)(\partial_\mu\phi)(x) \, \sqrt{|g|} \,d^nx,
\end{equation}
because $\phi$ has compact support. More generally, if $A$ is a linear differential operator on $\mathbb{R}^n$, then $AT$ is defined by
\begin{equation}
    (AT)[\phi]\equiv T[A^\dagger\phi],
\end{equation}
where $A^\dagger$ is the adjoint of $A$, determined by
\begin{equation}
    \int_{\mathbb{R}^n}(A\phi(x))\psi(x)\,d^n x
    =
    \int_{\mathbb{R}^n}\phi(x)(A^\dagger\psi(x))\,d^n x
\end{equation}
for all $\phi,\psi\in\mathcal{D}(\mathbb{R}^n)$. 

To study singularities of distributions, it is useful to analyse their behaviour in frequency space via the Fourier transform. Let $T\in\mathcal{D}'(\mathbb{R}^n)$ have compact support $K\subset\mathbb{R}^n$, meaning that $T[\phi]=0$ for all $\phi\in\mathcal{D}(\mathbb{R}^n)$ with $\supp\phi\cap K=\emptyset$. The \emph{Fourier transform of $T$} is given by
\begin{equation}
    \widehat{T}[\phi]\equiv T[\psi\widehat{\phi}],
\end{equation}
where $\psi\in\mathcal{D}(\mathbb{R}^n)$ is any test function satisfying $\psi|_K=1$. Compact support of $T$ implies that $\widehat{T}$ is represented by a smooth function on $\mathbb{R}^n$ \cite[Theorem IX.5]{Reed:1975uy}. Moreover, smoothness of $T$ is reflected in the decay of its Fourier transform \cite[Theorems IX.11 and IX.12]{Reed:1975uy}: $T$ is smooth if and only if $\widehat{T}$ decays faster than any inverse power of $|k|$, i.e., for every $N\in\mathbb{N}$ there exists $C_N>0$ such that
\begin{equation}
    |\widehat{T}(k)|\leq C_N(1+|k|)^{-N}.
\end{equation}
On a general manifold $M$, this criterion is applied locally in coordinate charts. The resulting notions are independent of the choice of coordinates.

The same Fourier-space decay can also be used to measure the strength of a singularity. This leads to the local Sobolev spaces. Let $s\in\mathbb{R}$. We say that a distribution $T\in\mathcal{D}'(M)$ lies in the \emph{local Sobolev space} $H^s_{\mathrm{loc}}(x)$ at a point $x\in M$ if there exists a test function $\psi\in\mathcal{D}(M)$ with $\psi(x)\neq 0$ such that, in local coordinates near $x$,
\begin{equation}\label{Sobolev}
    \int_{\mathbb{R}^n}(1+|k|^2)^s|\widehat{\psi T}(k)|^2\,d^n k<\infty.
\end{equation}
Larger values of $s$ correspond to stronger regularity. In particular, if $T\in H^s_{\mathrm{loc}}(x)$ with $s>\frac{n}{2}+\ell$, then $T$ is $C^\ell$ near $x$. Moreover,
\begin{equation}
    T\in H^s_{\mathrm{loc}}(x)\ \forall s\in\mathbb{R}
    \quad\Longleftrightarrow\quad
    T\text{ is smooth at }x.
\end{equation}
Conversely, more singular distributions have lower Sobolev regularity. For example, the Dirac delta distribution at $x$  belongs to $H^s_{\mathrm{loc}}(x)$ if and only if $s<-\frac{n}{2}$. 

The notion of singular support introduced above tells us \textit{where} the distribution fails to be smooth. However, in the study of partial differential equations, we often also wish to understand \textit{in which directions} a distribution is singular. Microlocal analysis provides a refinement of singular support by encoding both position and directional information. The basic idea is to localise a distribution near a point and analyse its Fourier transform: while smooth behaviour corresponds to rapid decay in all directions in frequency space, singularities manifest as directions in which this decay fails. This leads to the notion of the \textit{wave front set}, which is a subset of the cotangent bundle $T^*M$ and can be thought of as a directional refinement of the singular support.

Specifically, let $T \in \mathcal{D}'(M)$ and let $(x,k) \in T^*M \setminus  \mathbf{0}_M$. We say that $T$ is \emph{smooth at $(x,k)$} if there exists a test function $\psi \in \mathcal{D}(M)$ with $\psi(x) \neq 0$, and a conic neighbourhood $\Gamma \subset T^*_{x}M \setminus \{0\}$ of $k$, such that for all $N \in \mathbb{N}$ there exists $C_N > 0$ with
\begin{equation}
    |\widehat{\psi T}( k')| \leq C_N (1+|k'|)^{-N} \quad \text{for all } k' \in \Gamma.
\end{equation}
Here, conic means that $k' \in \Gamma$ implies $\lambda k' \in \Gamma$ for all $\lambda > 0$. We then define the \textit{wave front set} of $T$, denoted $\WF(T)$, as the set of points $(x, k) \in T^*M$, where  $k \in T^*_x M \setminus \{0\}$ is a cotangent direction in which the distribution fails to be smooth.
More precisely,
\begin{equation}
        (x, k) \in \WF(T) \iff k \neq 0 \text{ and $T$ is not smooth at $(x, k) \in T^*M $}. 
\end{equation}
It can be shown \cite[Theorem IX.44]{Reed:1975uy} that the notion of smoothness of $T$ at $(x, k)$ is independent of the choice of local coordinates at $x$ and is therefore well defined. Further, we have
\begin{equation}
\label{eq: WF equiv}
    x \in \singsupp T \iff \exists k \in T^*_xM  \text{ such that } (x, k) \in \WF(T).
\end{equation} 
In fact, $\singsupp(T) = \pi\left( \WF(T)\right)$, where $\pi \colon T^*M \to M$ is the canonical bundle projection. To give an example, consider the radial Heaviside distribution
\begin{equation}
T(x)=\Theta(|x|-1)
\end{equation}
on $\mathbb{R}^n$. This distribution is smooth away from the sphere $\mathbb{S}^{n-1}$, i.e.,
~it has $\singsupp(T)=\mathbb{S}^{n-1}.$ The jump is transverse to the sphere, while there is no singularity in the tangential directions -- the singular covectors are precisely the non-zero covectors normal to the sphere. Indeed, one can check that
\begin{equation}
\WF(T)= \left\{ (x, \lambda x_i \,dx^i) \in T^*\mathbb{R}^n ~|~ |x|=1, \, \lambda\in\mathbb{R}\setminus{0} \right\}.
\end{equation}
So the wave front set refines the singular support by saying not only that the singularity lies on the sphere, but also that its singular directions are normal to the sphere.

Similarly, we can refine the notion of local Sobolev regularity by incorporating directional information. Let $s \in \mathbb{R}$. We say that a distribution $T \in \mathcal{D}'(M)$ lies in the \emph{microlocal Sobolev space} $H^s_{\mathrm{loc}}(x,k)$ at the point $(x, k)$ if there exists another distribution $T_1$, such that
\begin{equation}
    T_1 \in H^s_{\mathrm{loc}}(x) \quad \text{and} \quad (x, k) \notin \WF(T - T_1).
\end{equation}
That is, $T_1$ has Sobolev regularity $s$ at $x$, while $T-T_1$ is smooth at $(x, k)$. Hence $T\in H^s_{\mathrm{loc}}(x,k)$ means that the singular behaviour of $T$ in the direction $k$ is no worse than Sobolev order $s$. This notion is compatible with ordinary local Sobolev regularity: one has \cite[Theorem 18.1.31]{Hormander:2007}
\begin{equation}
   T \in H^s_{\mathrm{loc}}(x) \quad \Longleftrightarrow \quad T \in H^s_{\mathrm{loc}}(x,k)
\ \text{for all } k \in T_x^*M \setminus \{ 0 \}. 
\end{equation}
Furthermore, for $(x,k) \in T^*M \setminus \mathbf{0}_M$, we have
\begin{equation}
 T \in H^s_{\mathrm{loc}}(x, k) \text{ for all } s \in \mathbb{R} \quad \Longleftrightarrow \quad 
T \text{ is smooth at } (x,k),  
\end{equation}
so microlocal smoothness is equivalent to microlocal Sobolev regularity of every order.

We continue with a discussion of linear partial differential operators. Let $A \colon C^\infty(M) \to C^\infty(M)$ be a linear partial differential operator of order $m$ on $M$, expressed in local coordinates as 
\begin{equation}
        A = \sum_{i=0}^m \alpha_{(i)}^{j_1 \ldots j_i} \nabla_{j_1} \ldots \nabla_{j_i}, 
\end{equation}
where $\nabla$ is some derivative operator on $M$ and  $\alpha_{(i)}^{j_1 \ldots j_i}$ smooth tensor fields on $M$. The \textit{principal symbol} $\sigma_A$ of $A$ is defined as a smooth function
\begin{equation}
        \sigma_A \colon T^*M \to \mathbb{R}, \quad \sigma_A(x, p) = \alpha_{(m)}^{j_1 \ldots j_m}(x) p_{j_1} \ldots p_{j_m}.
\end{equation}
One can check that $\sigma_A$ is independent of the choice of derivative operator $\nabla$. We then define the \textit{characteristic set} of $A$ as 
\begin{equation}
        \Char(A) \equiv\{ (x, p) \in T^*M \setminus \mathbf{0}_M \: | \: \sigma_A(x, p)=0 \}.
\end{equation}
There is an intuitive way to think about the above defined notions. First, note that there is a canonical \textit{symplectic structure} on $T^*M$ (denote the canonical symplectic form with $\omega$), so the cotangent bundle $T^*M$ can be viewed as a phase space of a dynamical system. Choosing $H=\sigma_A$ to be the Hamiltonian of the dynamical system, $\Char(A)$ consists of states in phase space with zero energy but non-vanishing momentum. Associated with $H$, we also have the Hamiltonian vector field $X_H$ on $T^*M$, which is defined by $\iota_{X_H}\omega = dH$ or, in local coordinates,
\begin{equation}
  X_H =
\frac{\partial H}{\partial p_i}\frac{\partial}{\partial x^i}
- \frac{\partial H}{\partial x^i}\frac{\partial}{\partial p_i}.  
\end{equation}
We then define the \textit{bicharacteristics} of a linear differential operator $A$ as the integral curves of the associated vector field $X_{\sigma_A}$ 
on $T^*M \setminus \mathbf{0}_M$, which start in $\Char(A)$. Since the Hamiltonian is constant along the integral curves of $X_{\sigma_A}$, it follows that the bicharacteristics of the corresponding differential operator $A$ are entirely contained in $\mathrm{Char}(A)$. For example, consider the wave operator $P=\partial_t^2-\partial_z^2$ on $\mathbb{R}^{1+1}$. Denote $x \equiv (t, z)$ and $p \equiv (\tau, \xi)$. Its principal symbol is $\sigma_P(x, p)=\tau^2-\xi^2$ and hence the characteristic set is
\begin{equation}
\Char(P)=\{ (x, p) \in T^*\mathbb{R}^2  ~|~ \tau^2-\xi^2=0, \,  (\tau,\xi)\neq 0\}.
\end{equation}
The corresponding bicharacteristics are given by the following lines in $\mathbb{R}^2 \times \mathbb{R}^2$:
\begin{equation}
\tau + \xi = 0, \, t + z = \mathrm{const.} \quad \text{and} \quad \tau - \xi = 0, \, t - z = \mathrm{const.} 
\end{equation}

We can now finally state the \textit{Propagation of Singularities Theorem}, which is used to show our claim regarding the singularities of the retarded Green's function. The following version is obtained by restricting Theorems 26.1.1 and 26.1.4 in \cite{Hormander:2009} from the case of pseudodifferential operators to
the simpler case of linear partial differential operators. 

\begin{theorem}
 \label{thm: propagation of singularities}
    Let $(M, g)$ be a smooth $n$-dimensional manifold. Let $P$ be a partial differential operator with real valued principal symbol of degree $m$ and suppose $T \in \mathcal{D}'(M)$ satisfies the equation $P T = 0$. Then
    \begin{enumerate}[label=\roman*)]
        \item $\WF(T) \subseteq \Char(P)$, 
        \item for any $(x, k) \in \Char(P)$ we have $T \in H^s_{\mathrm{loc}}(x, k)$ if and only if $T \in H^s_{\mathrm{loc}}(x', k') $ for all $(x', k') $ lying on the same bicharacteristic as $(x, k)$.
    \end{enumerate}
Thus, in particular, if $(x, k) \in \WF(T)$, then the whole bicharacteristic through $(x, k)$ lies in $\WF(T)$.
\end{theorem}   
The above theorem tells us that the solution $T$ to $PT=0$ has the same (ir)regularity along bicharacteristics. More precisely, suppose there is a point $x\in M$ at which the solution $T$ is not smooth. Then there is some $k \in T_x^*M$ such that $(x,k)\in\WF(T)$ and some $s\in \mathbb{R}$ such that $T\not\in H^s_{\mathrm{loc}}(x, k)$.  Now the first point of the theorem guarantees that $(x,k)\in\Char(P)$, while the second tells us that for any other point $(x',k')$ lying on the same bicharacteristic as $(x,k)$ we have $T\not\in H^s_{\mathrm{loc}}(x',k')$, i.e.~$T$ has the same “strength” of singularity (as measured by Sobolev space index $s$ in which the distribution fails to lie) at $x'$ as it does at $x$.

We can apply this theorem to the retarded Green's function viewed as a distribution $(G)_{x'}$ defined by 
\begin{equation}
    (G)_{x'}[\phi] \equiv \int G(x, x') \phi(x) \sqrt{|g|}\,d^n x.
\end{equation}
Observe that, in particular,  $(G)_{x'}$ is a solution to
\begin{equation}
    P (G)_{x'} = 0 \text{ on $M \setminus \{ x' \}$},
\end{equation}
where $P$ is a normally hyperbolic differential operator. We have $\sigma_P(x, p) = g^{\mu \nu} p_\mu p_\nu$ and it is straightforward to check that the bicharacteristics of $P$ are precisely the lifts of affinely parametrised null geodesics in $M$ to the cotangent bundle $T^*M$ with $p_\mu = \frac{1}{2}g_{\mu \nu} \dot{x}^\nu$. Next, let
\begin{equation}
    \Sigma^+ \equiv \{x\in \Omega \cap J^+(x') : \sigma(x,x')=0\},
\end{equation}
where $\Omega$ is a causal domain containing $x'$. From the local Hadamard form \eqref{thm: hadamard} and relation \eqref{eq: WF equiv}, it then immediately follows that for any $x \in \Sigma^+$, all points on any future-directed geodesic that starts at $x'$ and goes through $x$ lie in $\singsupp (G)_{x'}$. Thus, the local lightcone singularity supplied by the Hadamard form is propagated along the corresponding null geodesic, which is the content of the global statement \eqref{eq: thm global}.

%%%%%%%%%%%%%%%%%%%%%%%%%%%%%%%%%%%%%%%%%%%%%%%%%%%%%
%%%%%%%%%%%%%%%%%%%%%%%%%%%%%%%%%%%%%%%%%%%%%%%%%%%%%

\bibliography{draft}

\begin{thebibliography}{87}%
\makeatletter
\providecommand \@ifxundefined [1]{%
 \@ifx{#1\undefined}
}%
\providecommand \@ifnum [1]{%
 \ifnum #1\expandafter \@firstoftwo
 \else \expandafter \@secondoftwo
 \fi
}%
\providecommand \@ifx [1]{%
 \ifx #1\expandafter \@firstoftwo
 \else \expandafter \@secondoftwo
 \fi
}%
\providecommand \natexlab [1]{#1}%
\providecommand \enquote  [1]{``#1''}%
\providecommand \bibnamefont  [1]{#1}%
\providecommand \bibfnamefont [1]{#1}%
\providecommand \citenamefont [1]{#1}%
\providecommand \href@noop [0]{\@secondoftwo}%
\providecommand \href [0]{\begingroup \@sanitize@url \@href}%
\providecommand \@href[1]{\@@startlink{#1}\@@href}%
\providecommand \@@href[1]{\endgroup#1\@@endlink}%
\providecommand \@sanitize@url [0]{\catcode `\\12\catcode `\$12\catcode `\&12\catcode `\#12\catcode `\^12\catcode `\_12\catcode `\%12\relax}%
\providecommand \@@startlink[1]{}%
\providecommand \@@endlink[0]{}%
\providecommand \url  [0]{\begingroup\@sanitize@url \@url }%
\providecommand \@url [1]{\endgroup\@href {#1}{\urlprefix }}%
\providecommand \urlprefix  [0]{URL }%
\providecommand \Eprint [0]{\href }%
\providecommand \doibase [0]{http://dx.doi.org/}%
\providecommand \selectlanguage [0]{\@gobble}%
\providecommand \bibinfo  [0]{\@secondoftwo}%
\providecommand \bibfield  [0]{\@secondoftwo}%
\providecommand \translation [1]{[#1]}%
\providecommand \BibitemOpen [0]{}%
\providecommand \bibitemStop [0]{}%
\providecommand \bibitemNoStop [0]{.\EOS\space}%
\providecommand \EOS [0]{\spacefactor3000\relax}%
\providecommand \BibitemShut  [1]{\csname bibitem#1\endcsname}%
\let\auto@bib@innerbib\@empty
%</preamble>
\bibitem [{\citenamefont {Maldacena}(1998)}]{Maldacena:1997re}%
  \BibitemOpen
  \bibfield  {author} {\bibinfo {author} {\bibfnamefont {J.~M.}\ \bibnamefont {Maldacena}},\ }\href {\doibase 10.1023/A:1026654312961} {\bibfield  {journal} {\bibinfo  {journal} {Adv. Theor. Math. Phys.}\ }\textbf {\bibinfo {volume} {2}},\ \bibinfo {pages} {231} (\bibinfo {year} {1998})},\ \Eprint {http://arxiv.org/abs/hep-th/9711200} {arXiv:hep-th/9711200} \BibitemShut {NoStop}%
\bibitem [{\citenamefont {Gubser}\ \emph {et~al.}(1998)\citenamefont {Gubser}, \citenamefont {Klebanov},\ and\ \citenamefont {Polyakov}}]{Gubser:1998bc}%
  \BibitemOpen
  \bibfield  {author} {\bibinfo {author} {\bibfnamefont {S.~S.}\ \bibnamefont {Gubser}}, \bibinfo {author} {\bibfnamefont {I.~R.}\ \bibnamefont {Klebanov}}, \ and\ \bibinfo {author} {\bibfnamefont {A.~M.}\ \bibnamefont {Polyakov}},\ }\href {\doibase 10.1016/S0370-2693(98)00377-3} {\bibfield  {journal} {\bibinfo  {journal} {Phys. Lett. B}\ }\textbf {\bibinfo {volume} {428}},\ \bibinfo {pages} {105} (\bibinfo {year} {1998})},\ \Eprint {http://arxiv.org/abs/hep-th/9802109} {arXiv:hep-th/9802109} \BibitemShut {NoStop}%
\bibitem [{\citenamefont {Witten}(1998)}]{Witten:1998qj}%
  \BibitemOpen
  \bibfield  {author} {\bibinfo {author} {\bibfnamefont {E.}~\bibnamefont {Witten}},\ }\href {\doibase 10.4310/ATMP.1998.v2.n2.a2} {\bibfield  {journal} {\bibinfo  {journal} {Adv. Theor. Math. Phys.}\ }\textbf {\bibinfo {volume} {2}},\ \bibinfo {pages} {253} (\bibinfo {year} {1998})},\ \Eprint {http://arxiv.org/abs/hep-th/9802150} {arXiv:hep-th/9802150} \BibitemShut {NoStop}%
\bibitem [{\citenamefont {Fidkowski}\ \emph {et~al.}(2004)\citenamefont {Fidkowski}, \citenamefont {Hubeny}, \citenamefont {Kleban},\ and\ \citenamefont {Shenker}}]{Fidkowski:2003nf}%
  \BibitemOpen
  \bibfield  {author} {\bibinfo {author} {\bibfnamefont {L.}~\bibnamefont {Fidkowski}}, \bibinfo {author} {\bibfnamefont {V.}~\bibnamefont {Hubeny}}, \bibinfo {author} {\bibfnamefont {M.}~\bibnamefont {Kleban}}, \ and\ \bibinfo {author} {\bibfnamefont {S.}~\bibnamefont {Shenker}},\ }\href {\doibase 10.1088/1126-6708/2004/02/014} {\bibfield  {journal} {\bibinfo  {journal} {JHEP}\ }\textbf {\bibinfo {volume} {02}},\ \bibinfo {pages} {014} (\bibinfo {year} {2004})},\ \Eprint {http://arxiv.org/abs/hep-th/0306170} {arXiv:hep-th/0306170} \BibitemShut {NoStop}%
\bibitem [{\citenamefont {Festuccia}\ and\ \citenamefont {Liu}(2006)}]{Festuccia:2005pi}%
  \BibitemOpen
  \bibfield  {author} {\bibinfo {author} {\bibfnamefont {G.}~\bibnamefont {Festuccia}}\ and\ \bibinfo {author} {\bibfnamefont {H.}~\bibnamefont {Liu}},\ }\href {\doibase 10.1088/1126-6708/2006/04/044} {\bibfield  {journal} {\bibinfo  {journal} {JHEP}\ }\textbf {\bibinfo {volume} {04}},\ \bibinfo {pages} {044} (\bibinfo {year} {2006})},\ \Eprint {http://arxiv.org/abs/hep-th/0506202} {arXiv:hep-th/0506202} \BibitemShut {NoStop}%
\bibitem [{\citenamefont {\v{C}eplak}\ \emph {et~al.}(2024)\citenamefont {\v{C}eplak}, \citenamefont {Liu}, \citenamefont {Parnachev},\ and\ \citenamefont {Valach}}]{Ceplak:2024bja}%
  \BibitemOpen
  \bibfield  {author} {\bibinfo {author} {\bibfnamefont {N.}~\bibnamefont {\v{C}eplak}}, \bibinfo {author} {\bibfnamefont {H.}~\bibnamefont {Liu}}, \bibinfo {author} {\bibfnamefont {A.}~\bibnamefont {Parnachev}}, \ and\ \bibinfo {author} {\bibfnamefont {S.}~\bibnamefont {Valach}},\ }\href {\doibase 10.1007/JHEP10(2024)105} {\bibfield  {journal} {\bibinfo  {journal} {JHEP}\ }\textbf {\bibinfo {volume} {10}},\ \bibinfo {pages} {105} (\bibinfo {year} {2024})},\ \Eprint {http://arxiv.org/abs/2404.17286} {arXiv:2404.17286 [hep-th]} \BibitemShut {NoStop}%
\bibitem [{\citenamefont {Parisini}\ \emph {et~al.}(2024)\citenamefont {Parisini}, \citenamefont {Skenderis},\ and\ \citenamefont {Withers}}]{Parisini:2023nbd}%
  \BibitemOpen
  \bibfield  {author} {\bibinfo {author} {\bibfnamefont {E.}~\bibnamefont {Parisini}}, \bibinfo {author} {\bibfnamefont {K.}~\bibnamefont {Skenderis}}, \ and\ \bibinfo {author} {\bibfnamefont {B.}~\bibnamefont {Withers}},\ }\href {\doibase 10.1007/JHEP05(2024)296} {\bibfield  {journal} {\bibinfo  {journal} {JHEP}\ }\textbf {\bibinfo {volume} {05}},\ \bibinfo {pages} {296} (\bibinfo {year} {2024})},\ \Eprint {http://arxiv.org/abs/2312.03820} {arXiv:2312.03820 [hep-th]} \BibitemShut {NoStop}%
\bibitem [{\citenamefont {Afkhami-Jeddi}\ \emph {et~al.}(2025)\citenamefont {Afkhami-Jeddi}, \citenamefont {Caron-Huot}, \citenamefont {Chakravarty},\ and\ \citenamefont {Maloney}}]{Afkhami-Jeddi:2025wra}%
  \BibitemOpen
  \bibfield  {author} {\bibinfo {author} {\bibfnamefont {N.}~\bibnamefont {Afkhami-Jeddi}}, \bibinfo {author} {\bibfnamefont {S.}~\bibnamefont {Caron-Huot}}, \bibinfo {author} {\bibfnamefont {J.}~\bibnamefont {Chakravarty}}, \ and\ \bibinfo {author} {\bibfnamefont {A.}~\bibnamefont {Maloney}},\ }\href@noop {} {\  (\bibinfo {year} {2025})},\ \Eprint {http://arxiv.org/abs/2510.21673} {arXiv:2510.21673 [hep-th]} \BibitemShut {NoStop}%
\bibitem [{\citenamefont {{\v{C}}eplak}\ \emph {et~al.}(2026)\citenamefont {{\v{C}}eplak}, \citenamefont {Liu}, \citenamefont {Parnachev},\ and\ \citenamefont {Valach}}]{Ceplak:2025dds}%
  \BibitemOpen
  \bibfield  {author} {\bibinfo {author} {\bibfnamefont {N.}~\bibnamefont {{\v{C}}eplak}}, \bibinfo {author} {\bibfnamefont {H.}~\bibnamefont {Liu}}, \bibinfo {author} {\bibfnamefont {A.}~\bibnamefont {Parnachev}}, \ and\ \bibinfo {author} {\bibfnamefont {S.}~\bibnamefont {Valach}},\ }\href {\doibase 10.1007/JHEP05(2026)001} {\bibfield  {journal} {\bibinfo  {journal} {JHEP}\ }\textbf {\bibinfo {volume} {05}},\ \bibinfo {pages} {001} (\bibinfo {year} {2026})},\ \Eprint {http://arxiv.org/abs/2511.09638} {arXiv:2511.09638 [hep-th]} \BibitemShut {NoStop}%
\bibitem [{\citenamefont {Dodelson}\ \emph {et~al.}(2025)\citenamefont {Dodelson}, \citenamefont {Iossa},\ and\ \citenamefont {Karlsson}}]{Dodelson:2025jff}%
  \BibitemOpen
  \bibfield  {author} {\bibinfo {author} {\bibfnamefont {M.}~\bibnamefont {Dodelson}}, \bibinfo {author} {\bibfnamefont {C.}~\bibnamefont {Iossa}}, \ and\ \bibinfo {author} {\bibfnamefont {R.}~\bibnamefont {Karlsson}},\ }\href@noop {} {\  (\bibinfo {year} {2025})},\ \Eprint {http://arxiv.org/abs/2511.09616} {arXiv:2511.09616 [hep-th]} \BibitemShut {NoStop}%
\bibitem [{\citenamefont {Jia}\ and\ \citenamefont {Rangamani}(2025)}]{Jia:2025jbi}%
  \BibitemOpen
  \bibfield  {author} {\bibinfo {author} {\bibfnamefont {H.~F.}\ \bibnamefont {Jia}}\ and\ \bibinfo {author} {\bibfnamefont {M.}~\bibnamefont {Rangamani}},\ }\href@noop {} {\  (\bibinfo {year} {2025})},\ \Eprint {http://arxiv.org/abs/2512.15114} {arXiv:2512.15114 [hep-th]} \BibitemShut {NoStop}%
\bibitem [{\citenamefont {Ali~Ahmad}\ \emph {et~al.}(2026)\citenamefont {Ali~Ahmad}, \citenamefont {Almheiri},\ and\ \citenamefont {Lin}}]{AliAhmad:2026wem}%
  \BibitemOpen
  \bibfield  {author} {\bibinfo {author} {\bibfnamefont {S.}~\bibnamefont {Ali~Ahmad}}, \bibinfo {author} {\bibfnamefont {A.}~\bibnamefont {Almheiri}}, \ and\ \bibinfo {author} {\bibfnamefont {S.}~\bibnamefont {Lin}},\ }\href@noop {} {\  (\bibinfo {year} {2026})},\ \Eprint {http://arxiv.org/abs/2601.02354} {arXiv:2601.02354 [hep-th]} \BibitemShut {NoStop}%
\bibitem [{\citenamefont {Jia}\ and\ \citenamefont {Kulaxizi}(2026)}]{Jia:2026pmv}%
  \BibitemOpen
  \bibfield  {author} {\bibinfo {author} {\bibfnamefont {Y.}~\bibnamefont {Jia}}\ and\ \bibinfo {author} {\bibfnamefont {M.}~\bibnamefont {Kulaxizi}},\ }\href@noop {} {\  (\bibinfo {year} {2026})},\ \Eprint {http://arxiv.org/abs/2602.06558} {arXiv:2602.06558 [hep-th]} \BibitemShut {NoStop}%
\bibitem [{\citenamefont {Giombi}\ \emph {et~al.}(2026)\citenamefont {Giombi}, \citenamefont {Li},\ and\ \citenamefont {Shan}}]{Giombi:2026kdz}%
  \BibitemOpen
  \bibfield  {author} {\bibinfo {author} {\bibfnamefont {S.}~\bibnamefont {Giombi}}, \bibinfo {author} {\bibfnamefont {Y.-Z.}\ \bibnamefont {Li}}, \ and\ \bibinfo {author} {\bibfnamefont {J.}~\bibnamefont {Shan}},\ }\href@noop {} {\  (\bibinfo {year} {2026})},\ \Eprint {http://arxiv.org/abs/2603.11012} {arXiv:2603.11012 [hep-th]} \BibitemShut {NoStop}%
\bibitem [{\citenamefont {Arnaudo}\ and\ \citenamefont {Withers}(2026{\natexlab{a}})}]{Arnaudo:2026der}%
  \BibitemOpen
  \bibfield  {author} {\bibinfo {author} {\bibfnamefont {P.}~\bibnamefont {Arnaudo}}\ and\ \bibinfo {author} {\bibfnamefont {B.}~\bibnamefont {Withers}},\ }\href {\doibase 10.1007/JHEP06(2026)205} {\bibfield  {journal} {\bibinfo  {journal} {JHEP}\ }\textbf {\bibinfo {volume} {06}},\ \bibinfo {pages} {205} (\bibinfo {year} {2026}{\natexlab{a}})},\ \Eprint {http://arxiv.org/abs/2603.13469} {arXiv:2603.13469 [hep-th]} \BibitemShut {NoStop}%
\bibitem [{\citenamefont {Grozdanov}\ \emph {et~al.}(2026{\natexlab{a}})\citenamefont {Grozdanov}, \citenamefont {Valach},\ and\ \citenamefont {Vrbica}}]{Grozdanov:2026cut}%
  \BibitemOpen
  \bibfield  {author} {\bibinfo {author} {\bibfnamefont {S.}~\bibnamefont {Grozdanov}}, \bibinfo {author} {\bibfnamefont {S.}~\bibnamefont {Valach}}, \ and\ \bibinfo {author} {\bibfnamefont {M.}~\bibnamefont {Vrbica}},\ }\href {\doibase 10.1007/JHEP08(2026)022} {\bibfield  {journal} {\bibinfo  {journal} {JHEP}\ }\textbf {\bibinfo {volume} {08}},\ \bibinfo {pages} {022} (\bibinfo {year} {2026}{\natexlab{a}})},\ \Eprint {http://arxiv.org/abs/2603.15598} {arXiv:2603.15598 [hep-th]} \BibitemShut {NoStop}%
\bibitem [{\citenamefont {Jia}\ and\ \citenamefont {Rangamani}(2026)}]{Jia:2026ryl}%
  \BibitemOpen
  \bibfield  {author} {\bibinfo {author} {\bibfnamefont {H.~F.}\ \bibnamefont {Jia}}\ and\ \bibinfo {author} {\bibfnamefont {M.}~\bibnamefont {Rangamani}},\ }\href@noop {} {\  (\bibinfo {year} {2026})},\ \Eprint {http://arxiv.org/abs/2604.10803} {arXiv:2604.10803 [hep-th]} \BibitemShut {NoStop}%
\bibitem [{\citenamefont {Kovtun}\ and\ \citenamefont {Starinets}(2005)}]{Kovtun:2005ev}%
  \BibitemOpen
  \bibfield  {author} {\bibinfo {author} {\bibfnamefont {P.~K.}\ \bibnamefont {Kovtun}}\ and\ \bibinfo {author} {\bibfnamefont {A.~O.}\ \bibnamefont {Starinets}},\ }\href {\doibase 10.1103/PhysRevD.72.086009} {\bibfield  {journal} {\bibinfo  {journal} {Phys. Rev. D}\ }\textbf {\bibinfo {volume} {72}},\ \bibinfo {pages} {086009} (\bibinfo {year} {2005})},\ \Eprint {http://arxiv.org/abs/hep-th/0506184} {arXiv:hep-th/0506184} \BibitemShut {NoStop}%
\bibitem [{\citenamefont {Hartnoll}\ and\ \citenamefont {Kumar}(2005)}]{Hartnoll:2005ju}%
  \BibitemOpen
  \bibfield  {author} {\bibinfo {author} {\bibfnamefont {S.~A.}\ \bibnamefont {Hartnoll}}\ and\ \bibinfo {author} {\bibfnamefont {S.~P.}\ \bibnamefont {Kumar}},\ }\href {\doibase 10.1088/1126-6708/2005/12/036} {\bibfield  {journal} {\bibinfo  {journal} {JHEP}\ }\textbf {\bibinfo {volume} {12}},\ \bibinfo {pages} {036} (\bibinfo {year} {2005})},\ \Eprint {http://arxiv.org/abs/hep-th/0508092} {arXiv:hep-th/0508092} \BibitemShut {NoStop}%
\bibitem [{\citenamefont {Grozdanov}\ \emph {et~al.}(2016)\citenamefont {Grozdanov}, \citenamefont {Kaplis},\ and\ \citenamefont {Starinets}}]{Grozdanov:2016vgg}%
  \BibitemOpen
  \bibfield  {author} {\bibinfo {author} {\bibfnamefont {S.}~\bibnamefont {Grozdanov}}, \bibinfo {author} {\bibfnamefont {N.}~\bibnamefont {Kaplis}}, \ and\ \bibinfo {author} {\bibfnamefont {A.~O.}\ \bibnamefont {Starinets}},\ }\href {\doibase 10.1007/JHEP07(2016)151} {\bibfield  {journal} {\bibinfo  {journal} {JHEP}\ }\textbf {\bibinfo {volume} {07}},\ \bibinfo {pages} {151} (\bibinfo {year} {2016})},\ \Eprint {http://arxiv.org/abs/1605.02173} {arXiv:1605.02173 [hep-th]} \BibitemShut {NoStop}%
\bibitem [{\citenamefont {Casalderrey-Solana}\ \emph {et~al.}(2018)\citenamefont {Casalderrey-Solana}, \citenamefont {Grozdanov},\ and\ \citenamefont {Starinets}}]{Casalderrey-Solana:2018rle}%
  \BibitemOpen
  \bibfield  {author} {\bibinfo {author} {\bibfnamefont {J.}~\bibnamefont {Casalderrey-Solana}}, \bibinfo {author} {\bibfnamefont {S.}~\bibnamefont {Grozdanov}}, \ and\ \bibinfo {author} {\bibfnamefont {A.~O.}\ \bibnamefont {Starinets}},\ }\href {\doibase 10.1103/PhysRevLett.121.191603} {\bibfield  {journal} {\bibinfo  {journal} {Phys. Rev. Lett.}\ }\textbf {\bibinfo {volume} {121}},\ \bibinfo {pages} {191603} (\bibinfo {year} {2018})},\ \Eprint {http://arxiv.org/abs/1806.10997} {arXiv:1806.10997 [hep-th]} \BibitemShut {NoStop}%
\bibitem [{\citenamefont {Grozdanov}\ and\ \citenamefont {Starinets}(2019)}]{Grozdanov:2018gfx}%
  \BibitemOpen
  \bibfield  {author} {\bibinfo {author} {\bibfnamefont {S.}~\bibnamefont {Grozdanov}}\ and\ \bibinfo {author} {\bibfnamefont {A.~O.}\ \bibnamefont {Starinets}},\ }\href {\doibase 10.1007/JHEP04(2019)080} {\bibfield  {journal} {\bibinfo  {journal} {JHEP}\ }\textbf {\bibinfo {volume} {04}},\ \bibinfo {pages} {080} (\bibinfo {year} {2019})},\ \Eprint {http://arxiv.org/abs/1812.09288} {arXiv:1812.09288 [hep-th]} \BibitemShut {NoStop}%
\bibitem [{\citenamefont {Dodelson}(2025)}]{Dodelson:2024atp}%
  \BibitemOpen
  \bibfield  {author} {\bibinfo {author} {\bibfnamefont {M.}~\bibnamefont {Dodelson}},\ }\href {\doibase 10.21468/SciPostPhys.19.3.081} {\bibfield  {journal} {\bibinfo  {journal} {SciPost Phys.}\ }\textbf {\bibinfo {volume} {19}},\ \bibinfo {pages} {081} (\bibinfo {year} {2025})},\ \Eprint {http://arxiv.org/abs/2408.05790} {arXiv:2408.05790 [hep-th]} \BibitemShut {NoStop}%
\bibitem [{\citenamefont {Dodelson}\ \emph {et~al.}(2024)\citenamefont {Dodelson}, \citenamefont {Iossa}, \citenamefont {Karlsson},\ and\ \citenamefont {Zhiboedov}}]{Dodelson:2023vrw}%
  \BibitemOpen
  \bibfield  {author} {\bibinfo {author} {\bibfnamefont {M.}~\bibnamefont {Dodelson}}, \bibinfo {author} {\bibfnamefont {C.}~\bibnamefont {Iossa}}, \bibinfo {author} {\bibfnamefont {R.}~\bibnamefont {Karlsson}}, \ and\ \bibinfo {author} {\bibfnamefont {A.}~\bibnamefont {Zhiboedov}},\ }\href {\doibase 10.1007/JHEP01(2024)036} {\bibfield  {journal} {\bibinfo  {journal} {JHEP}\ }\textbf {\bibinfo {volume} {01}},\ \bibinfo {pages} {036} (\bibinfo {year} {2024})},\ \Eprint {http://arxiv.org/abs/2304.12339} {arXiv:2304.12339 [hep-th]} \BibitemShut {NoStop}%
\bibitem [{\citenamefont {Grozdanov}\ and\ \citenamefont {Vrbica}(2026)}]{Grozdanov:2025ulc}%
  \BibitemOpen
  \bibfield  {author} {\bibinfo {author} {\bibfnamefont {S.}~\bibnamefont {Grozdanov}}\ and\ \bibinfo {author} {\bibfnamefont {M.}~\bibnamefont {Vrbica}},\ }\href {\doibase 10.1007/JHEP02(2026)106} {\bibfield  {journal} {\bibinfo  {journal} {JHEP}\ }\textbf {\bibinfo {volume} {02}},\ \bibinfo {pages} {106} (\bibinfo {year} {2026})},\ \Eprint {http://arxiv.org/abs/2509.18074} {arXiv:2509.18074 [hep-th]} \BibitemShut {NoStop}%
\bibitem [{\citenamefont {Valach}(2025)}]{Valach:2025saf}%
  \BibitemOpen
  \bibfield  {author} {\bibinfo {author} {\bibfnamefont {S.}~\bibnamefont {Valach}},\ }\href@noop {} {\  (\bibinfo {year} {2025})},\ \Eprint {http://arxiv.org/abs/2508.17139} {arXiv:2508.17139 [hep-th]} \BibitemShut {NoStop}%
\bibitem [{\citenamefont {Araya}\ \emph {et~al.}(2026)\citenamefont {Araya}, \citenamefont {Esper}, \citenamefont {Jia}, \citenamefont {Kulaxizi},\ and\ \citenamefont {Parnachev}}]{Araya:2026shz}%
  \BibitemOpen
  \bibfield  {author} {\bibinfo {author} {\bibfnamefont {I.~J.}\ \bibnamefont {Araya}}, \bibinfo {author} {\bibfnamefont {C.}~\bibnamefont {Esper}}, \bibinfo {author} {\bibfnamefont {Y.}~\bibnamefont {Jia}}, \bibinfo {author} {\bibfnamefont {M.}~\bibnamefont {Kulaxizi}}, \ and\ \bibinfo {author} {\bibfnamefont {A.}~\bibnamefont {Parnachev}},\ }\href@noop {} {\  (\bibinfo {year} {2026})},\ \Eprint {http://arxiv.org/abs/2602.12893} {arXiv:2602.12893 [hep-th]} \BibitemShut {NoStop}%
\bibitem [{\citenamefont {Fitzpatrick}\ and\ \citenamefont {Huang}(2019)}]{Fitzpatrick:2019zqz}%
  \BibitemOpen
  \bibfield  {author} {\bibinfo {author} {\bibfnamefont {A.~L.}\ \bibnamefont {Fitzpatrick}}\ and\ \bibinfo {author} {\bibfnamefont {K.-W.}\ \bibnamefont {Huang}},\ }\href {\doibase 10.1007/JHEP08(2019)138} {\bibfield  {journal} {\bibinfo  {journal} {JHEP}\ }\textbf {\bibinfo {volume} {08}},\ \bibinfo {pages} {138} (\bibinfo {year} {2019})},\ \Eprint {http://arxiv.org/abs/1903.05306} {arXiv:1903.05306 [hep-th]} \BibitemShut {NoStop}%
\bibitem [{\citenamefont {Karlsson}\ \emph {et~al.}(2022)\citenamefont {Karlsson}, \citenamefont {Parnachev}, \citenamefont {Prilepina},\ and\ \citenamefont {Valach}}]{Karlsson:2022osn}%
  \BibitemOpen
  \bibfield  {author} {\bibinfo {author} {\bibfnamefont {R.}~\bibnamefont {Karlsson}}, \bibinfo {author} {\bibfnamefont {A.}~\bibnamefont {Parnachev}}, \bibinfo {author} {\bibfnamefont {V.}~\bibnamefont {Prilepina}}, \ and\ \bibinfo {author} {\bibfnamefont {S.}~\bibnamefont {Valach}},\ }\href {\doibase 10.1007/JHEP09(2022)234} {\bibfield  {journal} {\bibinfo  {journal} {JHEP}\ }\textbf {\bibinfo {volume} {09}},\ \bibinfo {pages} {234} (\bibinfo {year} {2022})},\ \Eprint {http://arxiv.org/abs/2206.05544} {arXiv:2206.05544 [hep-th]} \BibitemShut {NoStop}%
\bibitem [{\citenamefont {Huang}\ \emph {et~al.}(2023)\citenamefont {Huang}, \citenamefont {Karlsson}, \citenamefont {Parnachev},\ and\ \citenamefont {Valach}}]{Huang:2022vet}%
  \BibitemOpen
  \bibfield  {author} {\bibinfo {author} {\bibfnamefont {K.-W.}\ \bibnamefont {Huang}}, \bibinfo {author} {\bibfnamefont {R.}~\bibnamefont {Karlsson}}, \bibinfo {author} {\bibfnamefont {A.}~\bibnamefont {Parnachev}}, \ and\ \bibinfo {author} {\bibfnamefont {S.}~\bibnamefont {Valach}},\ }\href {\doibase 10.1007/JHEP05(2023)065} {\bibfield  {journal} {\bibinfo  {journal} {JHEP}\ }\textbf {\bibinfo {volume} {05}},\ \bibinfo {pages} {065} (\bibinfo {year} {2023})},\ \Eprint {http://arxiv.org/abs/2210.16274} {arXiv:2210.16274 [hep-th]} \BibitemShut {NoStop}%
\bibitem [{\citenamefont {Esper}\ \emph {et~al.}(2023)\citenamefont {Esper}, \citenamefont {Huang}, \citenamefont {Karlsson}, \citenamefont {Parnachev},\ and\ \citenamefont {Valach}}]{Esper:2023jeq}%
  \BibitemOpen
  \bibfield  {author} {\bibinfo {author} {\bibfnamefont {C.}~\bibnamefont {Esper}}, \bibinfo {author} {\bibfnamefont {K.-W.}\ \bibnamefont {Huang}}, \bibinfo {author} {\bibfnamefont {R.}~\bibnamefont {Karlsson}}, \bibinfo {author} {\bibfnamefont {A.}~\bibnamefont {Parnachev}}, \ and\ \bibinfo {author} {\bibfnamefont {S.}~\bibnamefont {Valach}},\ }\href {\doibase 10.1007/JHEP11(2023)107} {\bibfield  {journal} {\bibinfo  {journal} {JHEP}\ }\textbf {\bibinfo {volume} {11}},\ \bibinfo {pages} {107} (\bibinfo {year} {2023})},\ \Eprint {http://arxiv.org/abs/2306.00787} {arXiv:2306.00787 [hep-th]} \BibitemShut {NoStop}%
\bibitem [{\citenamefont {Buri{\'c}}\ \emph {et~al.}(2025)\citenamefont {Buri{\'c}}, \citenamefont {Gusev},\ and\ \citenamefont {Parnachev}}]{Buric:2025anb}%
  \BibitemOpen
  \bibfield  {author} {\bibinfo {author} {\bibfnamefont {I.}~\bibnamefont {Buri{\'c}}}, \bibinfo {author} {\bibfnamefont {I.}~\bibnamefont {Gusev}}, \ and\ \bibinfo {author} {\bibfnamefont {A.}~\bibnamefont {Parnachev}},\ }\href {\doibase 10.1007/JHEP09(2025)053} {\bibfield  {journal} {\bibinfo  {journal} {JHEP}\ }\textbf {\bibinfo {volume} {09}},\ \bibinfo {pages} {053} (\bibinfo {year} {2025})},\ \Eprint {http://arxiv.org/abs/2505.10277} {arXiv:2505.10277 [hep-th]} \BibitemShut {NoStop}%
\bibitem [{\citenamefont {Buri{\'c}}\ \emph {et~al.}(2026)\citenamefont {Buri{\'c}}, \citenamefont {Gusev},\ and\ \citenamefont {Parnachev}}]{Buric:2025fye}%
  \BibitemOpen
  \bibfield  {author} {\bibinfo {author} {\bibfnamefont {I.}~\bibnamefont {Buri{\'c}}}, \bibinfo {author} {\bibfnamefont {I.}~\bibnamefont {Gusev}}, \ and\ \bibinfo {author} {\bibfnamefont {A.}~\bibnamefont {Parnachev}},\ }\href {\doibase 10.1007/JHEP05(2026)059} {\bibfield  {journal} {\bibinfo  {journal} {JHEP}\ }\textbf {\bibinfo {volume} {05}},\ \bibinfo {pages} {059} (\bibinfo {year} {2026})},\ \Eprint {http://arxiv.org/abs/2508.08373} {arXiv:2508.08373 [hep-th]} \BibitemShut {NoStop}%
\bibitem [{\citenamefont {Barrat}\ \emph {et~al.}(2026)\citenamefont {Barrat}, \citenamefont {Bozkurt}, \citenamefont {Marchetto}, \citenamefont {Miscioscia},\ and\ \citenamefont {Pomoni}}]{Barrat:2025twb}%
  \BibitemOpen
  \bibfield  {author} {\bibinfo {author} {\bibfnamefont {J.}~\bibnamefont {Barrat}}, \bibinfo {author} {\bibfnamefont {D.~N.}\ \bibnamefont {Bozkurt}}, \bibinfo {author} {\bibfnamefont {E.}~\bibnamefont {Marchetto}}, \bibinfo {author} {\bibfnamefont {A.}~\bibnamefont {Miscioscia}}, \ and\ \bibinfo {author} {\bibfnamefont {E.}~\bibnamefont {Pomoni}},\ }\href {\doibase 10.1007/JHEP05(2026)180} {\bibfield  {journal} {\bibinfo  {journal} {JHEP}\ }\textbf {\bibinfo {volume} {05}},\ \bibinfo {pages} {180} (\bibinfo {year} {2026})},\ \Eprint {http://arxiv.org/abs/2510.20894} {arXiv:2510.20894 [hep-th]} \BibitemShut {NoStop}%
\bibitem [{\citenamefont {\v{C}eplak}\ and\ \citenamefont {Valach}(pear)}]{futurepap3}%
  \BibitemOpen
  \bibfield  {author} {\bibinfo {author} {\bibfnamefont {N.}~\bibnamefont {\v{C}eplak}}\ and\ \bibinfo {author} {\bibfnamefont {S.}~\bibnamefont {Valach}},\ }\href@noop {} {\enquote {\bibinfo {title} {{Where's Kasner? Towards the holographic manifestation of Kasner exponents}},}\ } (\bibinfo {year} {to appear})\BibitemShut {NoStop}%
\bibitem [{\citenamefont {\v{C}eplak}\ \emph {et~al.}(pear)\citenamefont {\v{C}eplak}, \citenamefont {Liu}, \citenamefont {Parnachev},\ and\ \citenamefont {Valach}}]{futureHONG}%
  \BibitemOpen
  \bibfield  {author} {\bibinfo {author} {\bibfnamefont {N.}~\bibnamefont {\v{C}eplak}}, \bibinfo {author} {\bibfnamefont {H.}~\bibnamefont {Liu}}, \bibinfo {author} {\bibfnamefont {A.}~\bibnamefont {Parnachev}}, \ and\ \bibinfo {author} {\bibfnamefont {S.}~\bibnamefont {Valach}},\ }\href@noop {} {\enquote {\bibinfo {title} {{Stringy splitting of black hole singularities}},}\ } (\bibinfo {year} {to appear})\BibitemShut {NoStop}%
\bibitem [{\citenamefont {Faruk}\ \emph {et~al.}(2024)\citenamefont {Faruk}, \citenamefont {Morvan},\ and\ \citenamefont {van~der Schaar}}]{Faruk:2023uzs}%
  \BibitemOpen
  \bibfield  {author} {\bibinfo {author} {\bibfnamefont {M.~M.}\ \bibnamefont {Faruk}}, \bibinfo {author} {\bibfnamefont {E.}~\bibnamefont {Morvan}}, \ and\ \bibinfo {author} {\bibfnamefont {J.~P.}\ \bibnamefont {van~der Schaar}},\ }\href {\doibase 10.1088/1475-7516/2024/05/118} {\bibfield  {journal} {\bibinfo  {journal} {JCAP}\ }\textbf {\bibinfo {volume} {05}},\ \bibinfo {pages} {118} (\bibinfo {year} {2024})},\ \Eprint {http://arxiv.org/abs/2312.06878} {arXiv:2312.06878 [gr-qc]} \BibitemShut {NoStop}%
\bibitem [{\citenamefont {Faruk}\ \emph {et~al.}(2025)\citenamefont {Faruk}, \citenamefont {Rost},\ and\ \citenamefont {van~der Schaar}}]{Faruk:2025bed}%
  \BibitemOpen
  \bibfield  {author} {\bibinfo {author} {\bibfnamefont {M.~M.}\ \bibnamefont {Faruk}}, \bibinfo {author} {\bibfnamefont {F.}~\bibnamefont {Rost}}, \ and\ \bibinfo {author} {\bibfnamefont {J.~P.}\ \bibnamefont {van~der Schaar}},\ }\href {\doibase 10.1007/JHEP07(2025)050} {\bibfield  {journal} {\bibinfo  {journal} {JHEP}\ }\textbf {\bibinfo {volume} {07}},\ \bibinfo {pages} {050} (\bibinfo {year} {2025})},\ \Eprint {http://arxiv.org/abs/2501.01388} {arXiv:2501.01388 [hep-th]} \BibitemShut {NoStop}%
\bibitem [{\citenamefont {Arnaudo}\ and\ \citenamefont {Withers}(2026{\natexlab{b}})}]{Arnaudo:2026tcy}%
  \BibitemOpen
  \bibfield  {author} {\bibinfo {author} {\bibfnamefont {P.}~\bibnamefont {Arnaudo}}\ and\ \bibinfo {author} {\bibfnamefont {B.}~\bibnamefont {Withers}},\ }\href@noop {} {\  (\bibinfo {year} {2026}{\natexlab{b}})},\ \Eprint {http://arxiv.org/abs/2605.16489} {arXiv:2605.16489 [gr-qc]} \BibitemShut {NoStop}%
\bibitem [{\citenamefont {Hadamard}(1923)}]{hadamard1923lectures}%
  \BibitemOpen
  \bibfield  {author} {\bibinfo {author} {\bibfnamefont {J.}~\bibnamefont {Hadamard}},\ }\href@noop {} {\emph {\bibinfo {title} {Lectures on Cauchy's problem in linear partial differential equations}}}\ (\bibinfo  {publisher} {Dover},\ \bibinfo {year} {1923})\BibitemShut {NoStop}%
\bibitem [{\citenamefont {Friedlander}(2010)}]{Friedlander:2010eqa}%
  \BibitemOpen
  \bibfield  {author} {\bibinfo {author} {\bibfnamefont {F.~G.}\ \bibnamefont {Friedlander}},\ }\href@noop {} {\emph {\bibinfo {title} {{The Wave Equation on a Curved Space-Time}}}}\ (\bibinfo  {publisher} {Cambridge University Press},\ \bibinfo {year} {2010})\BibitemShut {NoStop}%
\bibitem [{\citenamefont {Duistermaat}\ and\ \citenamefont {Hörmander}(1972)}]{Duistermaat:1972}%
  \BibitemOpen
  \bibfield  {author} {\bibinfo {author} {\bibfnamefont {J.~J.}\ \bibnamefont {Duistermaat}}\ and\ \bibinfo {author} {\bibfnamefont {L.}~\bibnamefont {Hörmander}},\ }\href {\doibase https://doi.org/10.1007/BF02392165} {\bibfield  {journal} {\bibinfo  {journal} {Acta Math.}\ }\textbf {\bibinfo {volume} {128}},\ \bibinfo {pages} {183–269} (\bibinfo {year} {1972})}\BibitemShut {NoStop}%
\bibitem [{\citenamefont {Bousso}\ and\ \citenamefont {Hawking}(1996)}]{Bousso:1996}%
  \BibitemOpen
  \bibfield  {author} {\bibinfo {author} {\bibfnamefont {R.}~\bibnamefont {Bousso}}\ and\ \bibinfo {author} {\bibfnamefont {S.~W.}\ \bibnamefont {Hawking}},\ }\href {\doibase 10.1103/PhysRevD.54.6312} {\bibfield  {journal} {\bibinfo  {journal} {Phys. Rev. D}\ }\textbf {\bibinfo {volume} {54}},\ \bibinfo {pages} {6312} (\bibinfo {year} {1996})},\ \Eprint {http://arxiv.org/abs/9606052} {arXiv:9606052 [gr-qc]} \BibitemShut {NoStop}%
\bibitem [{\citenamefont {Grewal}\ and\ \citenamefont {Law}(2025)}]{Grewal:2024jes}%
  \BibitemOpen
  \bibfield  {author} {\bibinfo {author} {\bibfnamefont {M.}~\bibnamefont {Grewal}}\ and\ \bibinfo {author} {\bibfnamefont {Y.~T.~A.}\ \bibnamefont {Law}},\ }\href {\doibase 10.1007/JHEP10(2025)052} {\bibfield  {journal} {\bibinfo  {journal} {JHEP}\ }\textbf {\bibinfo {volume} {10}},\ \bibinfo {pages} {052} (\bibinfo {year} {2025})},\ \Eprint {http://arxiv.org/abs/2403.06006} {arXiv:2403.06006 [hep-th]} \BibitemShut {NoStop}%
\bibitem [{\citenamefont {Amado}\ and\ \citenamefont {Hoyos-Badajoz}(2008)}]{Amado:2008hw}%
  \BibitemOpen
  \bibfield  {author} {\bibinfo {author} {\bibfnamefont {I.}~\bibnamefont {Amado}}\ and\ \bibinfo {author} {\bibfnamefont {C.}~\bibnamefont {Hoyos-Badajoz}},\ }\href {\doibase 10.1088/1126-6708/2008/09/118} {\bibfield  {journal} {\bibinfo  {journal} {JHEP}\ }\textbf {\bibinfo {volume} {09}},\ \bibinfo {pages} {118} (\bibinfo {year} {2008})},\ \Eprint {http://arxiv.org/abs/0807.2337} {arXiv:0807.2337 [hep-th]} \BibitemShut {NoStop}%
\bibitem [{\citenamefont {Festuccia}\ and\ \citenamefont {Liu}(2009)}]{Festuccia:2008zx}%
  \BibitemOpen
  \bibfield  {author} {\bibinfo {author} {\bibfnamefont {G.}~\bibnamefont {Festuccia}}\ and\ \bibinfo {author} {\bibfnamefont {H.}~\bibnamefont {Liu}},\ }\href {\doibase 10.1166/asl.2009.1029} {\bibfield  {journal} {\bibinfo  {journal} {Adv. Sci. Lett.}\ }\textbf {\bibinfo {volume} {2}},\ \bibinfo {pages} {221} (\bibinfo {year} {2009})},\ \Eprint {http://arxiv.org/abs/0811.1033} {arXiv:0811.1033 [gr-qc]} \BibitemShut {NoStop}%
\bibitem [{\citenamefont {Skinner}(2014)}]{Skinner}%
  \BibitemOpen
  \bibfield  {author} {\bibinfo {author} {\bibfnamefont {D.}~\bibnamefont {Skinner}},\ }\href@noop {} {\enquote {\bibinfo {title} {{Mathematical Methods, Lecture notes}},}\ }\bibinfo {howpublished} {\url{https://www.damtp.cam.ac.uk/user/dbs26/1BMethods/All.pdf}} (\bibinfo {year} {2014}),\ \bibinfo {note} {{Department of Applied Mathematics and Theoretical Physics, University of Cambridge}}\BibitemShut {NoStop}%
\bibitem [{\citenamefont {Son}\ and\ \citenamefont {Starinets}(2002)}]{Son:2002sd}%
  \BibitemOpen
  \bibfield  {author} {\bibinfo {author} {\bibfnamefont {D.~T.}\ \bibnamefont {Son}}\ and\ \bibinfo {author} {\bibfnamefont {A.~O.}\ \bibnamefont {Starinets}},\ }\href {\doibase 10.1088/1126-6708/2002/09/042} {\bibfield  {journal} {\bibinfo  {journal} {JHEP}\ }\textbf {\bibinfo {volume} {09}},\ \bibinfo {pages} {042} (\bibinfo {year} {2002})},\ \Eprint {http://arxiv.org/abs/hep-th/0205051} {arXiv:hep-th/0205051} \BibitemShut {NoStop}%
\bibitem [{\citenamefont {Solodukhin}(1999)}]{Solodukhin:1998ec}%
  \BibitemOpen
  \bibfield  {author} {\bibinfo {author} {\bibfnamefont {S.~N.}\ \bibnamefont {Solodukhin}},\ }\href {\doibase 10.1016/S0550-3213(98)00715-9} {\bibfield  {journal} {\bibinfo  {journal} {Nucl. Phys. B}\ }\textbf {\bibinfo {volume} {539}},\ \bibinfo {pages} {403} (\bibinfo {year} {1999})},\ \Eprint {http://arxiv.org/abs/hep-th/9806004} {arXiv:hep-th/9806004} \BibitemShut {NoStop}%
\bibitem [{\citenamefont {Minces}\ and\ \citenamefont {Rivelles}(2000)}]{Minces:1999eg}%
  \BibitemOpen
  \bibfield  {author} {\bibinfo {author} {\bibfnamefont {P.}~\bibnamefont {Minces}}\ and\ \bibinfo {author} {\bibfnamefont {V.~O.}\ \bibnamefont {Rivelles}},\ }\href {\doibase 10.1016/S0550-3213(99)00833-0} {\bibfield  {journal} {\bibinfo  {journal} {Nucl. Phys. B}\ }\textbf {\bibinfo {volume} {572}},\ \bibinfo {pages} {651} (\bibinfo {year} {2000})},\ \Eprint {http://arxiv.org/abs/hep-th/9907079} {arXiv:hep-th/9907079} \BibitemShut {NoStop}%
\bibitem [{\citenamefont {Witten}(2001)}]{Witten:2001ua}%
  \BibitemOpen
  \bibfield  {author} {\bibinfo {author} {\bibfnamefont {E.}~\bibnamefont {Witten}},\ }\href@noop {} {\  (\bibinfo {year} {2001})},\ \Eprint {http://arxiv.org/abs/hep-th/0112258} {arXiv:hep-th/0112258} \BibitemShut {NoStop}%
\bibitem [{\citenamefont {Minces}(2003)}]{Minces:2002wp}%
  \BibitemOpen
  \bibfield  {author} {\bibinfo {author} {\bibfnamefont {P.}~\bibnamefont {Minces}},\ }\href {\doibase 10.1103/PhysRevD.68.024027} {\bibfield  {journal} {\bibinfo  {journal} {Phys. Rev. D}\ }\textbf {\bibinfo {volume} {68}},\ \bibinfo {pages} {024027} (\bibinfo {year} {2003})},\ \Eprint {http://arxiv.org/abs/hep-th/0201172} {arXiv:hep-th/0201172} \BibitemShut {NoStop}%
\bibitem [{\citenamefont {Heemskerk}\ \emph {et~al.}(2009)\citenamefont {Heemskerk}, \citenamefont {Penedones}, \citenamefont {Polchinski},\ and\ \citenamefont {Sully}}]{Heemskerk:2009pn}%
  \BibitemOpen
  \bibfield  {author} {\bibinfo {author} {\bibfnamefont {I.}~\bibnamefont {Heemskerk}}, \bibinfo {author} {\bibfnamefont {J.}~\bibnamefont {Penedones}}, \bibinfo {author} {\bibfnamefont {J.}~\bibnamefont {Polchinski}}, \ and\ \bibinfo {author} {\bibfnamefont {J.}~\bibnamefont {Sully}},\ }\href {\doibase 10.1088/1126-6708/2009/10/079} {\bibfield  {journal} {\bibinfo  {journal} {JHEP}\ }\textbf {\bibinfo {volume} {10}},\ \bibinfo {pages} {079} (\bibinfo {year} {2009})},\ \Eprint {http://arxiv.org/abs/0907.0151} {arXiv:0907.0151 [hep-th]} \BibitemShut {NoStop}%
\bibitem [{\citenamefont {Faulkner}\ \emph {et~al.}(2011)\citenamefont {Faulkner}, \citenamefont {Liu},\ and\ \citenamefont {Rangamani}}]{Faulkner:2010jy}%
  \BibitemOpen
  \bibfield  {author} {\bibinfo {author} {\bibfnamefont {T.}~\bibnamefont {Faulkner}}, \bibinfo {author} {\bibfnamefont {H.}~\bibnamefont {Liu}}, \ and\ \bibinfo {author} {\bibfnamefont {M.}~\bibnamefont {Rangamani}},\ }\href {\doibase 10.1007/JHEP08(2011)051} {\bibfield  {journal} {\bibinfo  {journal} {JHEP}\ }\textbf {\bibinfo {volume} {08}},\ \bibinfo {pages} {051} (\bibinfo {year} {2011})},\ \Eprint {http://arxiv.org/abs/1010.4036} {arXiv:1010.4036 [hep-th]} \BibitemShut {NoStop}%
\bibitem [{\citenamefont {Grozdanov}(2012)}]{Grozdanov:2011aa}%
  \BibitemOpen
  \bibfield  {author} {\bibinfo {author} {\bibfnamefont {S.}~\bibnamefont {Grozdanov}},\ }\href {\doibase 10.1007/JHEP06(2012)079} {\bibfield  {journal} {\bibinfo  {journal} {JHEP}\ }\textbf {\bibinfo {volume} {06}},\ \bibinfo {pages} {079} (\bibinfo {year} {2012})},\ \Eprint {http://arxiv.org/abs/1112.3356} {arXiv:1112.3356 [hep-th]} \BibitemShut {NoStop}%
\bibitem [{\citenamefont {Wang}\ and\ \citenamefont {Yang}(2026)}]{Wang:2026esp}%
  \BibitemOpen
  \bibfield  {author} {\bibinfo {author} {\bibfnamefont {Y.}~\bibnamefont {Wang}}\ and\ \bibinfo {author} {\bibfnamefont {J.}~\bibnamefont {Yang}},\ }\href@noop {} {\  (\bibinfo {year} {2026})},\ \Eprint {http://arxiv.org/abs/2605.27641} {arXiv:2605.27641 [hep-th]} \BibitemShut {NoStop}%
\bibitem [{\citenamefont {Stein}\ and\ \citenamefont {Shakarchi}(2003)}]{book}%
  \BibitemOpen
  \bibfield  {author} {\bibinfo {author} {\bibfnamefont {E.}~\bibnamefont {Stein}}\ and\ \bibinfo {author} {\bibfnamefont {R.}~\bibnamefont {Shakarchi}},\ }\href@noop {} {\emph {\bibinfo {title} {Complex Analysis}}}\ (\bibinfo  {publisher} {Princeton University Press},\ \bibinfo {year} {2003})\BibitemShut {NoStop}%
\bibitem [{\citenamefont {Newton}(1982)}]{Newton:1982qc}%
  \BibitemOpen
  \bibfield  {author} {\bibinfo {author} {\bibfnamefont {R.~G.}\ \bibnamefont {Newton}},\ }\href@noop {} {\emph {\bibinfo {title} {{Scattering theory of waves and particles.}}}}\ (\bibinfo  {publisher} {Springer-Verlag New York},\ \bibinfo {year} {1982})\BibitemShut {NoStop}%
\bibitem [{\citenamefont {Festuccia}(2007)}]{festucciathesis}%
  \BibitemOpen
  \bibfield  {author} {\bibinfo {author} {\bibfnamefont {G.}~\bibnamefont {Festuccia}},\ }\emph {\bibinfo {title} {Black hole singularities in the framework of gauge/string duality}},\ \href@noop {} {Ph.D. thesis},\ \bibinfo  {school} {Massachusetts Institute of Technology} (\bibinfo {year} {2007})\BibitemShut {NoStop}%
\bibitem [{\citenamefont {Horowitz}\ and\ \citenamefont {Hubeny}(2000)}]{Horowitz:1999jd}%
  \BibitemOpen
  \bibfield  {author} {\bibinfo {author} {\bibfnamefont {G.~T.}\ \bibnamefont {Horowitz}}\ and\ \bibinfo {author} {\bibfnamefont {V.~E.}\ \bibnamefont {Hubeny}},\ }\href {\doibase 10.1103/PhysRevD.62.024027} {\bibfield  {journal} {\bibinfo  {journal} {Phys. Rev. D}\ }\textbf {\bibinfo {volume} {62}},\ \bibinfo {pages} {024027} (\bibinfo {year} {2000})},\ \Eprint {http://arxiv.org/abs/hep-th/9909056} {arXiv:hep-th/9909056} \BibitemShut {NoStop}%
\bibitem [{\citenamefont {Regge}\ and\ \citenamefont {Wheeler}(1957)}]{Regge:1957td}%
  \BibitemOpen
  \bibfield  {author} {\bibinfo {author} {\bibfnamefont {T.}~\bibnamefont {Regge}}\ and\ \bibinfo {author} {\bibfnamefont {J.~A.}\ \bibnamefont {Wheeler}},\ }\href {\doibase 10.1103/PhysRev.108.1063} {\bibfield  {journal} {\bibinfo  {journal} {Phys. Rev.}\ }\textbf {\bibinfo {volume} {108}},\ \bibinfo {pages} {1063} (\bibinfo {year} {1957})}\BibitemShut {NoStop}%
\bibitem [{\citenamefont {Zerilli}(1970)}]{Zerilli:1970se}%
  \BibitemOpen
  \bibfield  {author} {\bibinfo {author} {\bibfnamefont {F.~J.}\ \bibnamefont {Zerilli}},\ }\href {\doibase 10.1103/PhysRevLett.24.737} {\bibfield  {journal} {\bibinfo  {journal} {Phys. Rev. Lett.}\ }\textbf {\bibinfo {volume} {24}},\ \bibinfo {pages} {737} (\bibinfo {year} {1970})}\BibitemShut {NoStop}%
\bibitem [{\citenamefont {{Université de Grenoble. École d'été de physique théorique}}(1973)}]{BH:1973}%
  \BibitemOpen
  \bibfield  {author} {\bibinfo {author} {\bibnamefont {{Université de Grenoble. École d'été de physique théorique}}},\ }\href@noop {} {\emph {\bibinfo {title} {Black Holes: Les Astres Occlus}}},\ edited by\ \bibinfo {editor} {\bibfnamefont {C.}~\bibnamefont {DeWitt-Morette}}\ and\ \bibinfo {editor} {\bibfnamefont {B.~S.}\ \bibnamefont {DeWitt}}\ (\bibinfo  {publisher} {Gordon and Breach Science Publishers},\ \bibinfo {year} {1973})\BibitemShut {NoStop}%
\bibitem [{\citenamefont {Cardoso}\ and\ \citenamefont {Lemos}(2001)}]{Cardoso:2001bb}%
  \BibitemOpen
  \bibfield  {author} {\bibinfo {author} {\bibfnamefont {V.}~\bibnamefont {Cardoso}}\ and\ \bibinfo {author} {\bibfnamefont {J.~P.~S.}\ \bibnamefont {Lemos}},\ }\href {\doibase 10.1103/PhysRevD.64.084017} {\bibfield  {journal} {\bibinfo  {journal} {Phys. Rev. D}\ }\textbf {\bibinfo {volume} {64}},\ \bibinfo {pages} {084017} (\bibinfo {year} {2001})},\ \Eprint {http://arxiv.org/abs/gr-qc/0105103} {arXiv:gr-qc/0105103} \BibitemShut {NoStop}%
\bibitem [{\citenamefont {Lei}\ \emph {et~al.}(2021)\citenamefont {Lei}, \citenamefont {Wang},\ and\ \citenamefont {Jing}}]{Lei:2021kqv}%
  \BibitemOpen
  \bibfield  {author} {\bibinfo {author} {\bibfnamefont {Y.}~\bibnamefont {Lei}}, \bibinfo {author} {\bibfnamefont {M.}~\bibnamefont {Wang}}, \ and\ \bibinfo {author} {\bibfnamefont {J.}~\bibnamefont {Jing}},\ }\href {\doibase 10.1140/epjc/s10052-021-09942-8} {\bibfield  {journal} {\bibinfo  {journal} {Eur. Phys. J. C}\ }\textbf {\bibinfo {volume} {81}},\ \bibinfo {pages} {1129} (\bibinfo {year} {2021})},\ \Eprint {http://arxiv.org/abs/2108.04146} {arXiv:2108.04146 [gr-qc]} \BibitemShut {NoStop}%
\bibitem [{\citenamefont {Wang}\ \emph {et~al.}(2015)\citenamefont {Wang}, \citenamefont {Herdeiro},\ and\ \citenamefont {Sampaio}}]{Wang:2015goa}%
  \BibitemOpen
  \bibfield  {author} {\bibinfo {author} {\bibfnamefont {M.}~\bibnamefont {Wang}}, \bibinfo {author} {\bibfnamefont {C.}~\bibnamefont {Herdeiro}}, \ and\ \bibinfo {author} {\bibfnamefont {M.~O.~P.}\ \bibnamefont {Sampaio}},\ }\href {\doibase 10.1103/PhysRevD.92.124006} {\bibfield  {journal} {\bibinfo  {journal} {Phys. Rev. D}\ }\textbf {\bibinfo {volume} {92}},\ \bibinfo {pages} {124006} (\bibinfo {year} {2015})},\ \Eprint {http://arxiv.org/abs/1510.04713} {arXiv:1510.04713 [gr-qc]} \BibitemShut {NoStop}%
\bibitem [{\citenamefont {Brito}\ \emph {et~al.}(2015)\citenamefont {Brito}, \citenamefont {Cardoso},\ and\ \citenamefont {Pani}}]{Brito:2015oca}%
  \BibitemOpen
  \bibfield  {author} {\bibinfo {author} {\bibfnamefont {R.}~\bibnamefont {Brito}}, \bibinfo {author} {\bibfnamefont {V.}~\bibnamefont {Cardoso}}, \ and\ \bibinfo {author} {\bibfnamefont {P.}~\bibnamefont {Pani}},\ }\href {\doibase 10.1007/978-3-319-19000-6} {\bibfield  {journal} {\bibinfo  {journal} {Lect. Notes Phys.}\ }\textbf {\bibinfo {volume} {906}},\ \bibinfo {pages} {pp.1} (\bibinfo {year} {2015})},\ \Eprint {http://arxiv.org/abs/1501.06570} {arXiv:1501.06570 [gr-qc]} \BibitemShut {NoStop}%
\bibitem [{\citenamefont {Martel}\ and\ \citenamefont {Poisson}(2005)}]{Martel:2005ir}%
  \BibitemOpen
  \bibfield  {author} {\bibinfo {author} {\bibfnamefont {K.}~\bibnamefont {Martel}}\ and\ \bibinfo {author} {\bibfnamefont {E.}~\bibnamefont {Poisson}},\ }\href {\doibase 10.1103/PhysRevD.71.104003} {\bibfield  {journal} {\bibinfo  {journal} {Phys. Rev. D}\ }\textbf {\bibinfo {volume} {71}},\ \bibinfo {pages} {104003} (\bibinfo {year} {2005})},\ \Eprint {http://arxiv.org/abs/gr-qc/0502028} {arXiv:gr-qc/0502028} \BibitemShut {NoStop}%
\bibitem [{\citenamefont {Grozdanov}\ and\ \citenamefont {Vrbica}(2023)}]{Grozdanov:2023txs}%
  \BibitemOpen
  \bibfield  {author} {\bibinfo {author} {\bibfnamefont {S.}~\bibnamefont {Grozdanov}}\ and\ \bibinfo {author} {\bibfnamefont {M.}~\bibnamefont {Vrbica}},\ }\href {\doibase 10.1140/epjc/s10052-023-12273-5} {\bibfield  {journal} {\bibinfo  {journal} {Eur. Phys. J. C}\ }\textbf {\bibinfo {volume} {83}},\ \bibinfo {pages} {1103} (\bibinfo {year} {2023})},\ \Eprint {http://arxiv.org/abs/2303.15921} {arXiv:2303.15921 [hep-th]} \BibitemShut {NoStop}%
\bibitem [{\citenamefont {Grozdanov}\ and\ \citenamefont {Vrbica}(2024)}]{Grozdanov:2024wgo}%
  \BibitemOpen
  \bibfield  {author} {\bibinfo {author} {\bibfnamefont {S.}~\bibnamefont {Grozdanov}}\ and\ \bibinfo {author} {\bibfnamefont {M.}~\bibnamefont {Vrbica}},\ }\href {\doibase 10.1103/PhysRevLett.133.211601} {\bibfield  {journal} {\bibinfo  {journal} {Phys. Rev. Lett.}\ }\textbf {\bibinfo {volume} {133}},\ \bibinfo {pages} {211601} (\bibinfo {year} {2024})},\ \Eprint {http://arxiv.org/abs/2406.19790} {arXiv:2406.19790 [hep-th]} \BibitemShut {NoStop}%
\bibitem [{\citenamefont {Grozdanov}\ and\ \citenamefont {Vrbica}(2025)}]{Grozdanov:2025ner}%
  \BibitemOpen
  \bibfield  {author} {\bibinfo {author} {\bibfnamefont {S.}~\bibnamefont {Grozdanov}}\ and\ \bibinfo {author} {\bibfnamefont {M.}~\bibnamefont {Vrbica}},\ }\href {\doibase 10.1103/822n-ddzk} {\bibfield  {journal} {\bibinfo  {journal} {Phys. Rev. D}\ }\textbf {\bibinfo {volume} {112}},\ \bibinfo {pages} {066019} (\bibinfo {year} {2025})},\ \Eprint {http://arxiv.org/abs/2505.14229} {arXiv:2505.14229 [hep-th]} \BibitemShut {NoStop}%
\bibitem [{\citenamefont {Grozdanov}\ \emph {et~al.}(2018)\citenamefont {Grozdanov}, \citenamefont {Schalm},\ and\ \citenamefont {Scopelliti}}]{Grozdanov:2017ajz}%
  \BibitemOpen
  \bibfield  {author} {\bibinfo {author} {\bibfnamefont {S.}~\bibnamefont {Grozdanov}}, \bibinfo {author} {\bibfnamefont {K.}~\bibnamefont {Schalm}}, \ and\ \bibinfo {author} {\bibfnamefont {V.}~\bibnamefont {Scopelliti}},\ }\href {\doibase 10.1103/PhysRevLett.120.231601} {\bibfield  {journal} {\bibinfo  {journal} {Phys. Rev. Lett.}\ }\textbf {\bibinfo {volume} {120}},\ \bibinfo {pages} {231601} (\bibinfo {year} {2018})},\ \Eprint {http://arxiv.org/abs/1710.00921} {arXiv:1710.00921 [hep-th]} \BibitemShut {NoStop}%
\bibitem [{\citenamefont {Blake}\ \emph {et~al.}(2018{\natexlab{a}})\citenamefont {Blake}, \citenamefont {Lee},\ and\ \citenamefont {Liu}}]{Blake:2017ris}%
  \BibitemOpen
  \bibfield  {author} {\bibinfo {author} {\bibfnamefont {M.}~\bibnamefont {Blake}}, \bibinfo {author} {\bibfnamefont {H.}~\bibnamefont {Lee}}, \ and\ \bibinfo {author} {\bibfnamefont {H.}~\bibnamefont {Liu}},\ }\href {\doibase 10.1007/JHEP10(2018)127} {\bibfield  {journal} {\bibinfo  {journal} {JHEP}\ }\textbf {\bibinfo {volume} {10}},\ \bibinfo {pages} {127} (\bibinfo {year} {2018}{\natexlab{a}})},\ \Eprint {http://arxiv.org/abs/1801.00010} {arXiv:1801.00010 [hep-th]} \BibitemShut {NoStop}%
\bibitem [{\citenamefont {Blake}\ \emph {et~al.}(2018{\natexlab{b}})\citenamefont {Blake}, \citenamefont {Davison}, \citenamefont {Grozdanov},\ and\ \citenamefont {Liu}}]{Blake:2018leo}%
  \BibitemOpen
  \bibfield  {author} {\bibinfo {author} {\bibfnamefont {M.}~\bibnamefont {Blake}}, \bibinfo {author} {\bibfnamefont {R.~A.}\ \bibnamefont {Davison}}, \bibinfo {author} {\bibfnamefont {S.}~\bibnamefont {Grozdanov}}, \ and\ \bibinfo {author} {\bibfnamefont {H.}~\bibnamefont {Liu}},\ }\href {\doibase 10.1007/JHEP10(2018)035} {\bibfield  {journal} {\bibinfo  {journal} {JHEP}\ }\textbf {\bibinfo {volume} {10}},\ \bibinfo {pages} {035} (\bibinfo {year} {2018}{\natexlab{b}})},\ \Eprint {http://arxiv.org/abs/1809.01169} {arXiv:1809.01169 [hep-th]} \BibitemShut {NoStop}%
\bibitem [{\citenamefont {Grozdanov}(2019)}]{Grozdanov:2018kkt}%
  \BibitemOpen
  \bibfield  {author} {\bibinfo {author} {\bibfnamefont {S.}~\bibnamefont {Grozdanov}},\ }\href {\doibase 10.1007/JHEP01(2019)048} {\bibfield  {journal} {\bibinfo  {journal} {JHEP}\ }\textbf {\bibinfo {volume} {01}},\ \bibinfo {pages} {048} (\bibinfo {year} {2019})},\ \Eprint {http://arxiv.org/abs/1811.09641} {arXiv:1811.09641 [hep-th]} \BibitemShut {NoStop}%
\bibitem [{\citenamefont {Grozdanov}(2021)}]{Grozdanov:2020koi}%
  \BibitemOpen
  \bibfield  {author} {\bibinfo {author} {\bibfnamefont {S.}~\bibnamefont {Grozdanov}},\ }\href {\doibase 10.1103/PhysRevLett.126.051601} {\bibfield  {journal} {\bibinfo  {journal} {Phys. Rev. Lett.}\ }\textbf {\bibinfo {volume} {126}},\ \bibinfo {pages} {051601} (\bibinfo {year} {2021})},\ \Eprint {http://arxiv.org/abs/2008.00888} {arXiv:2008.00888 [hep-th]} \BibitemShut {NoStop}%
\bibitem [{\citenamefont {Grozdanov}\ \emph {et~al.}(2023)\citenamefont {Grozdanov}, \citenamefont {Lemut},\ and\ \citenamefont {Pedraza}}]{Grozdanov:2023tag}%
  \BibitemOpen
  \bibfield  {author} {\bibinfo {author} {\bibfnamefont {S.}~\bibnamefont {Grozdanov}}, \bibinfo {author} {\bibfnamefont {T.}~\bibnamefont {Lemut}}, \ and\ \bibinfo {author} {\bibfnamefont {J.~F.}\ \bibnamefont {Pedraza}},\ }\href {\doibase 10.1103/PhysRevD.108.L101901} {\bibfield  {journal} {\bibinfo  {journal} {Phys. Rev. D}\ }\textbf {\bibinfo {volume} {108}},\ \bibinfo {pages} {L101901} (\bibinfo {year} {2023})},\ \Eprint {http://arxiv.org/abs/2308.01371} {arXiv:2308.01371 [hep-th]} \BibitemShut {NoStop}%
\bibitem [{\citenamefont {Ahn}\ \emph {et~al.}(2025)\citenamefont {Ahn}, \citenamefont {Grozdanov}, \citenamefont {Jeong},\ and\ \citenamefont {Pedraza}}]{Ahn:2025exp}%
  \BibitemOpen
  \bibfield  {author} {\bibinfo {author} {\bibfnamefont {Y.}~\bibnamefont {Ahn}}, \bibinfo {author} {\bibfnamefont {S.}~\bibnamefont {Grozdanov}}, \bibinfo {author} {\bibfnamefont {H.-S.}\ \bibnamefont {Jeong}}, \ and\ \bibinfo {author} {\bibfnamefont {J.~F.}\ \bibnamefont {Pedraza}},\ }\href@noop {} {\  (\bibinfo {year} {2025})},\ \Eprint {http://arxiv.org/abs/2508.15589} {arXiv:2508.15589 [hep-th]} \BibitemShut {NoStop}%
\bibitem [{\citenamefont {Bhattacharya}\ \emph {et~al.}(2025)\citenamefont {Bhattacharya}, \citenamefont {Padhi}, \citenamefont {Sharma},\ and\ \citenamefont {Singha}}]{Bhattacharya:2025vyi}%
  \BibitemOpen
  \bibfield  {author} {\bibinfo {author} {\bibfnamefont {J.}~\bibnamefont {Bhattacharya}}, \bibinfo {author} {\bibfnamefont {N.}~\bibnamefont {Padhi}}, \bibinfo {author} {\bibfnamefont {A.}~\bibnamefont {Sharma}}, \ and\ \bibinfo {author} {\bibfnamefont {S.}~\bibnamefont {Singha}},\ }\href {\doibase 10.1007/JHEP08(2025)170} {\bibfield  {journal} {\bibinfo  {journal} {JHEP}\ }\textbf {\bibinfo {volume} {08}},\ \bibinfo {pages} {170} (\bibinfo {year} {2025})},\ \Eprint {http://arxiv.org/abs/2504.17781} {arXiv:2504.17781 [hep-th]} \BibitemShut {NoStop}%
\bibitem [{\citenamefont {Shahbazi-Moghaddam}(2025)}]{Shahbazi-Moghaddam:2024emr}%
  \BibitemOpen
  \bibfield  {author} {\bibinfo {author} {\bibfnamefont {A.}~\bibnamefont {Shahbazi-Moghaddam}},\ }\href {\doibase 10.1007/JHEP07(2025)218} {\bibfield  {journal} {\bibinfo  {journal} {JHEP}\ }\textbf {\bibinfo {volume} {07}},\ \bibinfo {pages} {218} (\bibinfo {year} {2025})},\ \Eprint {http://arxiv.org/abs/2411.11948} {arXiv:2411.11948 [hep-th]} \BibitemShut {NoStop}%
\bibitem [{\citenamefont {Banihashemi}\ \emph {et~al.}(2025)\citenamefont {Banihashemi}, \citenamefont {Shaghoulian},\ and\ \citenamefont {Shashi}}]{Banihashemi:2025qqi}%
  \BibitemOpen
  \bibfield  {author} {\bibinfo {author} {\bibfnamefont {B.}~\bibnamefont {Banihashemi}}, \bibinfo {author} {\bibfnamefont {E.}~\bibnamefont {Shaghoulian}}, \ and\ \bibinfo {author} {\bibfnamefont {S.}~\bibnamefont {Shashi}},\ }\href {\doibase 10.1088/1361-6382/adee72} {\bibfield  {journal} {\bibinfo  {journal} {Class. Quant. Grav.}\ }\textbf {\bibinfo {volume} {42}},\ \bibinfo {pages} {155004} (\bibinfo {year} {2025})},\ \Eprint {http://arxiv.org/abs/2503.17471} {arXiv:2503.17471 [hep-th]} \BibitemShut {NoStop}%
\bibitem [{\citenamefont {De~Clerck}\ \emph {et~al.}(2024)\citenamefont {De~Clerck}, \citenamefont {Hartnoll},\ and\ \citenamefont {Santos}}]{DeClerck:2023fax}%
  \BibitemOpen
  \bibfield  {author} {\bibinfo {author} {\bibfnamefont {M.}~\bibnamefont {De~Clerck}}, \bibinfo {author} {\bibfnamefont {S.~A.}\ \bibnamefont {Hartnoll}}, \ and\ \bibinfo {author} {\bibfnamefont {J.~E.}\ \bibnamefont {Santos}},\ }\href {\doibase 10.1007/JHEP07(2024)202} {\bibfield  {journal} {\bibinfo  {journal} {JHEP}\ }\textbf {\bibinfo {volume} {07}},\ \bibinfo {pages} {202} (\bibinfo {year} {2024})},\ \Eprint {http://arxiv.org/abs/2312.11622} {arXiv:2312.11622 [hep-th]} \BibitemShut {NoStop}%
\bibitem [{\citenamefont {Grozdanov}\ \emph {et~al.}(2026{\natexlab{b}})\citenamefont {Grozdanov}, \citenamefont {Movrin},\ and\ \citenamefont {Valach}}]{futurepaper2}%
  \BibitemOpen
  \bibfield  {author} {\bibinfo {author} {\bibfnamefont {S.}~\bibnamefont {Grozdanov}}, \bibinfo {author} {\bibfnamefont {V.}~\bibnamefont {Movrin}}, \ and\ \bibinfo {author} {\bibfnamefont {S.}~\bibnamefont {Valach}},\ }\href@noop {} {\  (\bibinfo {year} {2026}{\natexlab{b}})},\ \Eprint {http://arxiv.org/abs/2608.13643} {arXiv:2608.13643 [hep-th]} \BibitemShut {NoStop}%
\bibitem [{\citenamefont {Aguilar-Gutierrez}\ \emph {et~al.}(2024)\citenamefont {Aguilar-Gutierrez}, \citenamefont {Baiguera},\ and\ \citenamefont {Zenoni}}]{Aguilar-Gutierrez:2024rka}%
  \BibitemOpen
  \bibfield  {author} {\bibinfo {author} {\bibfnamefont {S.~E.}\ \bibnamefont {Aguilar-Gutierrez}}, \bibinfo {author} {\bibfnamefont {S.}~\bibnamefont {Baiguera}}, \ and\ \bibinfo {author} {\bibfnamefont {N.}~\bibnamefont {Zenoni}},\ }\href {\doibase 10.1007/JHEP05(2024)201} {\bibfield  {journal} {\bibinfo  {journal} {JHEP}\ }\textbf {\bibinfo {volume} {05}},\ \bibinfo {pages} {201} (\bibinfo {year} {2024})},\ \Eprint {http://arxiv.org/abs/2402.01357} {arXiv:2402.01357 [hep-th]} \BibitemShut {NoStop}%
\bibitem [{\citenamefont {Hörmander}(2007)}]{Hormander:2007}%
  \BibitemOpen
  \bibfield  {author} {\bibinfo {author} {\bibfnamefont {L.}~\bibnamefont {Hörmander}},\ }\href {\doibase https://doi.org/10.1007/978-3-540-49938-1} {\emph {\bibinfo {title} {{The Analysis of Linear Partial Differential Operators III}}}},\ Classics in Mathematics\ (\bibinfo  {publisher} {Springer Berlin, Heidelberg},\ \bibinfo {year} {2007})\BibitemShut {NoStop}%
\bibitem [{\citenamefont {Hörmander}(2009)}]{Hormander:2009}%
  \BibitemOpen
  \bibfield  {author} {\bibinfo {author} {\bibfnamefont {L.}~\bibnamefont {Hörmander}},\ }\href {\doibase https://doi.org/10.1007/978-3-642-00136-9} {\emph {\bibinfo {title} {{The Analysis of Linear Partial Differential Operators IV}}}},\ Classics in Mathematics\ (\bibinfo  {publisher} {Springer Berlin, Heidelberg},\ \bibinfo {year} {2009})\BibitemShut {NoStop}%
\bibitem [{\citenamefont {Reed}\ and\ \citenamefont {Simon}(1975)}]{Reed:1975uy}%
  \BibitemOpen
  \bibfield  {author} {\bibinfo {author} {\bibfnamefont {M.}~\bibnamefont {Reed}}\ and\ \bibinfo {author} {\bibfnamefont {B.}~\bibnamefont {Simon}},\ }\href@noop {} {\emph {\bibinfo {title} {{Methods of Modern Mathematical Physics. 2. Fourier Analysis, Self-adjointness}}}}\ (\bibinfo  {publisher} {Academic Press, Inc.},\ \bibinfo {year} {1975})\BibitemShut {NoStop}%
\end{thebibliography}%
\bibliographystyle{apsrev4-1}

\end{document}